\pdfoutput=1
\documentclass[12pt,a4paper]{article}
\usepackage{ifthen} % for conditional statements
\newboolean{pdflatex}
\setboolean{pdflatex}{true} % False for eps figures 

\newboolean{articletitles}
\setboolean{articletitles}{true} % False removes titles in references

\newboolean{uprightparticles}
\setboolean{uprightparticles}{false} %True for upright particle symbols

\def\paperauthors{LHCb collaboration} % Leave as is for PAPER, CONF and FIGURE
\def\paperasciititle{Measurement of the Bs->mumu decay properties and search for the B0->mumu and Bs->mumugamma decays} % Set ASCII title here !! MAKE sure it's only ASCII characters !! 
\def\papertitle{Measurement\\ of the $B^0_s\to\mu^+\mu^-$ decay properties\\ and search for the $B^0\to\mu^+\mu^-$\\ and $B^0_s\to\mu^+\mu^-\gamma$ decays} % Latex formatted title
\def\paperkeywords{{High Energy Physics}, {LHCb}, {Rare decays}} % Comma separated list
\def\papercopyright{\the\year\ CERN for the benefit of the LHCb collaboration} % new since 9/Apr/2018
\def\paperlicence{CC BY 4.0 licence}
\def\paperlicenceurl{https://creativecommons.org/licenses/by/4.0/}

\usepackage[top=1in, bottom=1.25in, left=1in, right=1in]{geometry}

\usepackage{microtype}
\usepackage{lineno}  % for line numbering during review
\usepackage{xspace} % To avoid problems with missing or double spaces after
\usepackage{caption} %these three command get the figure and table captions automatically small

\usepackage{graphicx}  % to include figures (can also use other packages)
\usepackage{color}
\usepackage{colortbl}
\graphicspath{{./figs/}} % Make Latex search fig subdir for figures
\usepackage{amsmath} % Adds a large collection of math symbols
\usepackage{amssymb}
\usepackage{amsfonts}
\usepackage{upgreek} % Adds in support for greek letters in roman typeset
\usepackage{subfigure}
\usepackage{booktabs}
\usepackage{lscape}

\newcommand*\patchAmsMathEnvironmentForLineno[1]{%
\expandafter\let\csname old#1\expandafter\endcsname\csname #1\endcsname
\expandafter\let\csname oldend#1\expandafter\endcsname\csname
end#1\endcsname
 \renewenvironment{#1}%
   {\linenomath\csname old#1\endcsname}%
   {\csname oldend#1\endcsname\endlinenomath}%
}
\newcommand*\patchBothAmsMathEnvironmentsForLineno[1]{%
  \patchAmsMathEnvironmentForLineno{#1}%
  \patchAmsMathEnvironmentForLineno{#1*}%
}
\AtBeginDocument{%
\patchBothAmsMathEnvironmentsForLineno{equation}%
\patchBothAmsMathEnvironmentsForLineno{align}%
\patchBothAmsMathEnvironmentsForLineno{flalign}%
\patchBothAmsMathEnvironmentsForLineno{alignat}%
\patchBothAmsMathEnvironmentsForLineno{gather}%
\patchBothAmsMathEnvironmentsForLineno{multline}%
\patchBothAmsMathEnvironmentsForLineno{eqnarray}%
}
\usepackage{feynmf}

\usepackage{tikz}
\usetikzlibrary{decorations.pathreplacing,decorations.pathmorphing,
decorations.markings,trees,calc,patterns}
\xdefinecolor{lhcbblue}{RGB}{0,86,166}
\definecolor{webbrown}{rgb}{.6,0,0}

\tikzset{
photon/.style={decorate, decoration={snake}, draw=webbrown},
higgs/.style={decorate, dashed, draw=webbrown},
particle/.style={draw=lhcbblue, postaction={decorate},decoration={markings,mark=at 
position .5 with {\arrow[draw=lhcbblue]{>}}}},
antiparticle/.style={draw=lhcbblue, 
postaction={decorate},decoration={markings,mark=at position .5 with 
{\arrow[draw=lhcbblue]{<}}}}, 
gluon/.style={decorate, draw=black,decoration={snake,amplitude=4pt, segment 
length=5pt}}, 
majorana/.style={draw=black, postaction={decorate},decoration={markings,mark=at 
position .48 with {\arrow[draw=black]{>}},mark=at position .52 with 
{\arrow[draw=black]{<}}}},
gluonloop/.style={circle, decorate, draw=black, 
decoration={coil,aspect=1.2,amplitude=2pt, segment length=4pt},minimum 
height=1.2em},
}

\usepackage{hyperxmp}

\usepackage[pdftex,
            pdfauthor={\paperauthors},
            pdftitle={\paperasciititle},
            pdfkeywords={\paperkeywords},
            pdfcopyright={Copyright (C) \papercopyright},
            pdflicenseurl={\paperlicenceurl}]{hyperref}
\usepackage[colorinlistoftodos,textsize=scriptsize]{todonotes}

\usepackage[bottom,flushmargin,hang,multiple]{footmisc}

\usepackage[all]{hypcap} % Internal hyperlinks to floats.

\usepackage{xspace} 
\usepackage{upgreek}

\def\lhcb   {\mbox{LHCb}\xspace}
\def\atlas  {\mbox{ATLAS}\xspace}
\def\cms    {\mbox{CMS}\xspace}

\def\babar  {\mbox{BaBar}\xspace}

\def\cleo   {\mbox{CLEO}\xspace}

\def\MagUp {\mbox{\em Mag\kern -0.05em Up}\xspace}

\ifthenelse{\boolean{uprightparticles}}%
{
 
 \def\Pgamma      {\ensuremath{\upgamma}\xspace}

 \def\Pmu         {\ensuremath{\upmu}\xspace}                 
 \def\Pnu         {\ensuremath{\upnu}\xspace}                 
                  
 \def\Ppi         {\ensuremath{\uppi}\xspace}

 \def\Pphi        {\ensuremath{\upphi}\xspace}

 \def\Ppsi        {\ensuremath{\uppsi}\xspace}

 \def\PDelta      {\ensuremath{\Delta}\xspace}                 
 \def\PXi         {\ensuremath{\Xi}\xspace}                 
 \def\PLambda     {\ensuremath{\Lambda}\xspace}                 
 \def\PSigma      {\ensuremath{\Sigma}\xspace}                 
 \def\POmega      {\ensuremath{\Omega}\xspace}                 
 \def\PUpsilon    {\ensuremath{\Upsilon}\xspace}

 \def\PB      {\ensuremath{\mathrm{B}}\xspace}                 
                  
 \def\PD      {\ensuremath{\mathrm{D}}\xspace}

 \def\PJ      {\ensuremath{\mathrm{J}}\xspace}                 
 \def\PK      {\ensuremath{\mathrm{K}}\xspace}

 \def\Pb      {\ensuremath{\mathrm{b}}\xspace}                 
 \def\Pc      {\ensuremath{\mathrm{c}}\xspace}

 \def\Pi      {\ensuremath{\mathrm{i}}\xspace}

 \def\Pp      {\ensuremath{\mathrm{p}}\xspace}

 \def\Ps      {\ensuremath{\mathrm{s}}\xspace}

 \def\thebaroffset{0.0em}
}
{
 
 \def\Pgamma      {\ensuremath{\gamma}\xspace}

 \def\Pmu         {\ensuremath{\mu}\xspace}                 
 \def\Pnu         {\ensuremath{\nu}\xspace}                 
                  
 \def\Ppi         {\ensuremath{\pi}\xspace}

 \def\Pphi        {\ensuremath{\phi}\xspace}

 \def\Ppsi        {\ensuremath{\psi}\xspace}                 
                  
 \mathchardef\PDelta="7101
 \mathchardef\PXi="7104
 \mathchardef\PLambda="7103
 \mathchardef\PSigma="7106
 \mathchardef\POmega="710A
 \mathchardef\PUpsilon="7107
                  
 \def\PB      {\ensuremath{B}\xspace}                 
                  
 \def\PD      {\ensuremath{D}\xspace}

 \def\PJ      {\ensuremath{J}\xspace}                 
 \def\PK      {\ensuremath{K}\xspace}

 \def\Pb      {\ensuremath{b}\xspace}                 
 \def\Pc      {\ensuremath{c}\xspace}

 \def\Pi      {\ensuremath{i}\xspace}

 \def\Pp      {\ensuremath{p}\xspace}

 \def\Ps      {\ensuremath{s}\xspace}

 \def\thebaroffset{0.18em}
}
\newcommand{\offsetoverline}[2][\thebaroffset]{\kern #1\overline{\kern -#1 #2}}%

\makeatletter
\ifcase \@ptsize \relax% 10pt
  \newcommand{\miniscule}{\@setfontsize\miniscule{4}{5}}% \tiny: 5/6
\or% 11pt
  \newcommand{\miniscule}{\@setfontsize\miniscule{5}{6}}% \tiny: 6/7
\or% 12pt
  \newcommand{\miniscule}{\@setfontsize\miniscule{5}{6}}% \tiny: 6/7
\fi
\makeatother

\DeclareRobustCommand{\optbar}[1]{\shortstack{{\miniscule (\rule[.5ex]{1.25em}{.18mm})}
  \\ [-.7ex] $#1$}}

\def\mup        {{\ensuremath{\Pmu^+}}\xspace}
\def\mun        {{\ensuremath{\Pmu^-}}\xspace} % muon negative (\mum is taken)

\def\mumu       {{\ensuremath{\Pmu^+\Pmu^-}}\xspace}

\def\neu        {{\ensuremath{\Pnu}}\xspace}
\def\neub       {{\ensuremath{\overline{\Pnu}}}\xspace}
\def\neum       {{\ensuremath{\neu_\mu}}\xspace}
\def\neumb      {{\ensuremath{\neub_\mu}}\xspace}
\def\squark    {{\ensuremath{\Ps}}\xspace}

\def\cquark    {{\ensuremath{\Pc}}\xspace}

\def\bquark    {{\ensuremath{\Pb}}\xspace}

\def\pion   {{\ensuremath{\Ppi}}\xspace}

\def\pip    {{\ensuremath{\pion^+}}\xspace}
\def\pim    {{\ensuremath{\pion^-}}\xspace}

\def\kaon    {{\ensuremath{\PK}}\xspace}
\def\KorKbar {\kern \thebaroffset\optbar{\kern -\thebaroffset \PK}{}\xspace}

\def\Kp      {{\ensuremath{\kaon^+}}\xspace}
\def\Km      {{\ensuremath{\kaon^-}}\xspace}

\def\D       {{\ensuremath{\PD}}\xspace}

\def\DorDbar {\kern \thebaroffset\optbar{\kern -\thebaroffset \PD}\xspace}
\def\Dz      {{\ensuremath{\D^0}}\xspace}

\def\Dp      {{\ensuremath{\D^+}}\xspace}
\def\Dm      {{\ensuremath{\D^-}}\xspace}

\def\DpDm    {\ensuremath{\Dp {\kern -0.16em \Dm}}\xspace}

\def\Dstarp  {{\ensuremath{\D^{*+}}}\xspace}

\def\B       {{\ensuremath{\PB}}\xspace}
\def\Bbar    {{\ensuremath{\offsetoverline{\PB}}}\xspace}

\def\BorBbar {\kern \thebaroffset\optbar{\kern -\thebaroffset \PB}\xspace}

\def\Bd      {{\ensuremath{\B^0}}\xspace}

\def\BdorBdbar {\kern \thebaroffset\optbar{\kern -\thebaroffset \Bd}\xspace}
\def\Bu      {{\ensuremath{\B^+}}\xspace}

\def\Bs      {{\ensuremath{\B^0_\squark}}\xspace}
\def\Bsb     {{\ensuremath{\Bbar{}^0_\squark}}\xspace}
\def\BsorBsbar {\kern \thebaroffset\optbar{\kern -\thebaroffset \Bs}\xspace}
\def\Bc      {{\ensuremath{\B_\cquark^+}}\xspace}

\def\Bds     {{\ensuremath{\B_{(\squark)}^0}}\xspace}
\def\Bdsb    {{\ensuremath{\Bbar{}_{(\squark)}^0}}\xspace}

\def\jpsi     {{\ensuremath{{\PJ\mskip -3mu/\mskip -2mu\Ppsi}}}\xspace}
\def\psitwos  {{\ensuremath{\Ppsi{(2S)}}}\xspace}

\def\Upsilonres  {{\ensuremath{\PUpsilon}}\xspace}
\def\Y#1S{\ensuremath{\Upsilonres{(#1S)}}\xspace}
\def\OneS  {{\Y1S}}
\def\TwoS  {{\Y2S}}
\def\ThreeS{{\Y3S}}

\def\proton      {{\ensuremath{\Pp}}\xspace}

\def\Lz          {{\ensuremath{\PLambda}}\xspace}

\def\LorLbar     {\kern \thebaroffset\optbar{\kern -\thebaroffset \PLambda}\xspace}

\def\Lc          {{\ensuremath{\Lz^+_\cquark}}\xspace}

\def\Lb           {{\ensuremath{\Lz^0_\bquark}}\xspace}

\def\BF         {{\ensuremath{\mathcal{B}}}\xspace}

\newcommand{\decay}[2]{\ensuremath{#1\!\to #2}\xspace} 

\def\to                 {\ensuremath{\rightarrow}\xspace}

\newcommand{\tauBs}{{\ensuremath{\tau_{\Bs}}}\xspace}

\def\CP                {{\ensuremath{C\!P}}\xspace}

\def\AT#1     {\ensuremath{A_{\mathrm{T}}^{#1}}\xspace}           % 2

\def\Bsmm     {\decay{\Bs}{\mup\mun}}
\def\Bdmm     {\decay{\Bd}{\mup\mun}}

\def\C#1      {\ensuremath{\mathcal{C}_{#1}}\xspace}                       % 9
\def\Cp#1     {\ensuremath{\mathcal{C}_{#1}^{'}}\xspace}                    % 7
\def\Ceff#1   {\ensuremath{\mathcal{C}_{#1}^{\mathrm{(eff)}}}\xspace}        % 9  
\def\Cpeff#1  {\ensuremath{\mathcal{C}_{#1}^{'\mathrm{(eff)}}}\xspace}       % 7
\def\Ope#1    {\ensuremath{\mathcal{O}_{#1}}\xspace}                       % 2
\def\Opep#1   {\ensuremath{\mathcal{O}_{#1}^{'}}\xspace}                    % 7

\newcommand{\nospaceunit}[1]{\ensuremath{\text{#1}}}       
\newcommand{\aunit}[1]{\ensuremath{\text{\,#1}}}       
\newcommand{\tev}{\aunit{Te\kern -0.1em V}\xspace}
\newcommand{\gev}{\aunit{Ge\kern -0.1em V}\xspace}
\newcommand{\mev}{\aunit{Me\kern -0.1em V}\xspace}
\newcommand{\kev}{\aunit{ke\kern -0.1em V}\xspace}
\newcommand{\ev}{\aunit{e\kern -0.1em V}\xspace}
 
\newcommand{\mevc}{\ensuremath{\aunit{Me\kern -0.1em V\!/}c}\xspace}
\newcommand{\gevc}{\ensuremath{\aunit{Ge\kern -0.1em V\!/}c}\xspace}
\newcommand{\mevcc}{\ensuremath{\aunit{Me\kern -0.1em V\!/}c^2}\xspace}
\newcommand{\gevcc}{\ensuremath{\aunit{Ge\kern -0.1em V\!/}c^2}\xspace}
\def\mum  {\ensuremath{\,\upmu\nospaceunit{m}}\xspace}

\def\fb   {\ensuremath{\aunit{fb}}\xspace}
\def\invfb   {\ensuremath{\fb^{-1}}\xspace}

\def\ps   {\ensuremath{\aunit{ps}}\xspace}

\newcommand{\stat}{\aunit{(stat)}\xspace}

\newcommand{\chisq}{\ensuremath{\chi^2}\xspace}

\newcommand{\chisqip}{\ensuremath{\chi^2_{\text{IP}}}\xspace}

\def\gsim{{~\raise.15em\hbox{$>$}\kern-.85em
          \lower.35em\hbox{$\sim$}~}\xspace}
\def\lsim{{~\raise.15em\hbox{$<$}\kern-.85em
          \lower.35em\hbox{$\sim$}~}\xspace}

\def\sPlot{\mbox{\em sPlot}\xspace}

\def\pt         {\ensuremath{p_{\mathrm{T}}}\xspace}

\def\ptot       {\ensuremath{p}\xspace}

\def\bcvegpy    {\mbox{\textsc{Bcvegpy}}\xspace}

\def\evtgen     {\mbox{\textsc{EvtGen}}\xspace}

\def\geant      {\mbox{\textsc{Geant4}}\xspace}

\def\photos     {\mbox{\textsc{Photos}}\xspace}

\def\pythia     {\mbox{\textsc{Pythia}}\xspace}

\def\tell1  {TELL1\xspace}
\def\ukl1   {UKL1\xspace}

\def\Bsbr{\ensuremath{\left(3.09^{\,+\,0.46\,+\,0.15}_{\,-\,0.43\,-\,0.11}\right)\times 10^{-9}}\xspace}
\def\Bdbr{\ensuremath{\left(1.20^{\,+\,0.83}_{\,-\,0.74}\pm 0.14\right)\times 10^{-10}}\xspace}
\def\Bsmmgbr{\ensuremath{\left(-2.5\pm1.4\pm0.8\right)\times 10^{-9}}\xspace}
\def\ratio{\ensuremath{0.039^{\,+\,0.030\,+\,0.006}_{\,-\,0.024\,-\,0.004}}\xspace}
\def\Bstau{\ensuremath{(2.07 \pm 0.29\pm 0.03)\,\ps}\xspace}
\def\Bdobslimitnf{\ensuremath{2.6\times 10^{-10}}\xspace} %  95% CL 
\def\Bsmmgobslimitnf{\ensuremath{2.0\times 10^{-9}}\xspace} %  95% CL 
\def\Bdobslimit{\ensuremath{2.3\,(2.6)\times10^{-10}}\xspace}
\def\Bsmmgobslimit{\ensuremath{1.5\,(2.0)\times10^{-9}}\xspace}
\def\ratiolimitnf{\ensuremath{0.095}\xspace} % 95%CL
\def\ratiolimit{\ensuremath{0.081\,(0.095)}\xspace} %  90%(95%) CL 

\def\Bdsigma{\ensuremath{1.7}\xspace}
\def\Bssigma{\ensuremath{10}\xspace}
\def\Bsmumugammasigma{\ensuremath{1.5}\xspace}

\def\ratioSM{\ensuremath{0.0281\pm 0.0016 \xspace}}

\def\Bsbrcomb{\ensuremath{\left(2.69 ^{\,+\,0.37}_{\,-\,0.35}\right)\times 10^{-9}}\xspace}
\def\Bdlimcomb{\ensuremath{1.9 \times 10^{-10}}\xspace}

\newcommand\runone{Run~1\xspace}
\newcommand\runtwo{Run~2\xspace}

\newcommand{\fsfd}{\ensuremath{f_s/f_d}\xspace}
\newcommand\qsquare{\ensuremath{q^2}\xspace}

\def\BDT{BDT\xspace}
\newcommand{\comment}[1]{}

\newcommand{\BRof}[1]{\ensuremath{{\cal B}(#1)}\xspace}

\def\BF         {{\ensuremath{\cal B}\xspace}}

\def\ys{\ensuremath{y_s}\xspace}
\def\ADeltaGamma{\ensuremath{A^{\mu\mu}_{\Delta\Gamma_s}}\xspace}

\newcommand{\tauBsmm}{\ensuremath{\tau_{\mup\mun}\xspace}}
\newcommand{\taugen}{\ensuremath{\tau_{\text{gen}}\xspace}}
\newcommand{\omegaj}{\ensuremath{\omega_{j}\xspace}}

\def\fBDT{\ensuremath{f_{\text{\BDT},i}}\xspace}
\def\fBDTmumuSim {\ensuremath{f_{\text{sim},i}^{\mu\mu}}\xspace}
\def\fBDTmumuPID {\ensuremath{f_{\text{PID},i}^{\mu\mu}}\xspace}
\def\fBDTmumutrig{\ensuremath{f_{\text{trig},i}^{\mu\mu}}\xspace}
\def\kBDTtau     {\ensuremath{k_{i}}\xspace}

\def\fBDThh{\ensuremath{f_{\text{\BDT},i}^\prime}\xspace}
\def\fBDThhData{\ensuremath{f_{\text{data},i}^{hh^\prime}}\xspace}
\def\fBDThhPID {\ensuremath{f_{\text{PID},i}^{hh^\prime}}\xspace}
\def\fBDThhtrig{\ensuremath{f_{\text{trig},i}^{hh^\prime}}\xspace}

\def\bdt {\ensuremath{\mbox{BDT}}}

\newcommand{\CLs}{\ensuremath{\textrm{CL}_{\textrm{s}}}\xspace}

\newcommand{\Bsmumu}{\decay{\Bs}{\mup \mun}}
\newcommand{\Bsbmumu}{\decay{\Bsb}{\mup \mun}}
\newcommand{\Bsmumutime}{\decay{\Bs(t)}{\mup \mun}}
\newcommand{\Bsbmumutime}{\decay{\Bsb(t)}{\mup \mun}}
\newcommand{\BsorBsbmumutime}{\decay{\B_s(t)}{\mup \mun}}
\newcommand{\Bdmumu}{\decay{\Bd}{\mup \mun}}
\newcommand{\Bsmumugamma}{\decay{\Bs}{\mup \mun \Pgamma}}
\newcommand{\Bdmumugamma}{\decay{\Bd}{\mup \mun \Pgamma}}
\newcommand{\Bdsmumugamma}{\decay{\Bds}{\mup \mun (\Pgamma)}}
\newcommand{\Bsmumug}{\decay{\Bs}{\mup \mun (\Pgamma)}}

\newcommand{\Bdsmumu}{\decay{\Bds}{\mup \mun}}

\newcommand{\BsKpi}{\decay{\Bs}{\Km \pip}}
\newcommand{\BsbarKpi}{\decay{\Bsb}{\Kp \pim }}

\newcommand{\BdKpi}{\decay{\Bd}{\Kp \pim }}

\newcommand{\Bpimumu}{\decay{\B^{0(+)}}{\pi^{0(+)} \mup \mun}}
\newcommand{\Bupimumu}{\decay{\B^+}{\pi^+ \mup \mun}}
\newcommand{\Bdpimumu}{\decay{\B^0}{\pi^0 \mup \mun}}
\newcommand{\BdPiMuNu}{\decay{\Bd}{\pim \mup \neum}}
\newcommand{\BsKMuNu}{\decay{\Bs}{\Km \mup \neum}}

\newcommand{\BuJpsiK}{\decay{\Bu}{\jpsi \Kp}}
\newcommand{\BuJpsiPi}{\decay{\Bu}{\jpsi \pip}}
\newcommand{\Bhh}{\decay{\Bds}{h^+ h^{\prime -} }}

\newcommand{\BsJpsiPhi}{\decay{\Bs}{\jpsi \Pphi}}

\newcommand{\BsKK}{\decay{\Bs}{\Kp \Km}}

\newcommand{\Lbph}{\decay{\Lb}{\proton {h}^{-}}}
\newcommand{\Lbpmunu}{\decay{\Lb}{\Pp \mun \neumb}} 
\newcommand{\BcJpsiMuNu}{\decay{\Bc}{\jpsi \mup \neum}}
\newcommand{\HbhMuNu}{\decay{H_b}{h^{+} \mun \neumb}}

\newcommand{\bsmumu}{\Bsmumu}
\newcommand{\bsmumugamma}{\Bsmumugamma}
\newcommand{\bdmumugamma}{\Bdmumugamma}
\newcommand{\bsmumug}{\Bsmumugamma}
\newcommand{\bdmumu}{\Bdmumu} 
\newcommand{\bdsmumu}{\Bdsmumu} 

\newcommand{\bupimumu}{\Bupimumu}
\newcommand{\bdpimumu}{\Bdpimumu}
\newcommand{\bdpimunu}{\BdPiMuNu}
\newcommand{\bskmunu}{\BsKMuNu}
\newcommand{\bcjpsimunu}{\BcJpsiMuNu}
\newcommand{\lbpmunu}{\Lbpmunu}
\newcommand{\bujpsik}{\BuJpsiK} 
\newcommand{\bujpsipi}{\BuJpsiPi} 
\newcommand{\bhh}{\Bhh}

\newcommand{\bbdim}{\ensuremath{b\bar{b}\to \mu^+ \mu^- X}\xspace}

\newcommand{\bsjpsiphi}{\BsJpsiPhi} 
\newcommand{\Bmm}{\Bdsmumu}
\newcommand{\bmumu}{\Bmm}

\newcommand{\bdkpi}{\mbox{\BdKpi}}

\newcommand\bskk{\BsKK}

\newcommand\bskpi{\BsKpi}

\def\BorBbars    {\kern 0.18em\optbar{\kern -0.18em B}}

\def\Y#1S{\ensuremath{\PUpsilon{(#1S)}}\xspace}% no space before {...}!
\def\OneS  {\Y1S}
\def\TwoS  {\Y2S}
\def\ThreeS{\Y3S}
\def\psitwos  {\ensuremath{\psi{(2S)}}\xspace}

\newcommand{\mevccspecial}{\ensuremath{\aunit{Me\kern -0.1em V\!/}c^2}} % Add in our symbols 

\usepackage{cite} % Allows for ranges in citations
\usepackage{mciteplus}
\mciteErrorOnUnknownfalse
\definecolor{darkred}{rgb}{0.6,0.0,0.0}
\definecolor{darkgreen}{rgb}{0.0,0.5,0.0}
\definecolor{lightgreen}{rgb}{0.75,1.0,0.75}
\definecolor{lightred}{rgb}{1.0,0.75,0.75}
\definecolor{lightblue}{rgb}{0.75,0.75,1.0}
\definecolor{darkblue}{RGB}{100,100,200}
\definecolor{verylightblue}{rgb}{0.9,0.9,1.0}
\definecolor{verylightred}{rgb}{1.0,0.9,0.9}
\definecolor{lightgray}{rgb}{0.9,0.9,0.9}
\definecolor{verylightgray}{rgb}{0.95,0.95,0.95}
\definecolor{darkgray}{rgb}{0.75,0.75,0.75}
\definecolor{orange}{rgb}{1.0,0.75,0.0}

\usepackage{longtable} % only for template; not usually to be used in PAPERs
\usepackage{multirow}
\usepackage{afterpage}

\usepackage{xcolor}
\usepackage{placeins}
\usepackage{float}

\begin{document}

%%%%%%%%%%%%%%%%%%%%%%%%%
%%%%% Title     %%%%%%%%%
%%%%%%%%%%%%%%%%%%%%%%%%%
\renewcommand{\thefootnote}{\fnsymbol{footnote}}
\setcounter{footnote}{1}

% %%%%%%% CHOOSE TITLE PAGE--------
%\onecolumn
%\input{title-LHCb-INT}
%\input{title-LHCb-ANA}
%\input{title-LHCb-CONF}
%\input{title-LHCb-FIGURE}
% ===============================================================================
% Purpose: LHCb-PAPER journal paper title page template
% Author: 
% Created on: 2010-09-25
% ===============================================================================

%%%%%%%%%%%%%%%%%%%%%%%%%
%%%%%  TITLE PAGE  %%%%%%
%%%%%%%%%%%%%%%%%%%%%%%%%
\begin{titlepage}
\pagenumbering{roman}

% Header ---------------------------------------------------
\vspace*{-1.5cm}
\centerline{\large EUROPEAN ORGANIZATION FOR NUCLEAR RESEARCH (CERN)}
\vspace*{1.5cm}
\noindent
\begin{tabular*}{\linewidth}{lc@{\extracolsep{\fill}}r@{\extracolsep{0pt}}}
\ifthenelse{\boolean{pdflatex}}% Logo format choice
{\vspace*{-1.5cm}\mbox{\!\!\!\includegraphics[width=.14\textwidth]{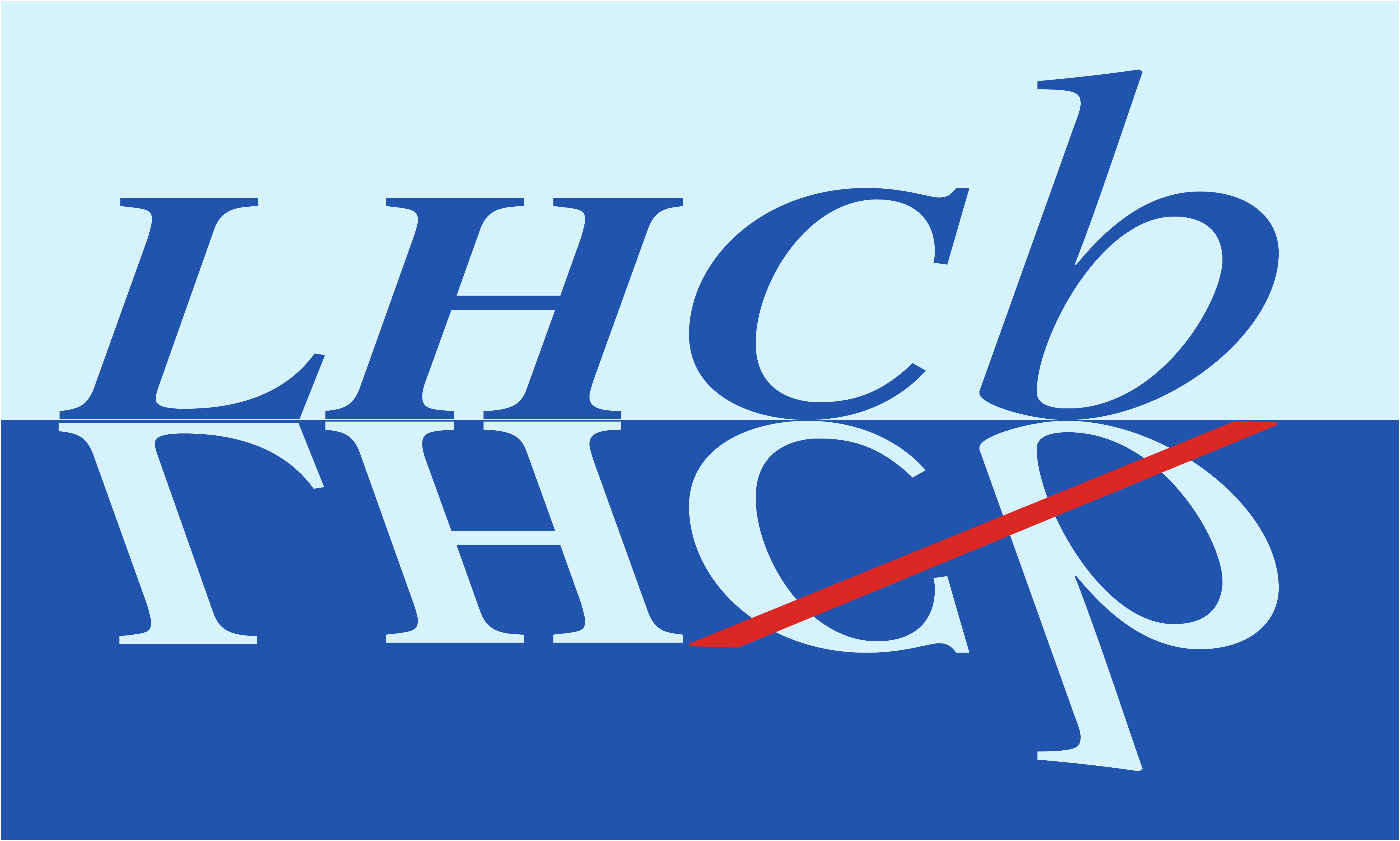}} & &}%
{\vspace*{-1.2cm}\mbox{\!\!\!\includegraphics[width=.12\textwidth]{figs/lhcb-logo.eps}} & &}%
\\
 & & CERN-EP-2021-133 \\  % ID 
 & & LHCb-PAPER-2021-008 \\  % ID 
% & & \today \\ % Date - Can also hardwire e.g.: 23 March 2010
 & & 16 February 2022\\
 & & \\
% not in paper \hline
\end{tabular*}

\vspace*{3.0cm}

% Title --------------------------------------------------
{\normalfont\bfseries\boldmath\huge
\begin{center}
% DO NOT EDIT HERE. Instead edit macro in main.tex to keep metadata correct
  \papertitle
\end{center}
}

\vspace*{2.0cm}

% Authors -------------------------------------------------
\begin{center}
% Edit macro in main.tex to keep metadata correct
\paperauthors\footnote{Authors are listed at the end of this paper.}
\end{center}

\vspace{\fill}

% Abstract -----------------------------------------------
\begin{abstract}
  \noindent
An improved measurement of the decay \mbox{\Bsmumu} and searches for the decays \mbox{\Bdmumu} and \Bsmumugamma are performed at the LHCb experiment using data collected in proton-proton collisions at $\sqrt{s} = 7,~8$ and $13\tev$, corresponding to integrated luminosities of 1, 2 and 6\invfb, respectively.
The \Bsmumu branching fraction and effective lifetime are measured to be \mbox{${\cal B}(\Bsmumu)=\Bsbr$} and \mbox{$\tau(\Bsmm)=\Bstau$}, respectively, where the uncertainties include both statistical and systematic contributions. No significant signal for \mbox{\Bdmumu} and \Bsmumugamma decays is found and the upper limits \mbox{$\BRof \Bdmumu < \Bdobslimitnf$} and $\BRof \Bsmumugamma < \Bsmmgobslimitnf$  at 95\% confidence level are determined, where the latter is limited to the range \mbox{$m_{\mu\mu} > 4.9 \gevcc$}. Additionally, the ratio between the \mbox{\Bdmumu} and \mbox{\Bsmumu} branching fractions is measured to be \mbox{$\mathcal{R}_{\mu^+\mu^-}<\ratiolimitnf$} at 95\% confidence level. The results are in agreement with the Standard Model predictions.
\end{abstract}

\vfill
\begin{center}
  Published in Phys. Rev. D105 (2022) 012010
\end{center}

\vspace{\fill}

{\footnotesize 
% Edit macro in main.tex to keep metadata correct
\centerline{\copyright~\papercopyright. \href{\paperlicenceurl}{\paperlicence}.}}
\vspace*{2mm}

\end{titlepage}

%%%%%%%%%%%%%%%%%%%%%%%%%%%%%%%%
%%%%%  EOD OF TITLE PAGE  %%%%%%
%%%%%%%%%%%%%%%%%%%%%%%%%%%%%%%%

%  empty page follows the title page ----
\newpage
\setcounter{page}{2}
\mbox{~}
%\newpage
%
%% Author List ----------------------------
%%  You need to get a new author list!
%\input{LHCb_authorlist.tex}
%
%The author list for journal publications is provided by the Membership Committee shortly after 'approval to go to paper' has been given.
%%It will be made available on the page
%%\verb!http://www.physik.uzh.ch/~strauman/forMemCo/LHCb-PAPER-XXXX-XXX/! .
%It will be sent to you by email shortly after a paper number has beens assigned.
%The author list should be included already at first circulation, 
%to allow new members of the collaboration to verify whether they have been included correctly.
%Occasionally a misspelled name is corrected or associated institutions become full members.
%In that case, a new author list will be sent to you.
%In case line numbering doesn't work well after including the authorlist, try moving the \verb!\bigskip! after the last author to a separate line.
%
%
%The authorship for Conference Reports should be ``The LHCb
%  collaboration'', with a footnote giving the name(s) of the contact
%  author(s), but without the full list of collaboration names.

%\twocolumn
% %%%%%%%%%%%%% ---------

\renewcommand{\thefootnote}{\arabic{footnote}}
\setcounter{footnote}{0}

%%%%%%%%%%%%%%%%%%%%%%%%%%%%%%%%
%%%%%  Table of Content   %%%%%%
%%%%%%%%%%%%%%%%%%%%%%%%%%%%%%%%
%%%% Uncomment if desired
%\tableofcontents
\cleardoublepage

%%%%%%%%%%%%%%%%%%%%%%%%%
%%%%% Main text %%%%%%%%%
%%%%%%%%%%%%%%%%%%%%%%%%%

\pagestyle{plain} % restore page numbers for the main text
\setcounter{page}{1}
\pagenumbering{arabic}

%% Uncomment during review phase. 
%% Comment before a final submission.
%\linenumbers

%% This is the main body
%% It is useful to have a single file so comemnts are not missed in overleaf.

\section{Introduction}
\label{sec:Introduction}
  Decays mediated by a quark flavour-changing neutral interaction are not allowed at tree level in the Standard Model (SM) of particle physics but can proceed through quantum loops, making them rare processes. The leptonic \Bdmm and \Bsmm decays (the inclusion of charge-conjugated processes is implied throughout this paper) are even rarer because they are additionally helicity-suppressed. As they are characterised by a purely leptonic final state, and thanks to the progress in lattice QCD calculations~\cite{Aoki:2019cca,Bazavov:2017lyh,Bussone:2016iua,Dowdall:2013tga,Hughes:2017spc}, their time-integrated branching fractions are predicted in the SM with small uncertainties to be \mbox{\BRof \Bsmumu = $(3.66 \pm 0.14) \times 10^{-9}$} and \mbox{\BRof \Bdmumu = $(1.03 \pm 0.05) \times 10^{-10}$}~\cite{Bobeth:2013uxa,Beneke:2019slt}. This makes these processes powerful probes for detecting deviations from the SM due to new physics (NP) contributions mediated, for instance, by heavy $Z^\prime$ gauge bosons, leptoquarks or non-SM Higgs bosons (see e.g. Ref.~\cite{Altmannshofer:2017wqy}).

\begin{figure}[t!]
    \centering
    \subfigure[]{\input{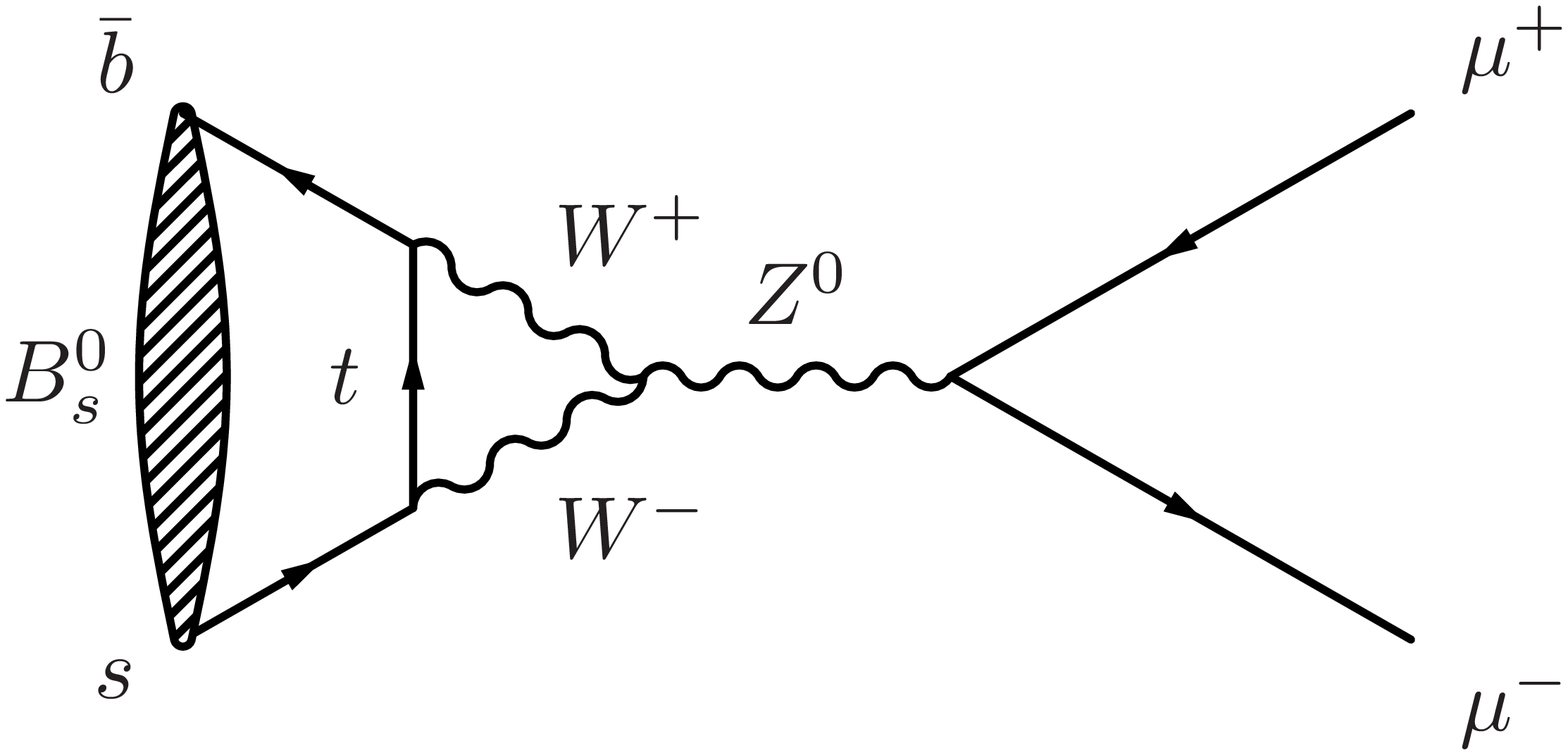}}
    \subfigure[]{\input{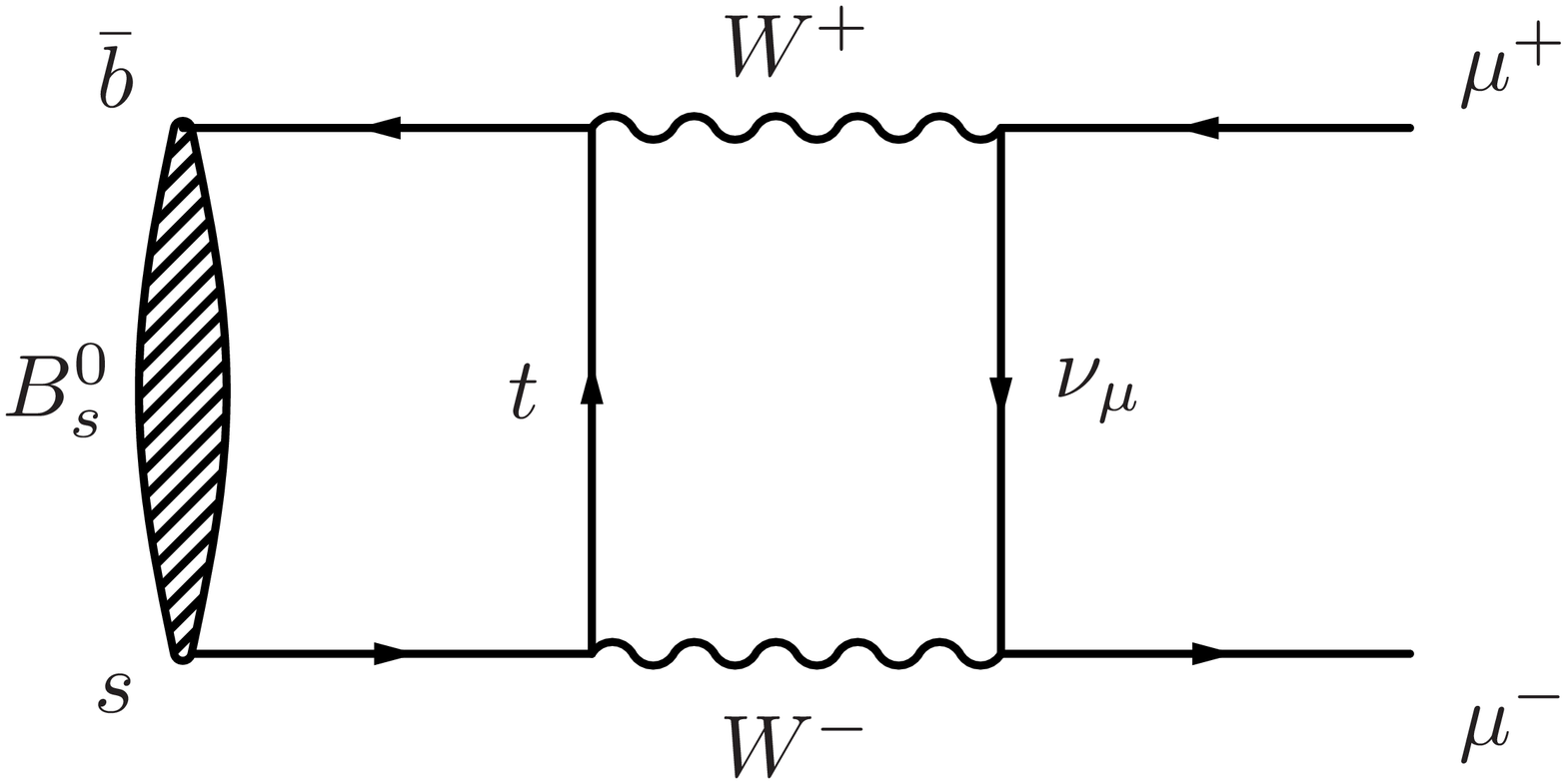}}
    \subfigure[]{\input{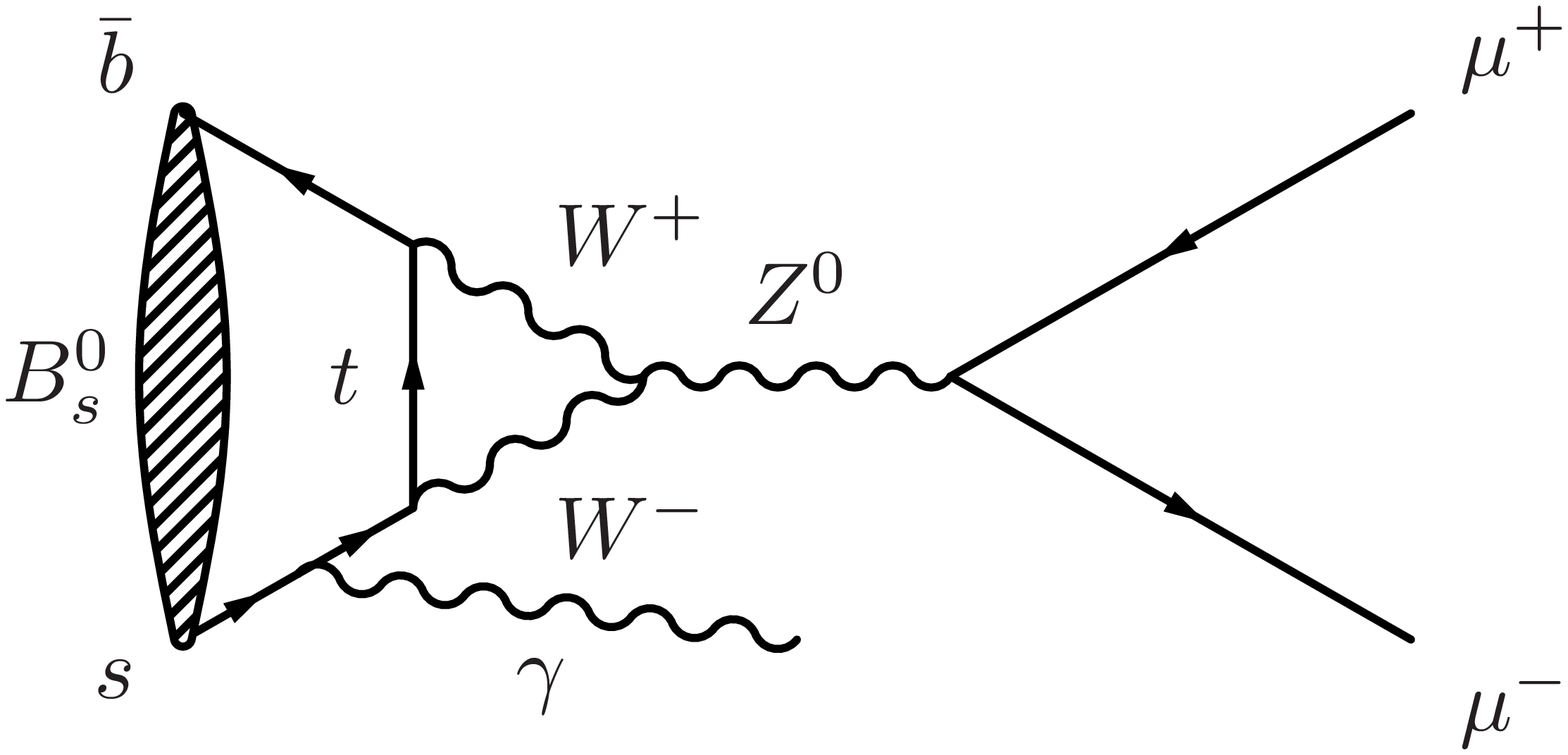}}
    \subfigure[]{\input{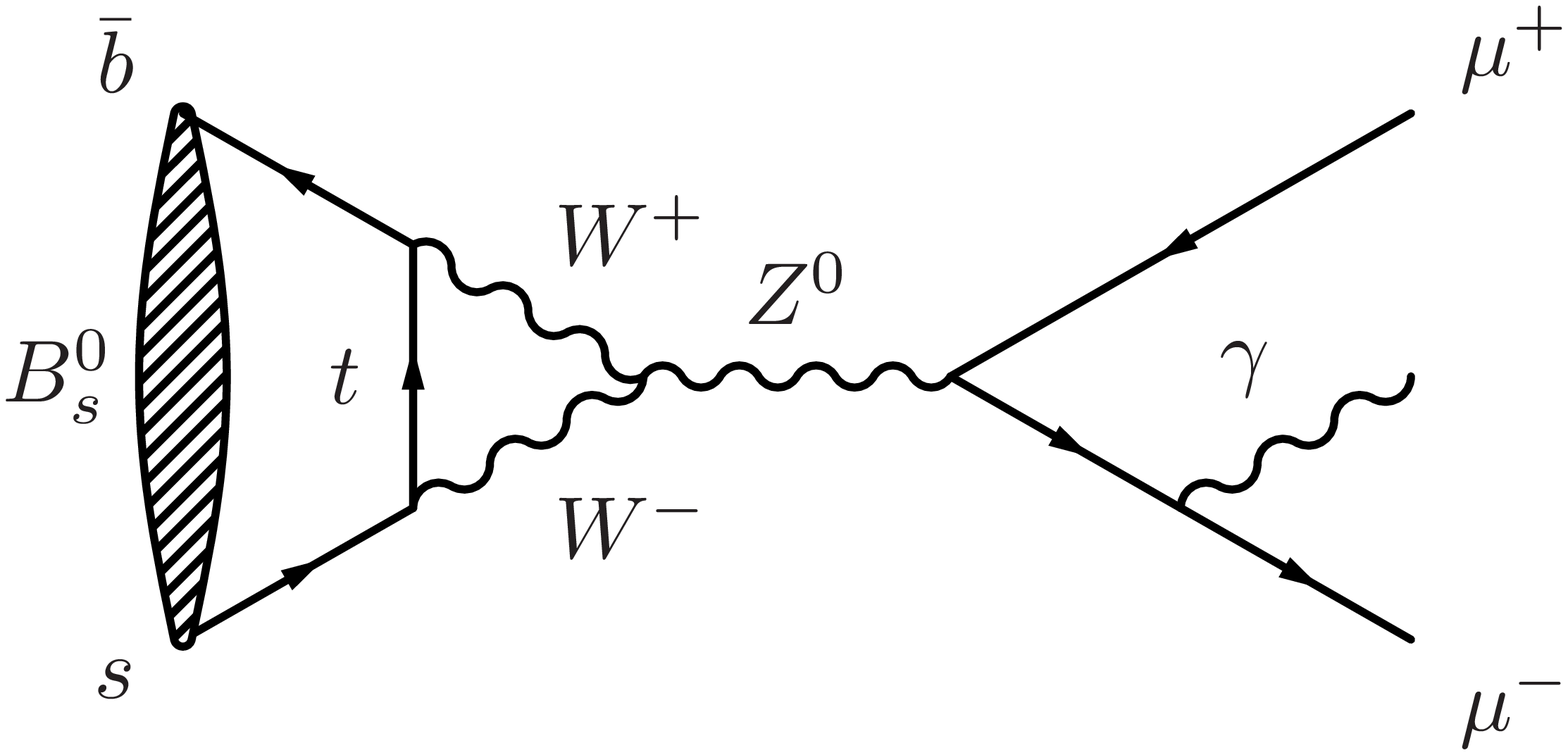}}
    \caption{SM Feynman diagrams mediating (top) the \Bsmumu and (bottom) the \bsmumugamma processes. Subpanels show (a) the so-called ``penguin'' diagram and (b) the ``box'' diagram for \Bsmumu, and (c) an ISR contribution and (d) an FSR contribution to \bsmumugamma.}
    \label{fig:diagrams}
\end{figure}

An effective field theory description of $\bquark \to \squark \mup \mun$ transitions makes it possible to tightly constrain the currents contributing to their amplitudes in a model-independent way. In this framework, the branching fractions of \Bdmm and \Bsmm decays are sensitive to axial-vector, scalar and pseudoscalar operators and their chirality-flipped counterparts~\cite{Logan_2000,Alonso_2014}. Of these, only the left-handed axial-vector current is present in the SM at a significant level. Examples of SM Feynman diagrams contributing to the \Bsmm amplitude are shown in Fig.~\ref{fig:diagrams}(a) and~\ref{fig:diagrams}(b).

Given the low rate of the \Bdmm and \Bsmm decays, their branching fractions are measured without distinguishing between \Bds and \Bdsb at production.  Both \Bd and \Bs mesons oscillate into their antiparticles but, in contrast to the \Bd system, the light and heavy mass eigenstates of \Bs mesons are characterised by a sizeable difference between their decay widths, $\Delta\Gamma_s=0.085\pm 0.004\ps^{-1}$~\cite{HFLAV18}, and thus have different lifetimes. This gives rise to the relation~\cite{DeBruyn:2012wk}
 \begin{equation}
 \mathcal{B}(\bsmumu) = \left[ \frac{1+ \ADeltaGamma y_s }{1 - y_s^2} \right] \mathcal{B}(\bsmumu)_{ t=0}\ ,
 \label{eq:brcorr}
 \end{equation}
between the flavour-untagged and time-integrated branching fraction and the value at decay time $t=0$. The theoretical prediction mentioned at the beginning of this section includes this correction. In Eq.~\ref{eq:brcorr}, \mbox{$y_s \equiv \Delta \Gamma_s/ (2\Gamma_s)=0.065\pm0.003$}~\cite{HFLAV18} and the parameter \ADeltaGamma is defined as $\ADeltaGamma\equiv-2\Re(\lambda)/(1+|\lambda|^2)$, with \mbox{$\lambda=(q/p)(A(\Bsbmumu)/A(\Bsmm))$}. The complex coefficients $p$ and $q$ relate the mass and \CP eigenstates of the $\Bs-\Bsb$ system with the flavour eigenstates \Bs and \Bsb~(see, e.g., Ref.~\cite{HFLAV18}), and $A$ is the amplitude of the process. In the SM, only the \CP-odd eigenstate (which, except for small modifications from \CP violation, corresponds to the heavy mass eigenstate) decays to $\mu^+\mu^-$ and the quantity \ADeltaGamma is equal to unity. However, in the presence of NP contributions it can assume any value in the range $-1 \leq \ADeltaGamma \leq 1$~\cite{DeBruyn:2012wk}. Thus the \Bsmm branching fraction might differ from the SM prediction in either of the two factors in the right-hand side of Eq.~\ref{eq:brcorr}.

The \Bsmm effective lifetime is defined as
\begin{equation}
\label{eq:tau_deltaagamma}
    \tauBsmm \equiv\frac{\int^\infty_0t\, \Gamma\!\left(\BsorBsbmumutime \right) \text{d}t}{\int^\infty_0  \Gamma\!\left(\BsorBsbmumutime \right) \text{d}t}
    =\frac{\tau_{\Bs}}{1-\ys^2}\left[\frac{1+2\ADeltaGamma\ys+\ys^2}{1+\ADeltaGamma\ys}\right],
\end{equation}
where $t$ is the decay time of the \Bs or \Bsb meson, the decay-time distribution $\Gamma\!\left(\BsorBsbmumutime \right)$ for \Bsmm and \Bsbmumu decays with and without oscillations is defined as \mbox{$\Gamma\!\left(\BsorBsbmumutime \right) \equiv\Gamma\!\left(\Bsmumutime\right)+\Gamma\!\left(\Bsbmumutime\right)$}, and $\tauBs=1.515\pm0.004\ps$~\cite{HFLAV18} is the mean \Bs lifetime. By measuring the \Bsmm effective lifetime, the contribution of each mass eigenstate, and thus the \CP structure of the interaction involved in the decay, can be inferred, and a direct evaluation of \ADeltaGamma can be performed. The lifetime thus makes it possible to discriminate between contributions from scalar or pseudoscalar interactions in a complementary way to the branching ratio. Similar effects are not significant for \bdmumu decays due to the negligible decay width difference of the \Bd mass eigenstates.

The ratio of the \bdmumu and \Bsmumu branching fractions also provides powerful discrimination between NP theories~\cite{Buras:2003td}. This quantity is theoretically more precise than the two individual branching fractions due to the cancellation of common theoretical uncertainties. It can be obtained as
\begin{equation}
\label{eq:ratio}
    \mathcal{R}_{\mu^+\mu^-} \equiv \frac{\BRof \Bdmumu}{\BRof \Bsmumu} = \frac{\tau_{\Bd}}{1/\Gamma_H^s}\left( \frac{f_\Bd}{f_\Bs}\right)^2 \left\vert \frac{V_{td}}{V_{ts}}\right\vert^2 \frac{\sqrt{M_\Bd^2-4m_\mu^2}}{\sqrt{M_\Bs^2-4m_\mu^2}},
\end{equation}
where $\tau_\Bd$ is the lifetime of the \Bd, $\Gamma_H^s$ is the width of the heavy-mass eigenstate of the \Bs meson, $M_\Bd$ and $M_\Bs$ are the masses of the \Bds mesons, $f_\Bd$ and $f_\Bs$ are the \Bds meson decay constants, $V_{td}$ and $V_{ts}$ are Cabibbo-Kobayashi-Maskawa (CKM) matrix elements and $m_\mu$ is the mass of the muon. In the SM, $\mathcal{R}_{\mu^+\mu^-}$ is predicted to be \ratioSM~\cite{Beneke:2019slt} and it assumes the same value in all NP models with the same flavour structure as the SM~\cite{DAmbrosio:2002vsn}.

The \bsmumug decay is also rare in the SM. Compared to the \Bsmm amplitude, the additional suppression arising from the photon is compensated by the fact that the amplitude is no longer helicity suppressed, increasing the total predicted branching fraction to $\mathcal{O}(10^{-8})$~\cite{Eilam:1996vg,Aliev:1996ud,Geng:2000fs,Melikhov:2004mk,Kozachuk:2017mdk,Dubnicka:2018gqg,Beneke:2020fot}. Two groups of amplitudes contribute to this decay: those where the photon is emitted from the initial state (initial-state radiation or ISR), an example of which is shown in Fig.~\ref{fig:diagrams}(c), and those in which it is emitted from the final state (final-state radiation, FSR), as in Fig.~\ref{fig:diagrams}(d). Their interference is evaluated to be negligible due to their combined helicity and kinematic suppression~\cite{Melikhov:2004mk,Kozachuk:2017mdk,Dettori:2016zff}. The FSR contribution to the \bsmumugamma process is experimentally included in the \bsmumu decay through the description of the radiative tail in its mass distribution. The ISR component is sensitive to a wider range of interactions and is treated as a separate contribution to the mass fit. In the mass region of interest its contribution decreases as the mass increases, becoming null for values larger than the \Bs mass. Similar to other multibody $b\to s \ell\ell$ decays, the sensitivity to different interactions depends on the dimuon mass squared, \qsquare, of the decay. At low \qsquare, the decay is mostly sensitive to tensor and pseudotensor interactions, while at high \qsquare vector and axial-vector contributions dominate~\cite{Aliev:2001bw,Guadagnoli:2017quo}. This makes the ISR \bsmumugamma decay at high \qsquare an ideal place to probe the same interactions that drive the anomalies seen in some $b\to s \ell\ell$ decays~\cite{LHCb-PAPER-2017-013,LHCb-PAPER-2019-009,LHCb-PAPER-2020-002,LHCb-PAPER-2020-041}. In the rest of this paper, \bsmumugamma refers only to the ISR process.  

Measurements of \Bdmm and \Bsmm processes have attracted considerable experimental interest since the first search for these decays at the \cleo experiment~\cite{PhysRevD.30.2279} almost forty years ago. The first evidence for the \bsmumu decay was obtained at \lhcb~\cite{LHCb-PAPER-2012-043} with data corresponding to 2\invfb of proton-proton ($pp$) collisions, and the decay was then observed through a combined analysis of data taken by the \lhcb and \cms experiments~\cite{LHCb-PAPER-2014-049}. Subsequent measurements were performed by the \lhcb collaboration~\cite{LHCb-PAPER-2017-001} with 4.4\invfb, by the \atlas collaboration~\cite{Aaboud:2018mst} with 51.3\invfb, and by the \cms collaboration~\cite{Sirunyan:2019xdu} with 63\invfb. 
These last three measurements are combined in Ref.~\cite{LHCb-CONF-2020-002}, yielding \mbox{$\BRof \Bsmumu = \Bsbrcomb$} and an upper limit on the \bdmumu  decay of \mbox{$\BRof \Bdmumu < \Bdlimcomb$} at 95\% confidence level (CL). In the two-dimensional plane of \BRof \Bsmumu and \BRof \Bdmumu, the consistency of the profile likelihood minimum with the SM prediction is measured to be 2.1 standard deviations~($\sigma$). To date, no experimental search has been performed for the \bsmumugamma decay, while the corresponding \Bd decay has been probed by the \babar experiment, yielding $\BRof\bdmumugamma < 1.5\times 10^{-7}$ at 90\% CL in the whole \qsquare region~\cite{Aubert:2007up}, which is well above the SM prediction. 

This paper presents improved measurements of the \Bsmm time-integrated branching fraction and effective lifetime, as well as a search for the \Bdmm decay, superseding the results in Ref.~\cite{LHCb-PAPER-2017-001}. Moreover, a first search for the \bsmumug decay at high dimuon mass is also presented. These results, also reported in Ref.~\cite{LHCb-PAPER-2021-007}, are based on data collected with the \lhcb detector, corresponding to an integrated luminosity of 1\invfb  of $pp$ collisions at a centre-of-mass energy $\sqrt{s}=7\tev$, 2\invfb at $\sqrt{s}=8\tev$ and 6\invfb at $\sqrt{s}=13\tev$. The first two data sets are referred to as Run 1 and the latter as Run 2. Throughout this paper, \Bmm candidates include \Bsmumu, \Bdmumu or \mbox{\Bsmumugamma} decays with the dimuon pair selected in the mass range \mbox{[4900, 6000]\mevcc} and the photon not reconstructed.

\section{Analysis Strategy}
\label{sec:Strategy}
The signature of \Bmm decays in the LHCb detector consists of two oppositely charged muons with a dimuon mass in the \Bs or \Bd mass region, and a decay vertex displaced with respect to any $pp$ interaction vertex as a result of the significant average flight distance of the \B mesons. The \bsmumugamma channel is searched for with the same signature, without reconstructing the photon, as was proposed in Ref.~\cite{Dettori:2016zff}.

The main background can be divided into two categories: combinatorial background arising from random combinations of muons from two distinct \bquark-hadron decays in the same event, and physical background comprising \bquark-hadron decays where one or more final state particles has either been misidentified as a muon or not reconstructed. Combinatorial background candidates are distributed across the entire search region from low to high mass, while physical background contributions tend to populate the region below the \Bs mass. The dominant physical background sources are: \bhh decays where the hadrons $h,h^{\prime}=K,\pi$ are misidentified as muons, which mainly contribute to the \Bd mass region; and partially reconstructed \bquark-hadron decays, which populate the same lower dimuon mass region as the \bsmumugamma signal. The most important partially reconstructed background sources are semileptonic \HbhMuNu (where $H_b$ is a \bquark hadron), \bcjpsimunu (with $\jpsi \to \mumu$) and \Bpimumu decays.

Combinatorial background is separated from the \Bmm signal by exploiting differences between their topologies and the relative isolation from other tracks in the event of the muons forming the \B candidate. This information is combined in a multivariate classifier based on a boosted decision tree~\cite{Breiman}, the output response of which, BDT, is used to classify the events as described in Sec.~\ref{sec:Selection}. A stringent particle identification (PID) requirement is used to suppress physical background from \bhh and semileptonic decays, described in Sec.~\ref{sec:Backgrounds}. 

As presented in Sec.~\ref{sec:BFfit}, the signal yields are estimated using an extended unbinned maximum-likelihood fit to the dimuon mass distribution, which is performed simultaneously in intervals of the BDT response to increase the sensitivity of the measurement. The BDT and mass distributions of the signals are calibrated and validated using data, as detailed in Sec.~\ref{sec:Calibration}. 

To measure the branching fractions, the yields of \mbox{\Bsmumu}, \mbox{\Bdmumu} and \mbox{\bsmumugamma} decays are normalised relative to those of \BdKpi and \bujpsik decays, with $\jpsi\to \mu^+ \mu^-$, reported in Sec.~\ref{sec:Normalisation}. 

The measurement of the \bsmumu effective lifetime uses a similar selection strategy, which is optimised to achieve the highest sensitivity. After the selection, a maximum-likelihood fit is performed to the dimuon mass distribution in two BDT regions to subtract the background. The decay-time acceptance in each BDT region is calibrated on corrected simulation samples and validated by applying the analysis procedure to \Bhh candidates from data. Finally, the effective lifetime is extracted using a maximum-likelihood fit to the background-subtracted decay-time distribution, performed simultaneously across both BDT regions, as presented in Sec.~\ref{sec:Lifetime}.

\section{Detector and simulation}
\label{sec:DetectorSimulation}
The \lhcb detector~\cite{LHCb-DP-2008-001,LHCb-DP-2014-002} is a single-arm forward spectrometer covering the \mbox{pseudorapidity} range $2<\eta <5$, designed for the study of particles containing \bquark or \cquark quarks. The detector includes a high-precision tracking system consisting of a silicon-strip vertex detector surrounding the $pp$ interaction region, a large-area silicon-strip detector located upstream of a dipole magnet with a bending power of about $4{\mathrm{\,Tm}}$, and three stations of silicon-strip detectors and straw drift tubes placed downstream of the magnet. The tracking system provides a measurement of the momentum, \ptot, of charged particles with a relative uncertainty that varies from 0.5\% at low momentum to 1.0\% at 200\gevc. The minimum distance of a track to a primary $pp$ collision vertex (PV), the impact parameter (IP), is measured with a resolution of $(15+29/\pt)\mum$, where \pt is the component of the momentum transverse to the beam, in\,\gevc. Different types of charged hadrons are distinguished using information from two ring-imaging Cherenkov detectors. Photons, electrons and hadrons are identified by a calorimeter system consisting of scintillating-pad and preshower detectors, an electromagnetic and a hadronic calorimeter. Muons are identified by a system composed of alternating layers of iron and multiwire proportional chambers. The online event selection is performed by a trigger, which consists of a hardware stage, based on information from the calorimeter and muon systems, followed by two software stages. The first software stage performs a preliminary event reconstruction using only part of the available event information, while the second stage performs a full event reconstruction.

Simulation is used to estimate the acceptance, reconstruction and selection efficiencies and to optimise the analysis strategy. The $pp$ collisions are generated using \pythia~\cite{Sjostrand:2006za,*Sjostrand:2007gs} with a specific \lhcb configuration~\cite{LHCb-PROC-2010-056}. Decays of particles are described by \evtgen~\cite{Lange:2001uf}. Decays of \Bc meson are generated using the dedicated \bcvegpy generator~\cite{Chang:2003cq,Chang:2015qea}.Final-state radiation in the decay of particles is simulated using \photos~\cite{Davidson:2010ew}, which is observed to agree with a full quantum electrodynamics calculation at the level of $1\%$~\cite{Bordone:2016gaq}. The interaction of the generated particles with the detector, and its response, are implemented using the \geant toolkit~\cite{Allison:2006ve, *Agostinelli:2002hh}, as described in Ref.~\cite{LHCb-PROC-2011-006}.

\section{Signal selection}
\label{sec:Selection}
In the online event selection, signal candidates are first required to pass the hardware trigger, which selects events with at least one muon with high transverse momentum, followed by a two-level software stage, which applies a full event reconstruction. The software stage imposes minimum requirements on the muon transverse momentum and impact parameter with respect to all PV.
However, to maximise the signal selection efficiency, events triggered by particles not related to the signal candidates are also retained for further analysis. 

Candidate \Bmm decays are selected offline by combining two well-reconstructed oppositely charged particles identified as muons~\cite{LHCb-DP-2013-001}, with transverse momentum in the range $0.25<p_{\rm T}<40\gevc$, and momentum $p< 500\gevc$. The muon candidates are required to form a secondary vertex (SV) with a vertex-fit $\chi^2$ per degree of freedom smaller than 9 and separated from any PV with a significance greater than 15. Only muon candidate tracks with $\chisqip>25$ for any PV are selected, where \chisqip is defined as the difference between the vertex-fit $\chi^2$ of the PV formed with and without the particle in question.

The resulting \Bds candidates must have a decay time lower than $13.25\ps$, $\chisqip<25$ with respect to the PV for which the \chisqip is minimal (henceforth referred to as the PV associated with the \Bds candidate) and $p_{\rm T} > 0.5 \gevc$. To suppress the \BcJpsiMuNu background, a \Bds candidate is rejected if either of the two candidate muons combined with any other oppositely charged muon candidate in the event has a mass within 30\mevcc of the \jpsi mass~\cite{PDG2020} (\jpsi veto). Further requirements on the particle identification (PID) information of the two muons are imposed in order to reject misidentified hadronic background. PID identification uses multivariate techniques to combine information from different subsystems taking correlations into account~\cite{LHCb-DP-2014-002}. 

Candidate \Bmm decays used in the branching fraction measurements are selected in the dimuon mass range $4900\leq m(\mumu)\leq6000\,\mevcc$, while those used in the lifetime measurement are selected in a narrower range, \mbox{$5320\leq m(\mumu)\leq6000\,\mevcc$}. The reduced mass range used in the lifetime measurement excludes most of the \Bdmm and physical background decays that populate the lower dimuon mass region, greatly simplifying the fit and making it possible to impose less stringent PID requirements used to reject misidentified background, thus increasing the signal selection efficiency. To avoid potential biases, the candidates in the mass region \mbox{$5200\leq m(\mumu)\leq5445\,\mevcc$}, where \Bdsmumu candidates peak, were not examined until the analysis procedure was finalised.

In addition to the signal channels, \mbox{\bhh} and \mbox{\bujpsik} decays  are selected as normalisation and control channels, and \bsjpsiphi as control channel. Candidate \mbox{\bhh} decays are selected using the same requirements as the signal channels, except that the muon identification criteria are replaced with hadron identification, the events are triggered independently of the decay final state, and the \jpsi veto is not applied.
Candidate \mbox{\bujpsik} and \bsjpsiphi decays are formed by combining a muon pair, with mass close to the \jpsi mass~\cite{PDG2020}, with one track (\bujpsik) or two oppositely charged tracks consistent with originating from a $\phi$ decay (\bsjpsiphi), with the kaon mass hypothesis assigned. All tracks forming \mbox{\bujpsik} and \mbox{\bsjpsiphi} candidates are selected with the same requirements as those applied to select the \Bmm candidates, except for the dimuon mass range and the particle identification criteria. The muons are only required to pass the muon system identification criteria~\cite{LHCb-DP-2013-001} and no kaon identification criteria are required, given the already excellent signal purity achieved. The same trigger strategy as for the signal decays is used for these two channels. 

Background events are further rejected using a loose requirement on the response of a boosted decision tree~\cite{Breiman,Adaboost,TMVA4}, which was first described in Ref.~\cite{LHCb-PAPER-2012-007} and has remained unchanged. The classifier takes as input: the angle between the direction of the momentum of the \Bds candidate and the direction defined by the vector joining the primary and the secondary vertices; the \Bds candidate IP and its vertex $\chi^2$; the minimum IP of the muons with respect to any PV; the minimum distance between the two muon tracks; the $\chi^2$ of the SV. This classifier is also applied to the control and normalisation channels, where for the \bujpsik and \bsjpsiphi modes the \chisq of the SV is replaced with that of the \jpsi vertex. The selected sample of \Bdsmumu candidates is dominated by random combinations of two muons (combinatorial background), mainly originating from semileptonic decays of two different $b$ hadrons. The reconstruction and selection efficiencies for the signal and normalisation modes are reported in Sec.~\ref{sec:efficiencies}.

Isolation variables, which quantify the possibility that other tracks in the event originated from the same hadron decay as the signal muon candidates, are constructed in order to further reject background. Most combinatorial background candidates arise from semileptonic $b$-hadron decays, where other charged particles produced in the decay may be reconstructed close to the signal muon candidate. Two isolation variables are designed to recognise these particles, each considering a different type of track: one uses additional tracks that have been reconstructed both before and after the magnet (long tracks), while the other considers tracks reconstructed only in the vertex detector (VELO tracks). These isolation variables are defined based on the proximity of the two muons forming the \Bds candidate to other tracks in the event.

The closeness of each muon candidate to either a long track or a VELO track is measured using two dedicated multivariate classifiers that take the following quantities as inputs: the minimum $\chi^2_{\rm IP}$ of the track with respect to any PV; the signed distance between the muon-track vertex and the PV associated to the \Bds candidate; the signed distance between the muon-track vertex and the \Bds decay vertex; the distance of closest approach between the track and the muon; the angular separation between the track and the muon; a quantity that measures the compatibility of the muon-track system with having originated from the PV associated to the \Bds candidate. The long-track isolation classifier takes three additional variables as input: the absolute difference between the azimuthal angles of the track and the muon; the absolute difference between the pseudorapidities of the track and the muon; and the track $p_{\rm T}$. The classifiers are trained on collections of track-muon pairs from simulated \Bsmm decays and from simulated \bbdim events. The latter sample includes decays with two muons originating from two different $b$-hadrons or from the same $b$-hadron. In both cases the muons can either originate directly from the hadron containing the $b$-quark or from intermediate resonances.
 Only tracks originating from the same $b$-hadron decay as the muon candidate are considered to train the classifier with the simulated \bbdim events. The output value of the classifiers for a given track-muon pair is defined to be higher when the track is ``closer'' to the muon. Defining $I(\mu^\pm)$ as the maximum value of a given classifier over all the track-muon pairs in the event, the long and VELO-track isolation variables are each defined as $I(\mup)+I(\mun)$.

\begin{figure}[tb]
    \centering
     \includegraphics[width=0.49\linewidth]{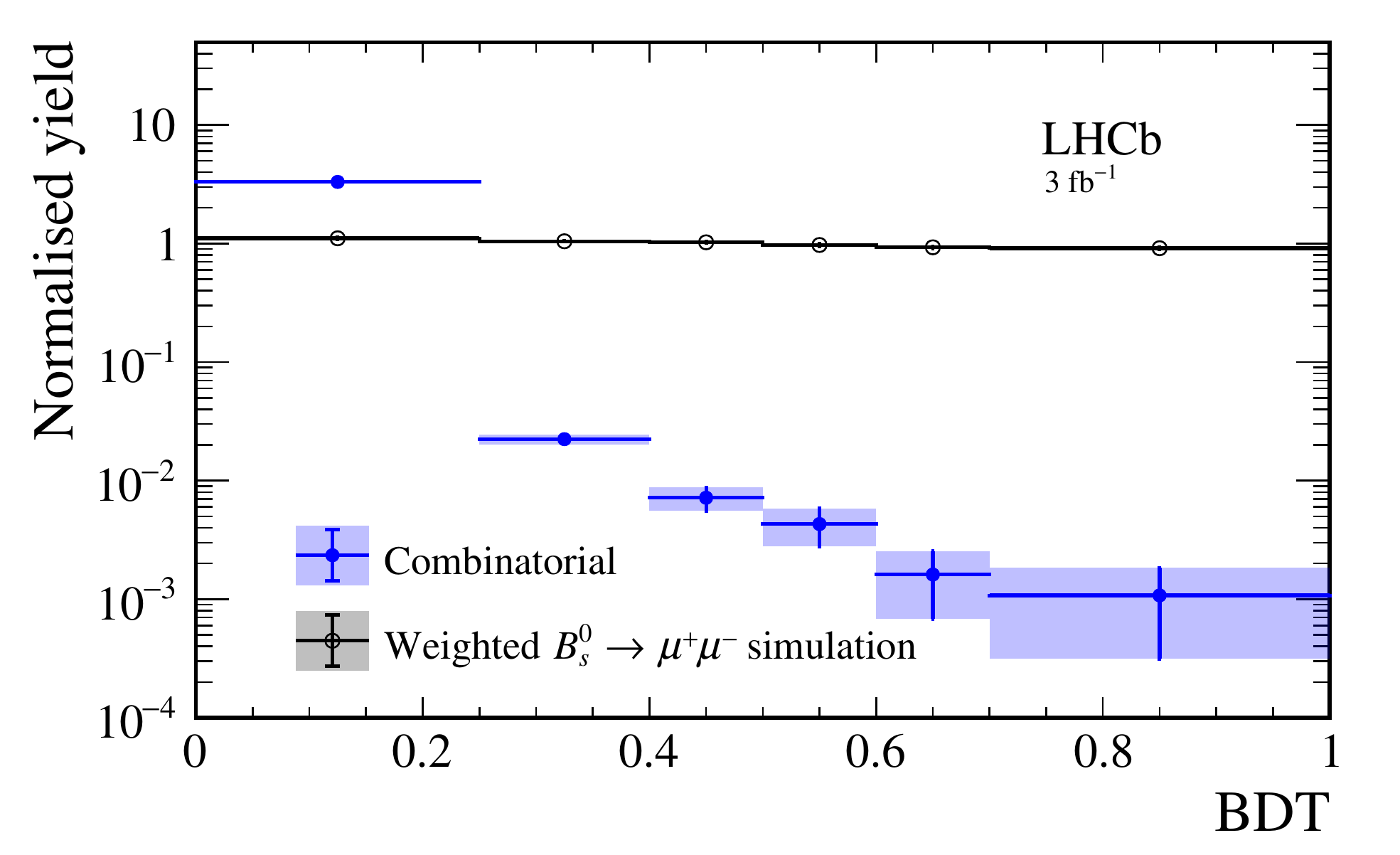}
     \includegraphics[width=0.49\linewidth]{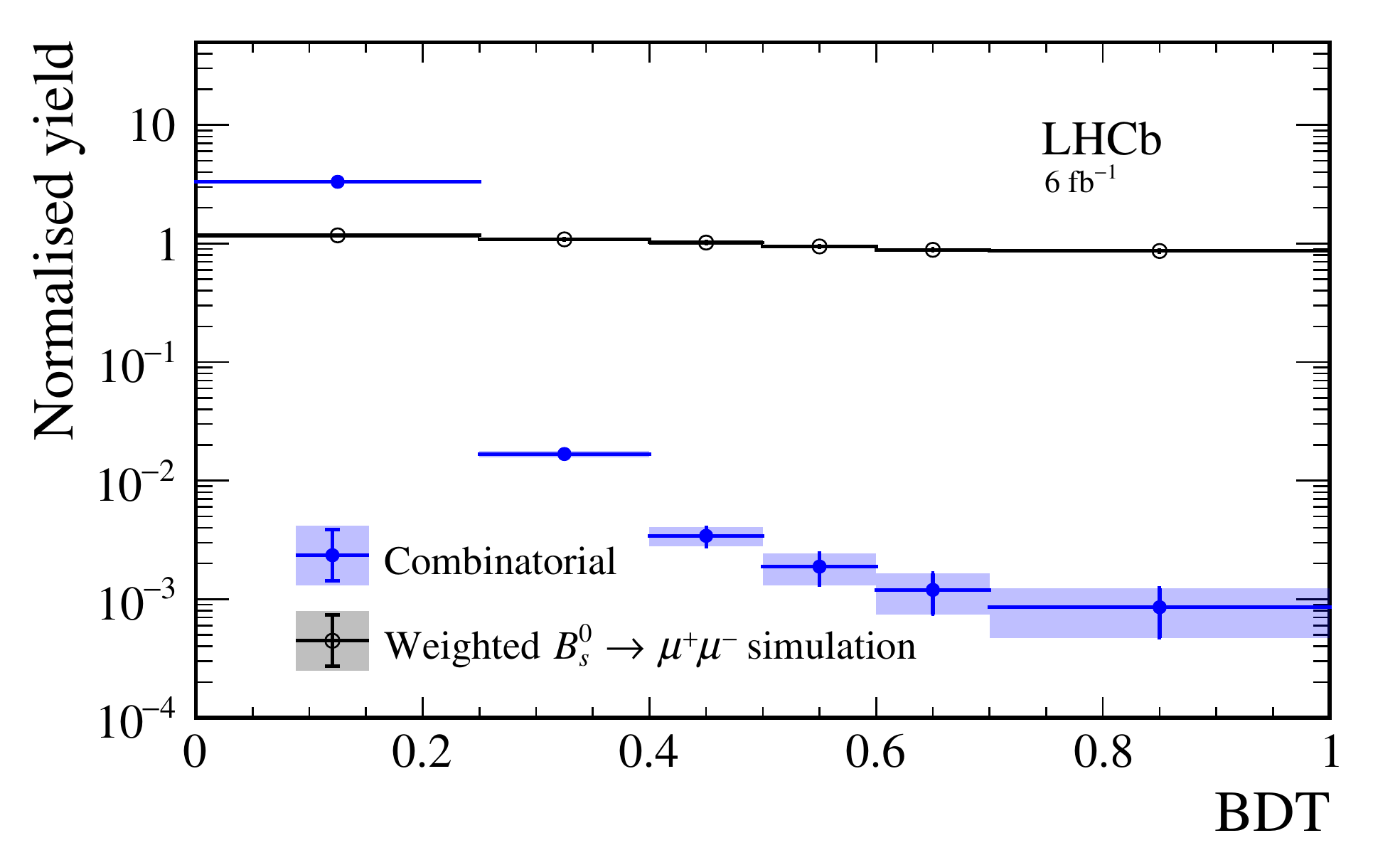}
\caption{BDT distribution calibrated using corrected simulated \bsmumu decays (black circles) and combinatorial background from high dimuon-mass data sidebands (blue filled circles) in (left) \runone and (right) \runtwo data. Blue error bands represent the statistical uncertainty.}
\label{fig:BDTshapeComb}
\end{figure}

Signal and background events are separated using a final boosted decision tree classifier that combines kinematic, topological and isolation information, defined as in the previous measurement~\cite{LHCb-PAPER-2017-001}. The BDT response is used to divide the data into samples of varying signal purity, which are then fitted simultaneously as described in Secs.~\ref{sec:BFfit} and \ref{sec:Lifetime}. The classifier is trained using simulated samples of \Bsmm decays as signal and of inclusive \bbdim events as proxy for the combinatorial background. It combines information from the following input variables: $\sqrt{\Delta\phi^2+\Delta\eta^2}$, where $\Delta\phi$ and $\Delta\eta$ are the azimuthal angle and pseudorapidity differences between the two muon candidates; the minimum $\chi^2_{\rm IP}$ of the two muons with respect to the \Bds associated PV; the angle between the \Bds candidate momentum and the vector joining the \Bds decay vertex and \Bds associated PV; the \Bds candidate vertex-fit $\chi^2$; the \Bds impact parameter significance with respect to the \Bds associated PV; the long- and VELO-track isolation variables. The BDT classifier response is defined to have an approximately uniform distribution in the range $0\leq\rm{BDT}\leq1$ for signal, and to peak at zero for background. Its correlation with the dimuon mass is below 5\%. The branching fraction measurement is performed by dividing the Run 1 and Run 2 data samples into six subsets each, based on regions in the BDT response with boundaries 0, 0.25, 0.4, 0.5, 0.6, 0.7 and 1. Figure~\ref{fig:BDTshapeComb} shows the expected BDT distribution for \bsmumu decays, as determined in Sec.~\ref{sec:bdt_calibration}, and combinatorial background. 
The sample with $0\leq\rm{BDT}<0.25$ is discarded as it is dominated by background. The data used in the measurement of the \Bsmm effective lifetime are split into two regions in the BDT response with the ranges $0.35\leq\rm{BDT}<0.55$ and $0.55\leq\rm{BDT}\leq1$, which are chosen to minimise the expected statistical uncertainty on the effective lifetime based on the results of pseudoexperiments.

\section{Signal calibration}
\label{sec:Calibration}

The dimuon mass and the \BDT classifier response are used to separate signal from background in the determination of the branching fractions and the \Bsmm effective lifetime. It is therefore essential that these variables are accurately calibrated in order to account for possible discrepancies between data and simulation. The calibration procedures for these two variables are described in the following sections.

\subsection{Mass shape calibration}
\label{sec:mass_calib}
The mass shape of the \Bsmm and \Bdmumu signals is described with a double-sided Crystal Ball (DSCB) function~\cite{Skwarnicki:1986xj}
\begin{equation}    
    f(m|\mu,\sigma,\alpha_l,n_l,\alpha_r,n_r) = N
        \begin{cases}     
       (\frac{n_l}{\alpha_l})^{n_l} \text{exp}[- \frac{\alpha_l^2}{2}]  (-\frac{m - \mu}{\sigma} + \frac{n_l}{\alpha_l} - \alpha_l )^{-n_l}, & \text{if}\ \frac{m - \mu}{\sigma} < - \alpha_l \\
        (\frac{n_r}{\alpha_r})^{n_r} \text{exp}[- \frac{\alpha_r^2}{2}]  (\frac{m - \mu}{\sigma} + \frac{n_r}{\alpha_r} - \alpha_r )^{-n_r}, & \text{if}\ \frac{m - \mu}{\sigma} > \alpha_r \\
        \text{exp}[-\frac{(m - \mu)^{2}}{2 \sigma^2}], & \text{otherwise,} \\
        \end{cases}  
        \label{eq:dscb}           
\end{equation}
where $m$ is the dimuon mass and all the parameters are positive. The function has a Gaussian core with mean $\mu$ and resolution $\sigma$ and power-law tails on both sides defined by two starting points in units of $\sigma$, $\alpha_l$ and $\alpha_r$, and two slopes, $n_l$ and $n_r$ for the left and right side, respectively. 

\begin{figure}[tb]
    \centering
    \includegraphics[width=0.45\textwidth]{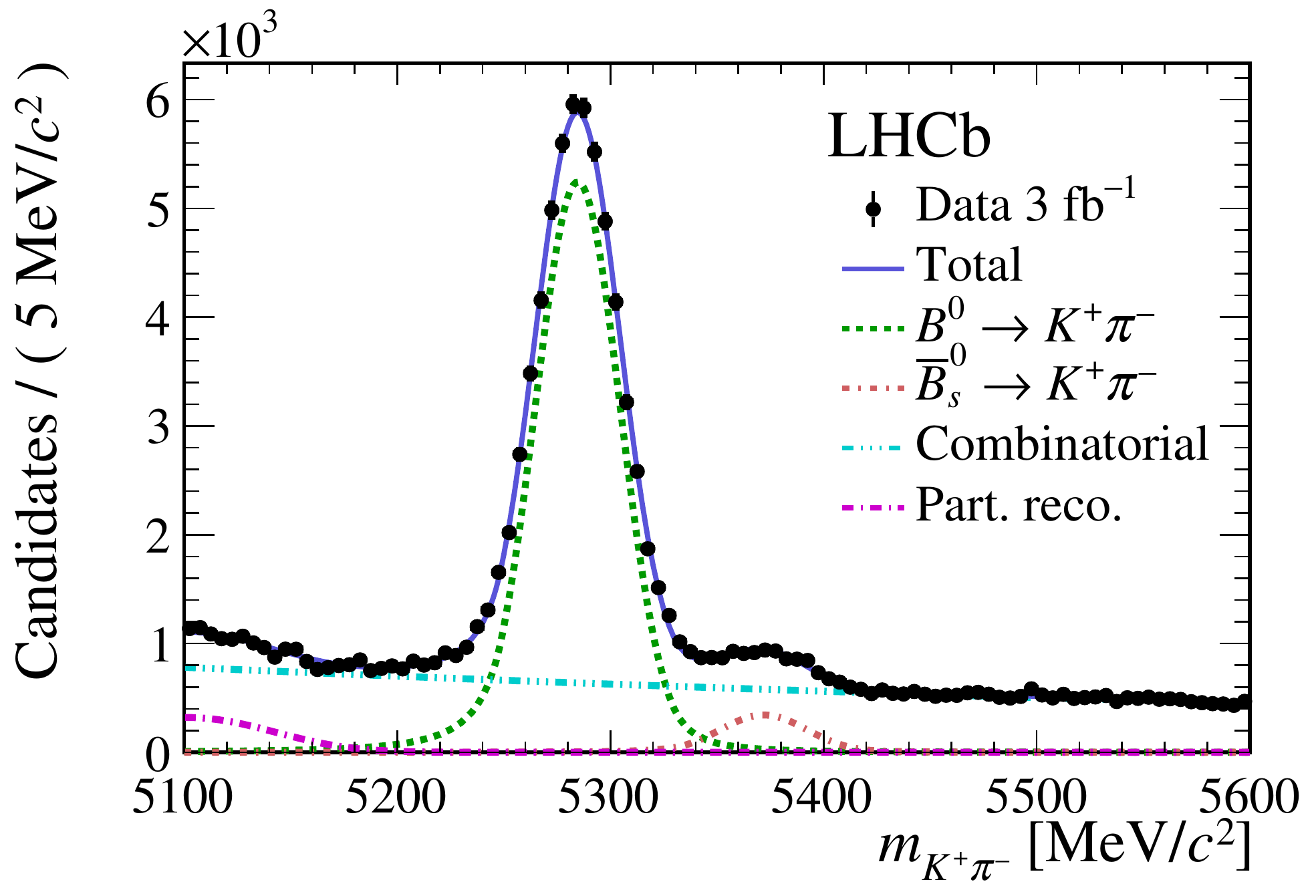}
    \includegraphics[width=0.45\textwidth]{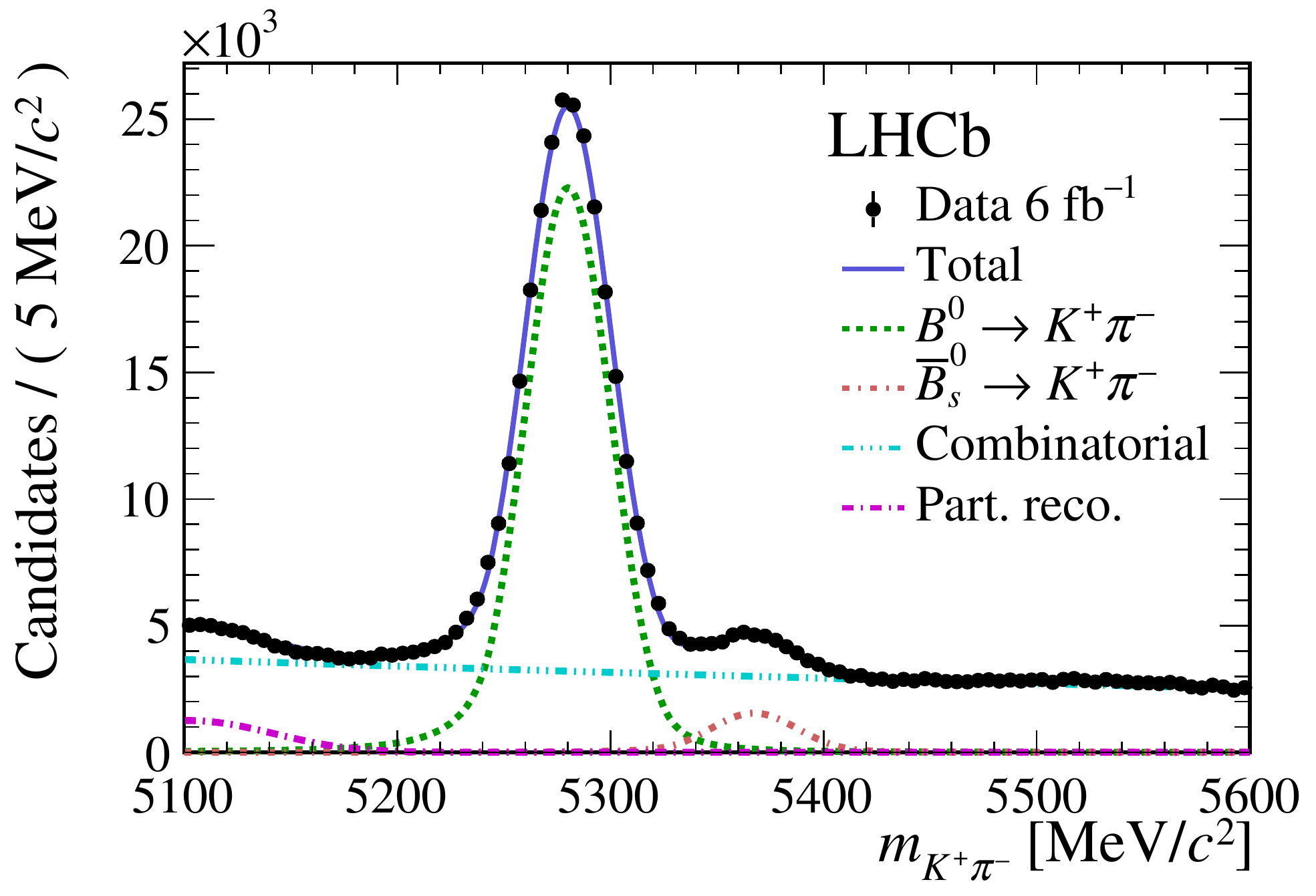}
    \includegraphics[width=0.45\textwidth]{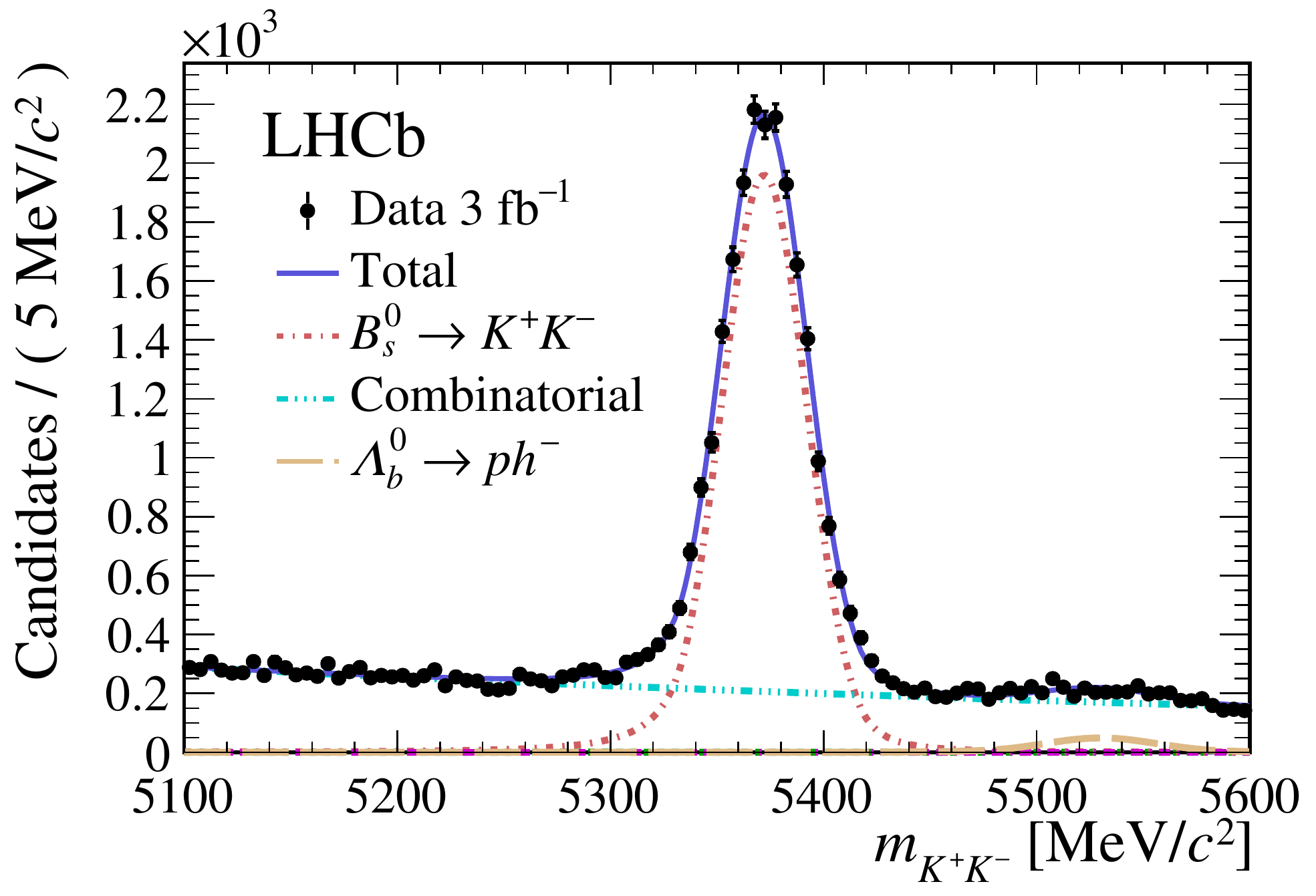}
    \includegraphics[width=0.45\textwidth]{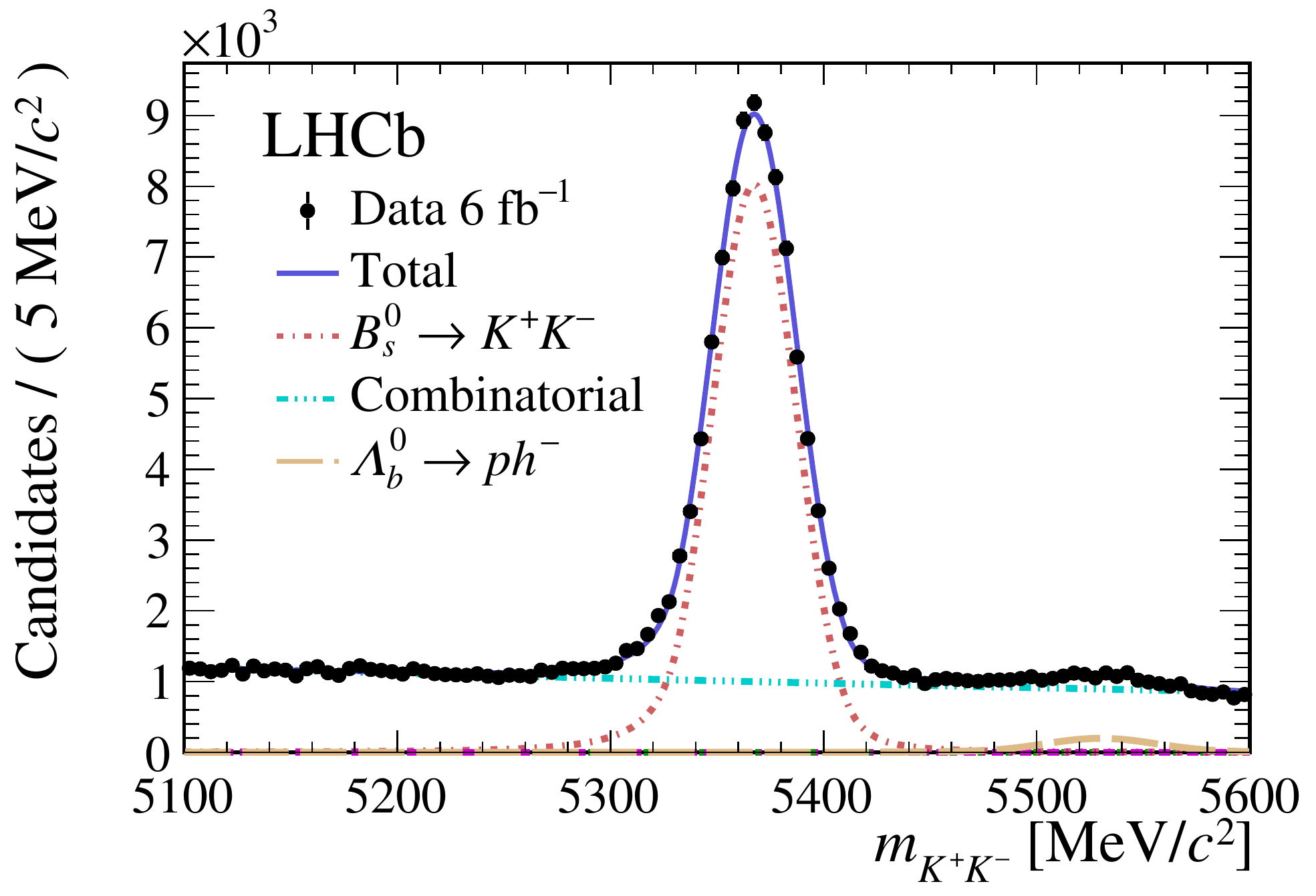}
    \caption{Mass distributions of selected (top) \BdKpi and (bottom) \BsKK candidates in (left) \runone and (right) \runtwo data. The results of the fits used to determine the means of the \bdsmumu mass distributions are overlaid and the different components are detailed in the legends.}
    \label{fig:masscalib_mean}
\end{figure}

The mean of the \Bsmumu and \Bdmumu signal peaks are calibrated with data samples containing \bskk and \BdKpi decays, respectively. Besides the contamination from combinatorial background, the \BdKpi sample contains contributions from \bskpi decays and partially reconstructed background decays, while the \bskk sample contains contribution from misidentified \Lbph decays. The $m_{K^+\pi^-}$ and $m_{K^+K^-}$ mass distributions of the \mbox{\BdKpi} and \mbox{\bskk} decays, shown in Fig.~\ref{fig:masscalib_mean}, are modelled with a DSCB function. The difference between the \Bd and \Bs mass values, taken from Refs.~\cite{PDG2020,LHCb-PAPER-2015-010,LHCb-PAPER-2011-035,Acosta:2005mq}, is used to constrain the \bskpi mean with respect to the \BdKpi mode. The mass resolutions for the two modes are constrained using the mass resolution calibration described below. The combinatorial background is modelled with an exponential shape with its slope parameter left to vary freely. The partially reconstructed background component is described by an ARGUS shape~\cite{argus}, while the component for \Lbph decays is modelled as the sum of two Crystal Ball (CB) functions~\cite{Skwarnicki:1986xj}, with all parameters, except the total yield, fixed from simulation. The results of the fits are shown in Fig.~\ref{fig:masscalib_mean}. To check the correlation between the PID selection and the mean of the \Bhh signal peak, and to study the effect of possible contamination from misidentified background, the fit is repeated after tightening the PID requirements. The variation of the mean value is assigned as a systematic uncertainty. The mean is found to be uncorrelated with the BDT response. 

\begin{figure}[tb]
    \centering
    \includegraphics[width=0.45\textwidth]{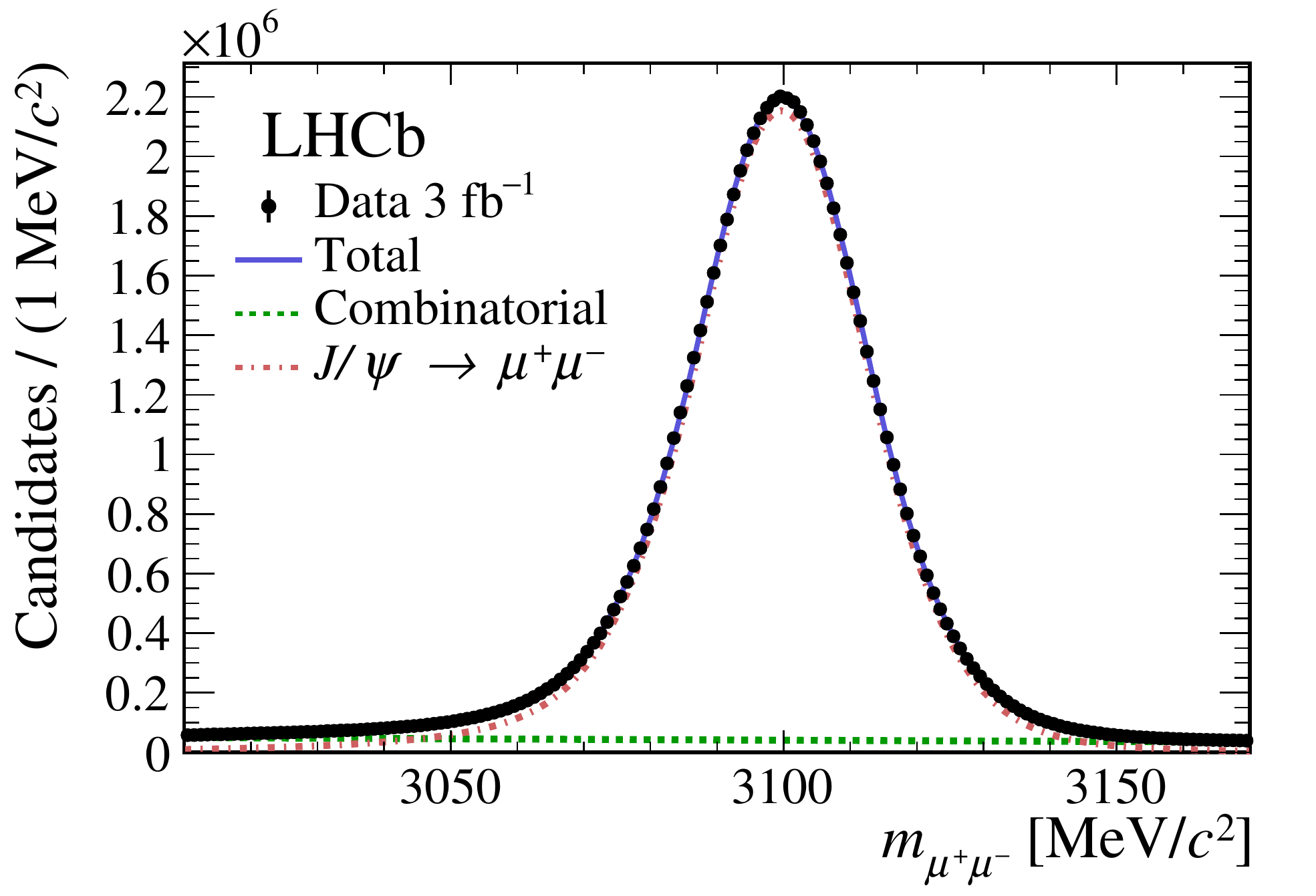}
    \includegraphics[width=0.45\textwidth]{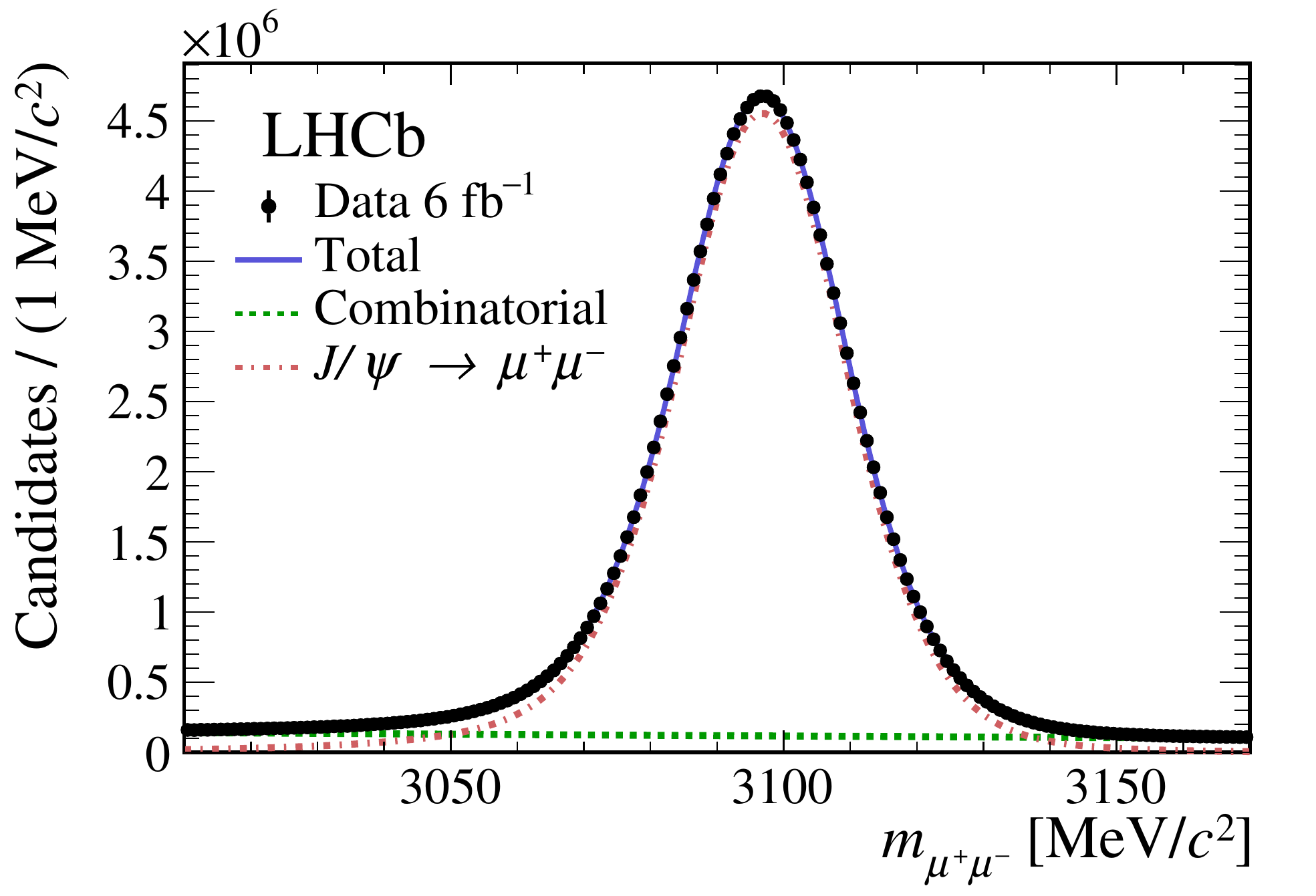}
    \includegraphics[width=0.45\textwidth]{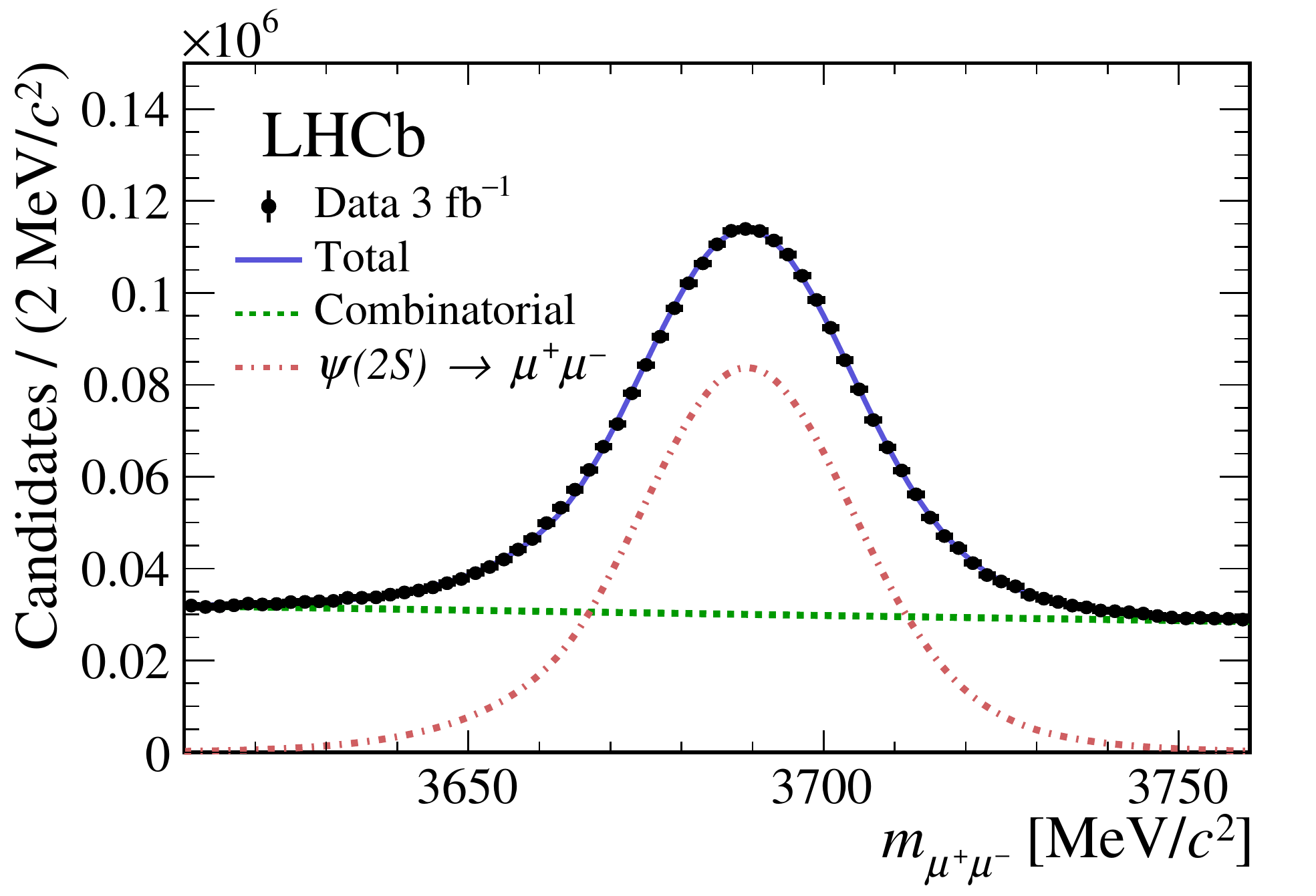}
    \includegraphics[width=0.45\textwidth]{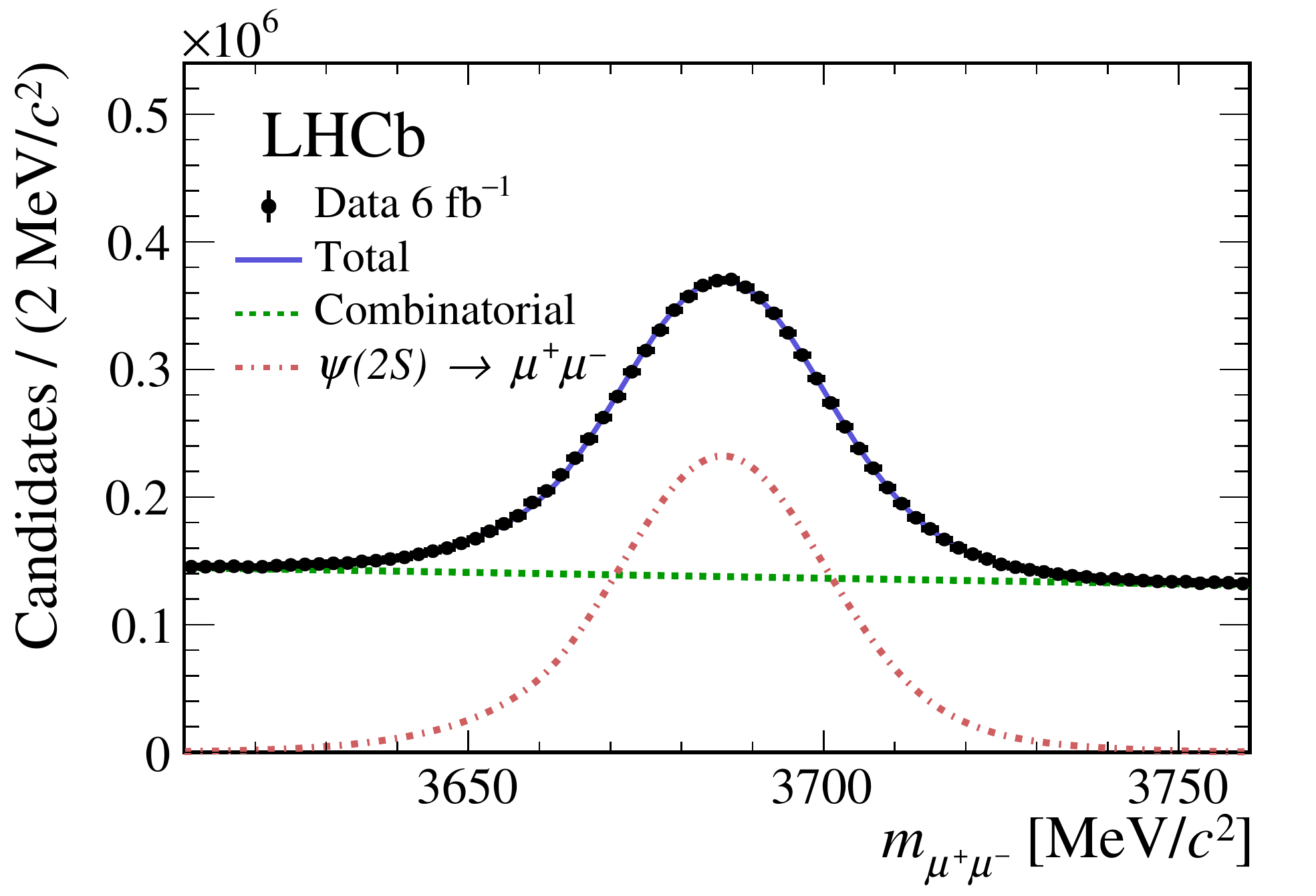}
    \includegraphics[width=0.45\textwidth]{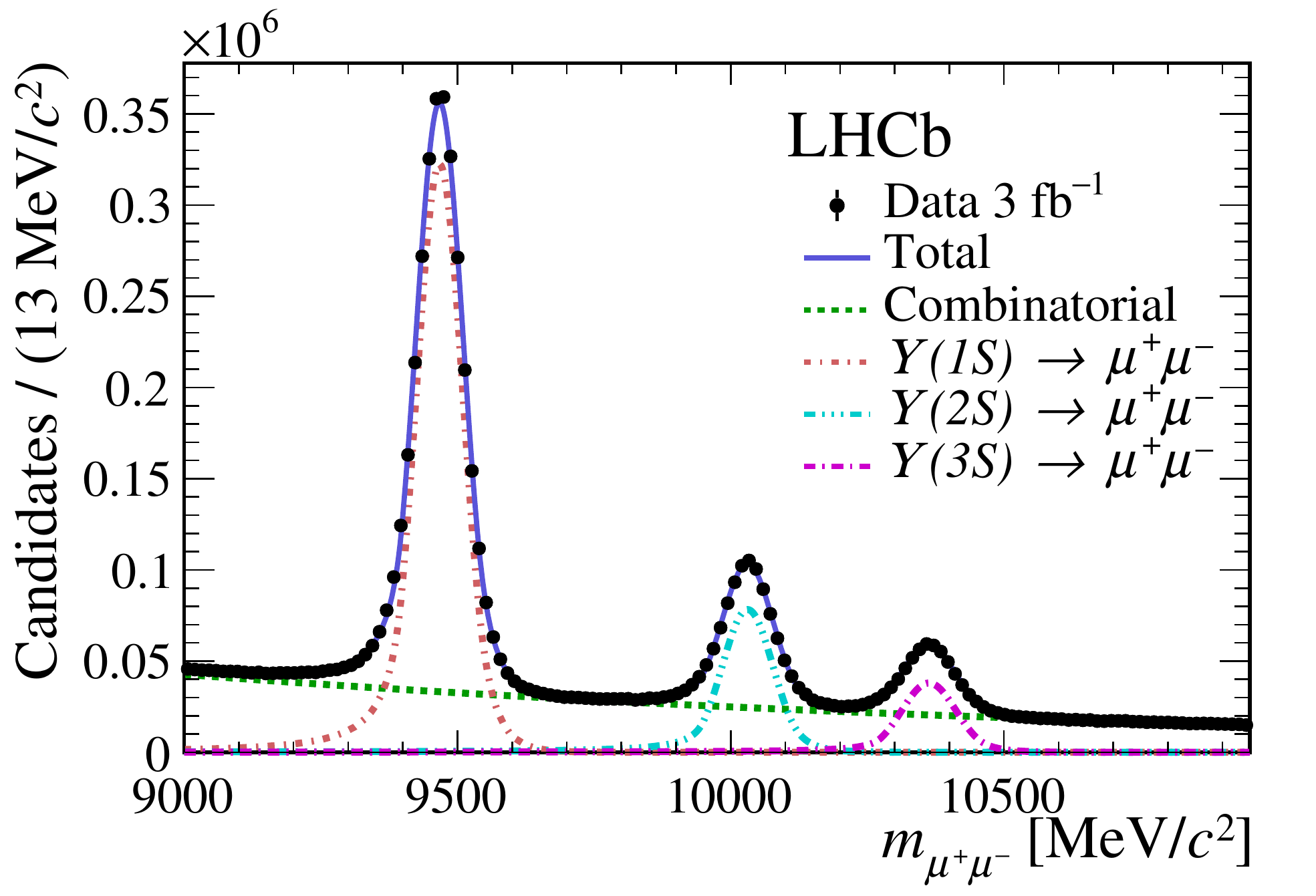}
    \includegraphics[width=0.45\textwidth]{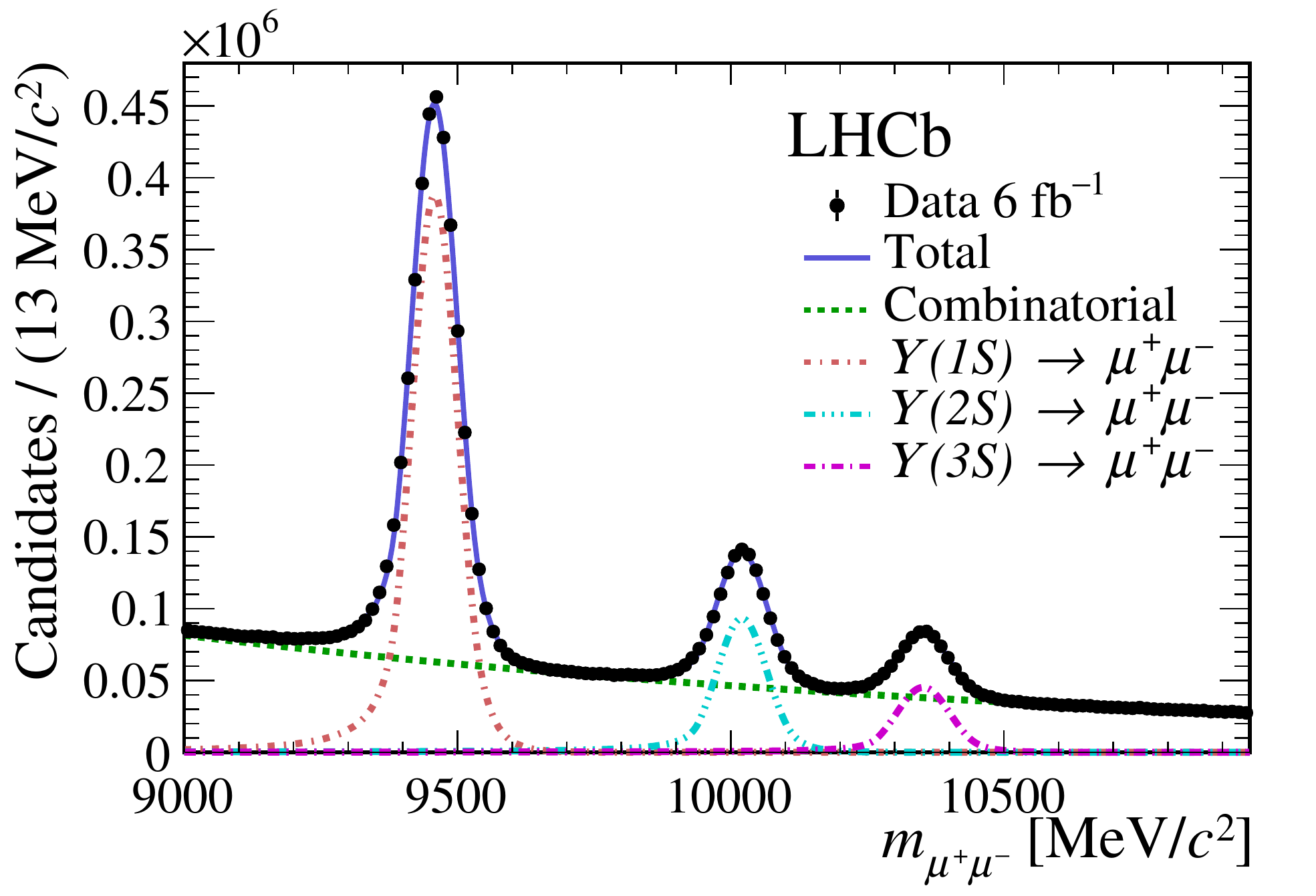}    
    \caption{Mass distributions of (top) $\jpsi\to\mu^+\mu^-$, (centre) $\psitwos\to\mu^+\mu^-$, (bottom) $\varUpsilon(1S,2S,3S)\to\mu^+\mu^-$ candidates in (left) \runone and (right) \runtwo data. The result from the fit to determine the mass resolutions to each sample is overlaid, and the components are detailed in the legend.}
    \label{fig:quarkonia}
\end{figure}

\begin{figure}[tb]
    \centering
    \includegraphics[width=0.45\textwidth]{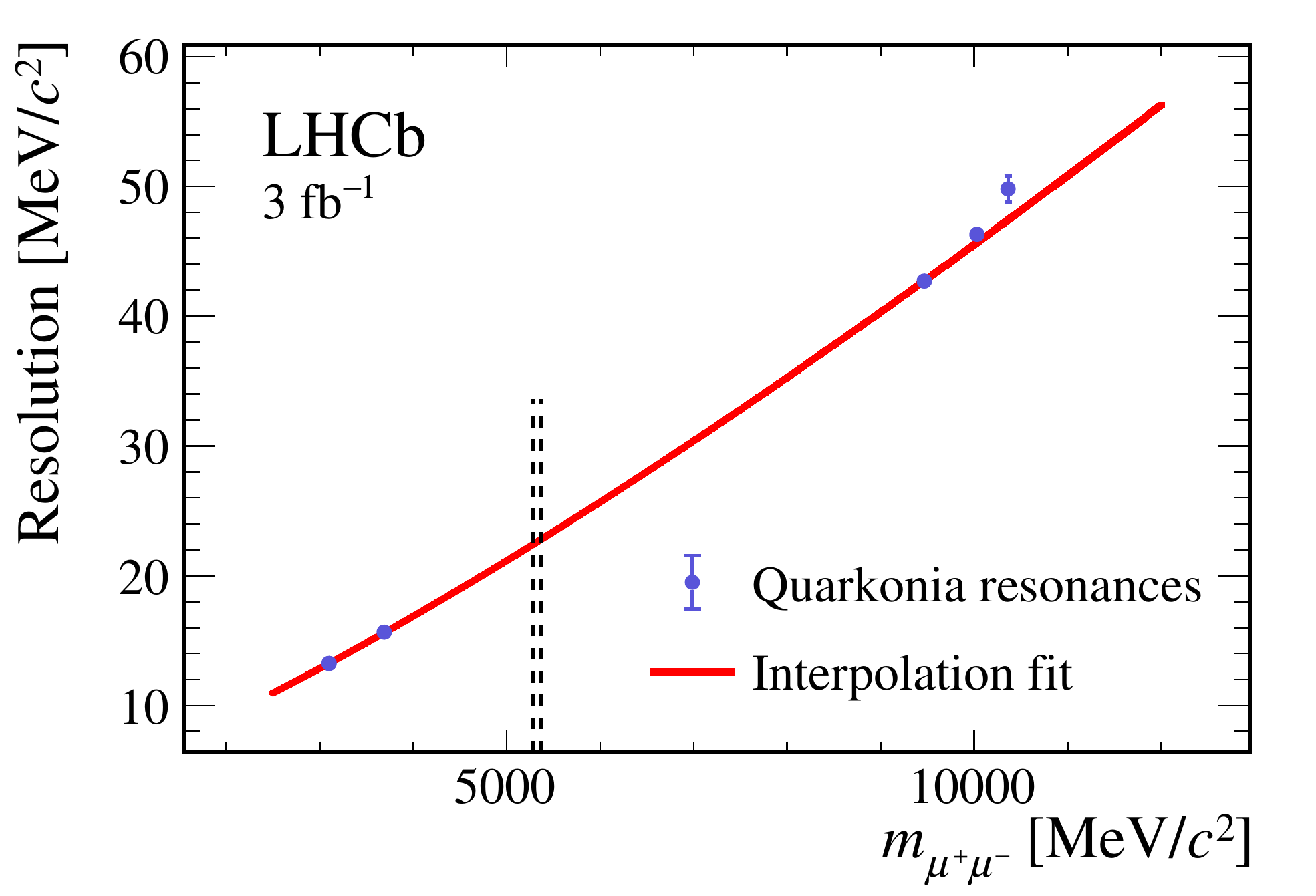}
    \includegraphics[width=0.45\textwidth]{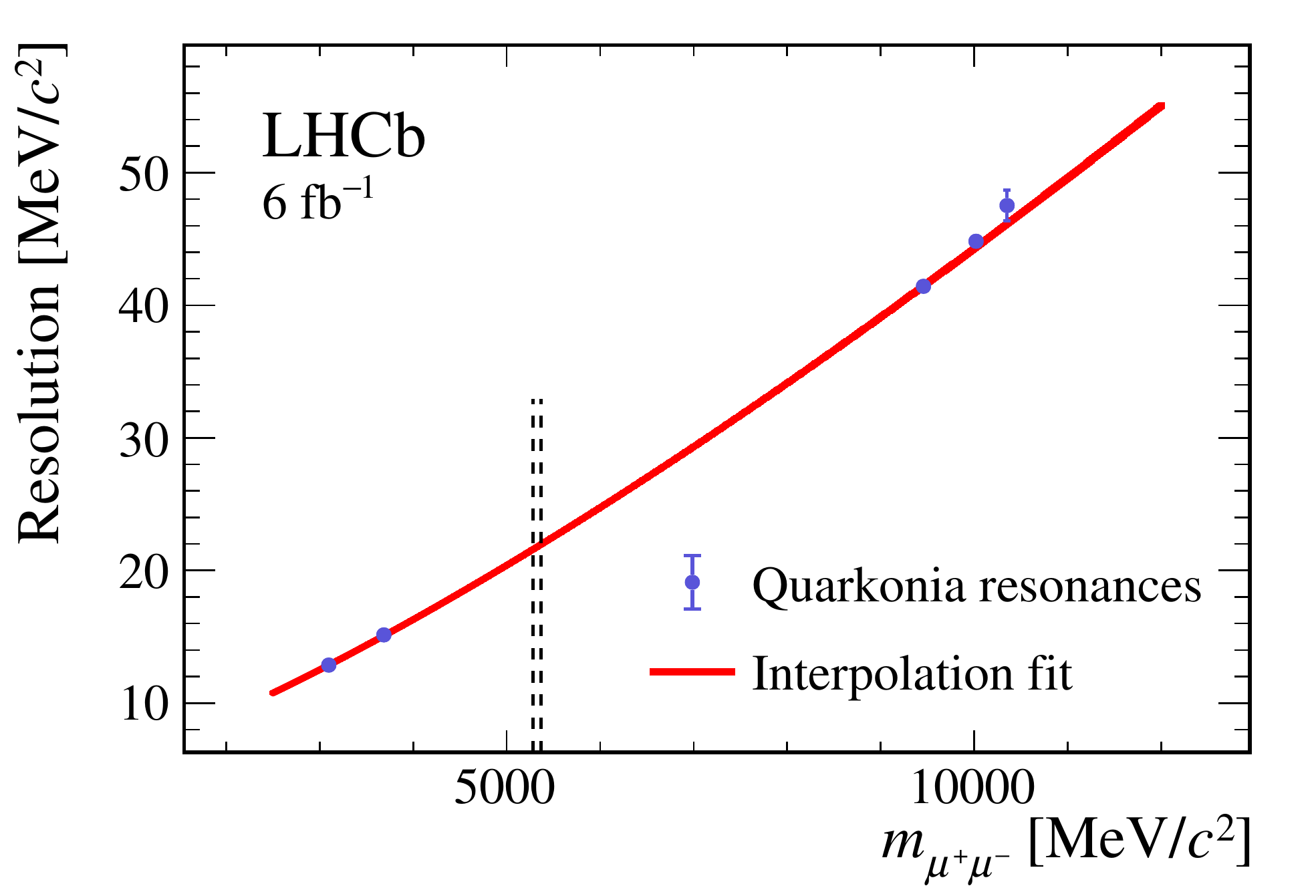}
    \caption{Fit to the measured mass resolutions of quarkonia resonances (blue dots) to obtain the mass resolution at the \Bd and \Bs masses as indicated by the two dashed lines in (left) \runone and (right) \runtwo data. }
    \label{fig:widthcalib}
\end{figure}

The mass resolution is calibrated with data samples containing charmonium (\jpsi, \psitwos) and bottomonium (\OneS, \TwoS and \ThreeS) resonances decaying into two muons, selected similarly to the signal. The natural widths of all these resonances are negligible compared to the mass resolution of the \lhcb experiment. The resolution of each resonance is obtained from a mass fit to the data. The distributions of the dimuon mass, $m_{\mu^+\mu^-}$, of quarkonium decays, shown in Fig.~\ref{fig:quarkonia}, are modelled with a DSCB function. The combinatorial background is modelled with an exponential shape. In the bottomonium fits, the tails are constrained from simulation. A second-order Chebychev polynomial is used as alternative shape for the combinatorial background. The difference between the mass resolutions measured with the two background descriptions is taken as systematic uncertainty. The power-law function $\sigma_{\mu^+\mu^-} (m_{\mu^+\mu^-}) = a_0 + a_1 \cdot (m_{\mu^+\mu^-})^{a_2}$ is found to describe the mass resolution of simulated Drell-Yan events accurately. This function is fitted to the measured resolutions of the quarkonia, including their systematic uncertainties, and used to determine the mass resolution at the \Bs and \Bd mass, as shown in Fig~\ref{fig:widthcalib}. The mass resolution in Run 2 is found to be slightly better than in Run 1, which is explained by improvements in the track reconstruction. The \ThreeS resolution is larger than expected from the power-law function and thus an interpolation with a third-order polynomial was also performed. The differences of the interpolated \Bd and \Bs mass resolutions with respect to their default widths are assigned as a systematic uncertainty.

The four tail parameters of the \Bmm signal shape are in common between the \Bd and \Bs decays. They are  determined from the mass distributions in simulation, after convoluting them with a Gaussian function to match their core resolutions with the values found in data as described in the previous paragraph.

Each step of the mass calibration is performed separately for each data-taking year. Subsequently, average results for the mass shape parameters for \runone and \runtwo are calculated by weighting the year-by-year values by the integrated luminosity in each year. These averages are then used in the final mass fit, with their values shown in Table~\ref{tab:masscalib}. A small but significant dependence of the mass resolution with the BDT response is found, for which correction factors ranging from $0.97$ to $1.03$ are applied. For the left-hand tail parameters, BDT-dependent correction factors ranging from $0.9$ to $1.2$ are obtained.

\begin{table}[b]
\small
 \caption{\label{tab:masscalib} Luminosity-weighted signal mass shape parameter combinations per data set, including propagated uncertainties. Where appropriate, statistical and systematic uncertainties are added in quadrature. As the tail parameters determined in \runone and \runtwo are consistent, they are combined into common estimates. }
 \begin{center}
  \begin{tabular}{l|c c c}
  \toprule
    & Run 1 & Run 2     & Common \\
    \midrule
  \Bd mean $(\!\mevcc)$ & $5284.61\pm 0.18$ & $5280.13 \pm 0.16$ & -\\
  \Bs mean (\!\mevcc) & $5372.27\pm 0.36$ & $5367.54 \pm 0.26$ & -\\
  \midrule
  \Bd width (\!\mevcc) & $22.4 \pm 0.7$ & $21.6 \pm 0.6$ & -\\
  \Bs width (\!\mevcc) & $22.8 \pm 0.7$ & $22.0 \pm 0.6$ & -\\
  \midrule
 $n_l$ tail & $1.55 \pm 0.06$ & $1.49 \pm 0.02$ & $1.50 \pm 0.04$ \\
 $n_r$ tail  & $5.86 \pm 0.31$ & $5.80 \pm 0.24$ & $5.81 \pm 0.26$ \\
 $\alpha_l$ tail  & $1.79 \pm 0.03$ & $1.79 \pm 0.01$ & $1.79 \pm 0.02$ \\
 $\alpha_r$ tail  & $2.12 \pm 0.04$ & $2.14 \pm 0.03$ & $2.14 \pm 0.04$ \\
    \bottomrule
  \end{tabular}
 \end{center}
\end{table}

\subsection{BDT calibration}
\label{sec:bdt_calibration}

The BDT response is determined using simulated decays, to which corrections are applied to account for possible discrepancies between simulation and data. This calibration procedure is performed in three steps. Firstly, the \B-meson kinematics and the detector occupancy of the simulation are corrected using control channels in data. Secondly, the effect of the PID and trigger selections are evaluated on data and used to correct the BDT response. Finally, since the BDT response for real \bsmumu decays is strongly correlated with their decay-time distribution and hence with the \bsmumu effective lifetime, which is \textit{a priori} unknown, an additional correction is applied under different effective lifetime hypotheses. Taking these three correction factors into account, the fraction of signal decays in each BDT region, \fBDT, can be expressed as
\begin{equation}
    \fBDT=\fBDTmumuSim \cdot \fBDTmumuPID \cdot \fBDTmumutrig ~ (\cdot ~ \kBDTtau),% out-symbol fixed
    \label{eq:bdtMC}
\end{equation}
where \fBDTmumuSim is the fraction of events per BDT region in the corrected simulation, \fBDTmumuPID and \fBDTmumutrig are the weights used to correct for the PID and trigger selections, respectively, and the effective lifetime correction \kBDTtau is used for \bsmumu and \bsmumugamma decays. No such correction is needed for \bdmumu decays due to the small width difference between the \Bd mass eigenstates.

The \B meson kinematics of the simulated signals are corrected using a gradient boosting reweighter. This technique consists of training a boosted decision tree classifier to align two samples, in this case data and simulation, as described in Ref~\cite{Rogozhnikov:2016bdp}. The transverse momentum \pt, the pseudorapidity $\eta$, and the \chisqip of the \B~candidate are used as input variables for the gradient boosting reweighter, as these are the variables required to correct the simulation. The weights obtained from this procedure are applied to all simulation samples used for calibration and normalisation. The kinematic distributions for \Bd and \Bs mesons differ, as determined in hadronisation fraction measurements~\cite{LHCb-PAPER-2020-046}, thus they are corrected with a sample of \bujpsik and \bsjpsiphi decays, respectively.

An additional correction to the BDT distribution shape stems from the event occupancy, measured as the number of tracks in the event. As this affects the muon track isolation variables, which are important inputs to the BDT classifier, the correction is determined in four intervals of the total number of reconstructed tracks. The correction weights are determined by comparing the relative number of \bujpsik decays in background-subtracted data and simulated samples in these intervals. It is ensured that the input variable distributions of \bujpsik candidates match those of \bmumu candidates as closely as possible by evaluating the BDT response based on the final state muons and the \Bu candidate, with two exceptions: the final state kaon is excluded from the calculation of the isolation variables, and the decay vertex \chisq is determined on the \jpsi candidate, to match the number of degrees of freedom of the signal.

The PID efficiency correction per BDT region for \Bmm decays, \fBDTmumuPID, is determined on dedicated calibration samples and convolved with the muon kinematics of simulated signal per region, as described in Sec.~\ref{sec:efficiencies}. As the total PID efficiency is part of the normalisation (Sec.~\ref{sec:Normalisation}), the PID efficiency correction of the BDT response is determined as the relative PID efficiency per BDT region; no uncertainty is assigned on this correction, as it is already included in the total efficiency.

A similar procedure is adopted for the trigger efficiency per BDT regions, \fBDTmumutrig. The trigger efficiencies are determined on data samples containing \bujpsik decays with the trigger calibration method (see Sec.~\ref{sec:efficiencies})~\cite{LHCb-PUB-2014-039}. They are calculated in ranges of the maximum \pt of the two muons and the product of the \pt of the two muons, which are the variables used for the muon and dimuon hardware trigger. The trigger efficiency per BDT region is determined by the convolution of the obtained efficiencies with the kinematics of simulated signal. The details of this method, also employed for the full efficiency determination, are given in Sec.~\ref{sec:efficiencies}.

In simulation, \Bsmumug decays are generated using the mean \Bs lifetime, while the effective lifetime can have any value between the lifetime of the light and the heavy mass eigenstates. As the BDT classifier is correlated with the \Bs candidate decay time, an additional correction is included for the \Bsmumug BDT response distributions. The correction is evaluated for $\ADeltaGamma = -1$, 0 and 1, corresponding to $\tauBsmm = 1.423$, $1.527$ and $1.620\ps$, and covering the full physically allowed range. Simulated candidates selected with the procedure described in Sec.~\ref{sec:Selection} are weighted according to
\begin{equation}
  \omegaj=\frac{\taugen}{\tauBsmm}e^{-t_j\left(1/\tauBsmm-1/\taugen\right)},
\end{equation}
where $t_j$ is the reconstructed decay time of the candidate $j$, $\taugen$ is the lifetime used for generation, and $\tauBsmm$ is the effective lifetime calculated from \ys, \tauBs and \ADeltaGamma. A correction factor, \kBDTtau, is then calculated for each BDT region according to
\begin{equation}
    \kBDTtau=\sum_{j=1}^{N_i}\frac{\omegaj}{N_i}
\end{equation}
where $N_i$ is the number of signal decays in each BDT region.

\begin{figure}[tb]
    \centering
    \includegraphics[width=0.49\textwidth]{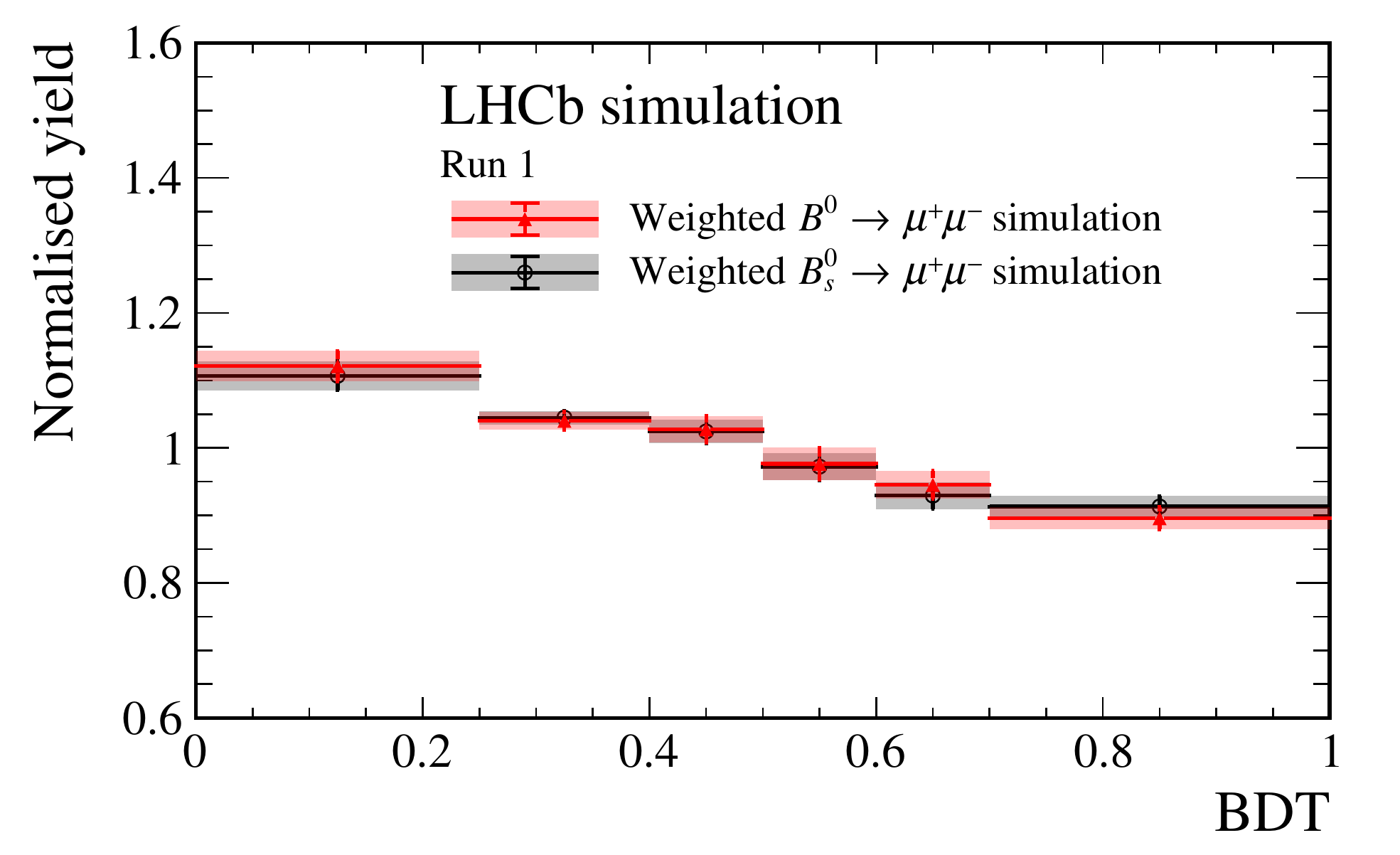}
    \includegraphics[width=0.49\textwidth]{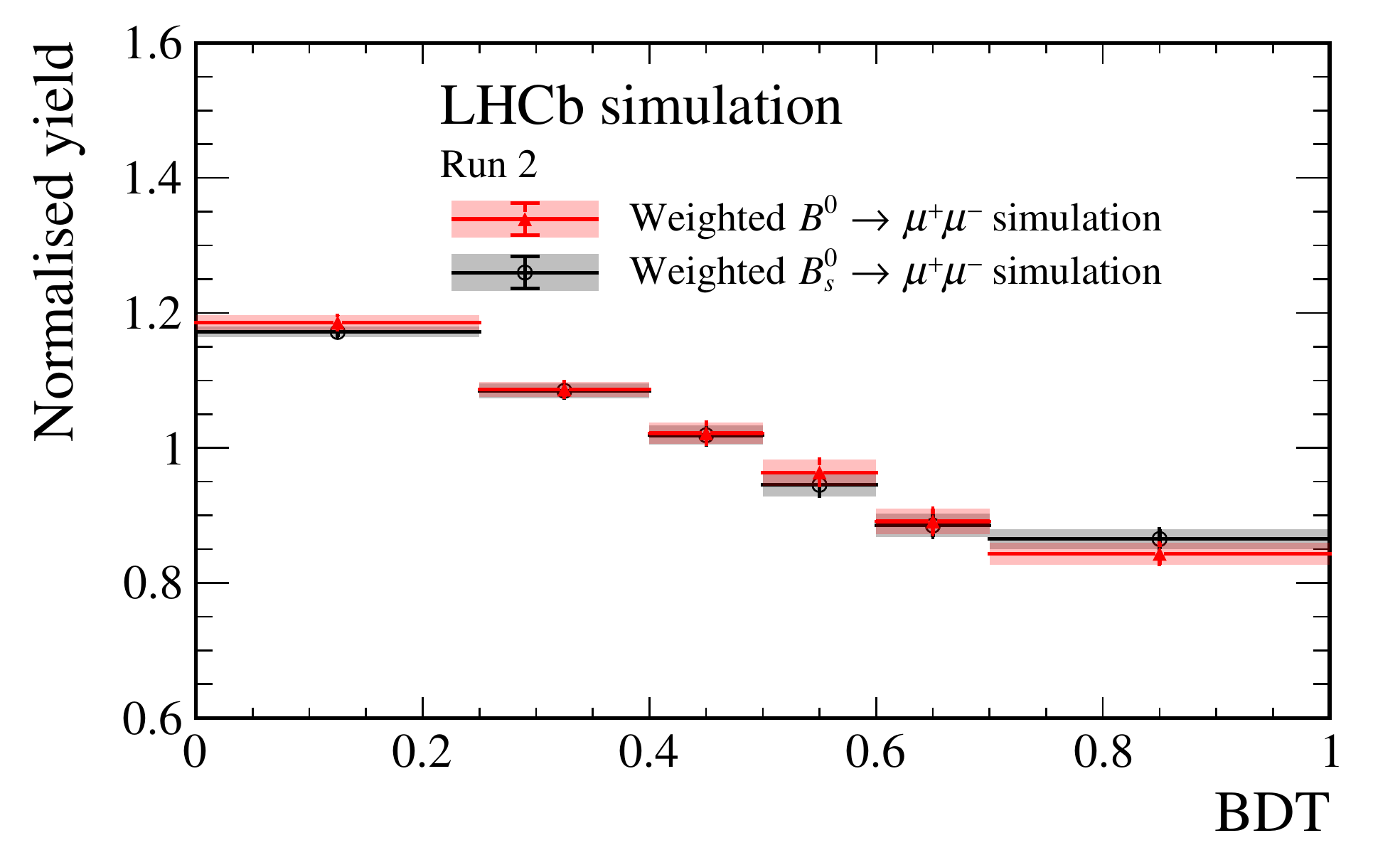}
    \caption{The calibrated BDT distribution for \bsmumu decays with $\ADeltaGamma=1$  (black) and \bdmumu decays (red) for (left) Run 1 and (right) Run 2, including the total uncertainty on the fraction per BDT region. The \bsmumu and \bdmumu distributions are determined on corrected simulated samples, as described in the text.}
    \label{fig:bdt_calibration}
\end{figure}

The calibrated BDT response for \Bsmumu decays under the SM hypothesis $\ADeltaGamma = 1$ and for \Bdmumu decays are shown in Fig.~\ref{fig:bdt_calibration}. The main systematic uncertainties on the calibrated BDT distribution arise from the limited samples used for the trigger efficiency and the event occupancy corrections; they are summed in quadrature with each other and with the statistical uncertainties to determine the total uncertainty on the BDT distribution. The systematic uncertainty on the trigger efficiency correction has been evaluated using \bujpsik as a control channel and is related to possible discrepancies between the signal decay and the control channel and to mismodelling in the simulation used in the TISTOS method described in Sec.~\ref{sec:efficiencies}. The systematic uncertainty due to the event occupancy is obtained by comparing the correction obtained with the default and an alternative interval scheme used to determine the occupancy weights. The systematic uncertainties from the PID selection correction and kinematic reweighting are found to be negligible.

To validate the BDT calibration procedure, an alternative calibration is performed using a data sample containing \bhh decays. While this procedure directly measures the BDT distribution on data and the BDT response is expected to be very similar for any two-body \B decay, \Bhh decays require significant corrections to be compared to \Bmm decays, making it less precise than the default strategy.

The two most frequent \B-meson decays into two hadrons, \BdKpi and \mbox{\bskk}, are considered for the BDT calibration cross-check of \bdmumu and \bsmumu, respectively. The same selection is required for \bhh candidates as for the signal, except for the trigger and the particle identification requirements. A trigger selection independent of the \Bds decay products is applied to the candidates to select \Bhh decays in order to avoid selection biases. Then, a PID selection is applied to separate \BdKpi and \bskk decays from other \bhh decays. 

The fraction of \Bhh decays in each BDT region, \fBDThhData, is determined by fitting the mass distribution of the two hadrons with the corresponding mass hypothesis. A detailed description of the mass fit for \bhh decays is given in Sec.~\ref{sec:mass_calib}.

To avoid correlated uncertainties, the BDT distribution from \Bhh data corrected for PID and trigger efficiencies is compared with the distribution of \mbox{\Bmm} decays from corrected simulation samples. The \Bhh PID efficiencies are determined with a dedicated procedure~\cite{LHCb-PUB-2016-021}, while the trigger efficiency is evaluated on \bujpsik decays with the trigger calibration method reported in Sec.~\ref{sec:efficiencies} for the hardware trigger and first software trigger requirements, and determined with \bhh simulated samples for the second software trigger selection.

\begin{figure}[tb]
    \centering
    \includegraphics[width=0.49\textwidth]{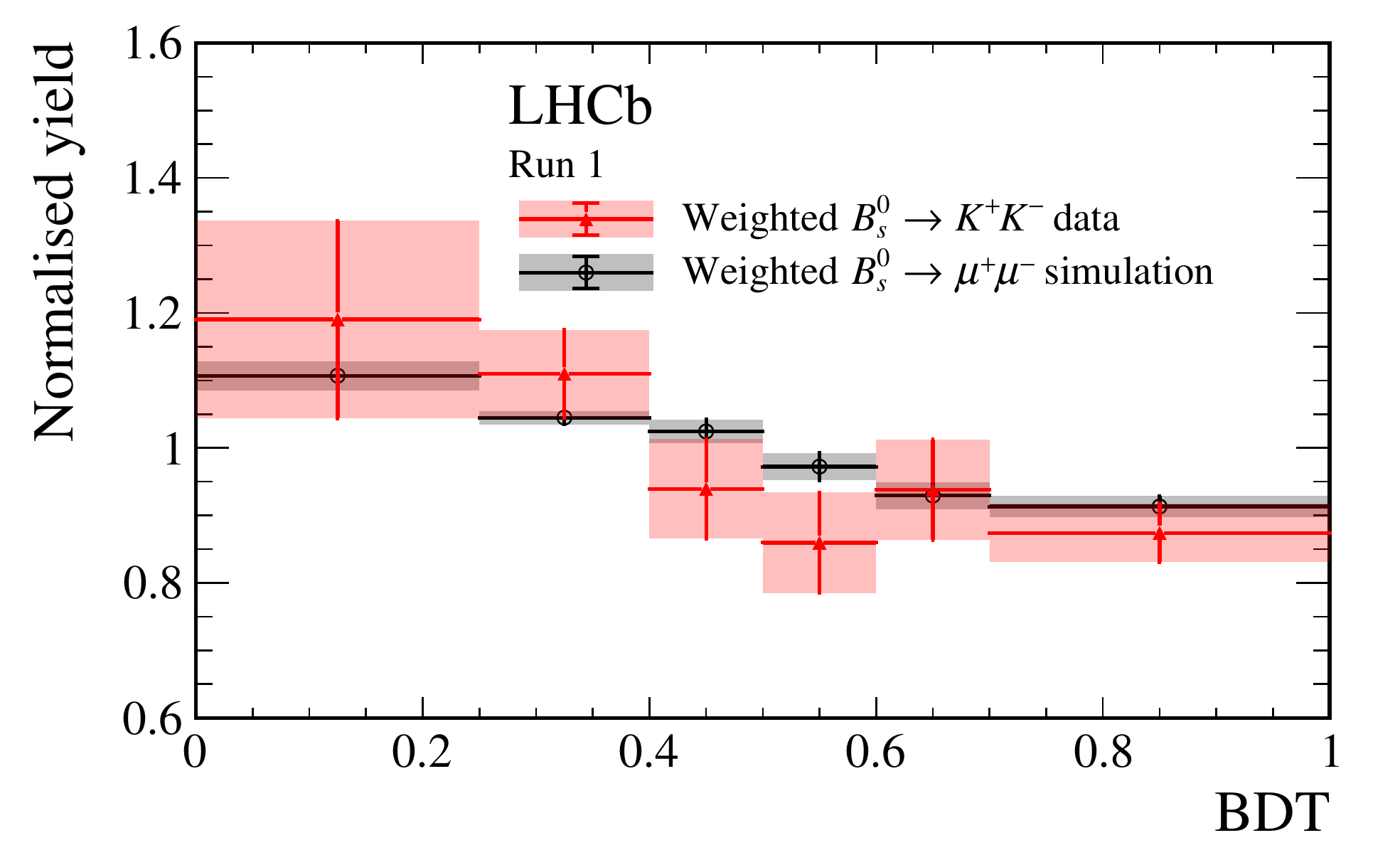}
    \includegraphics[width=0.49\textwidth]{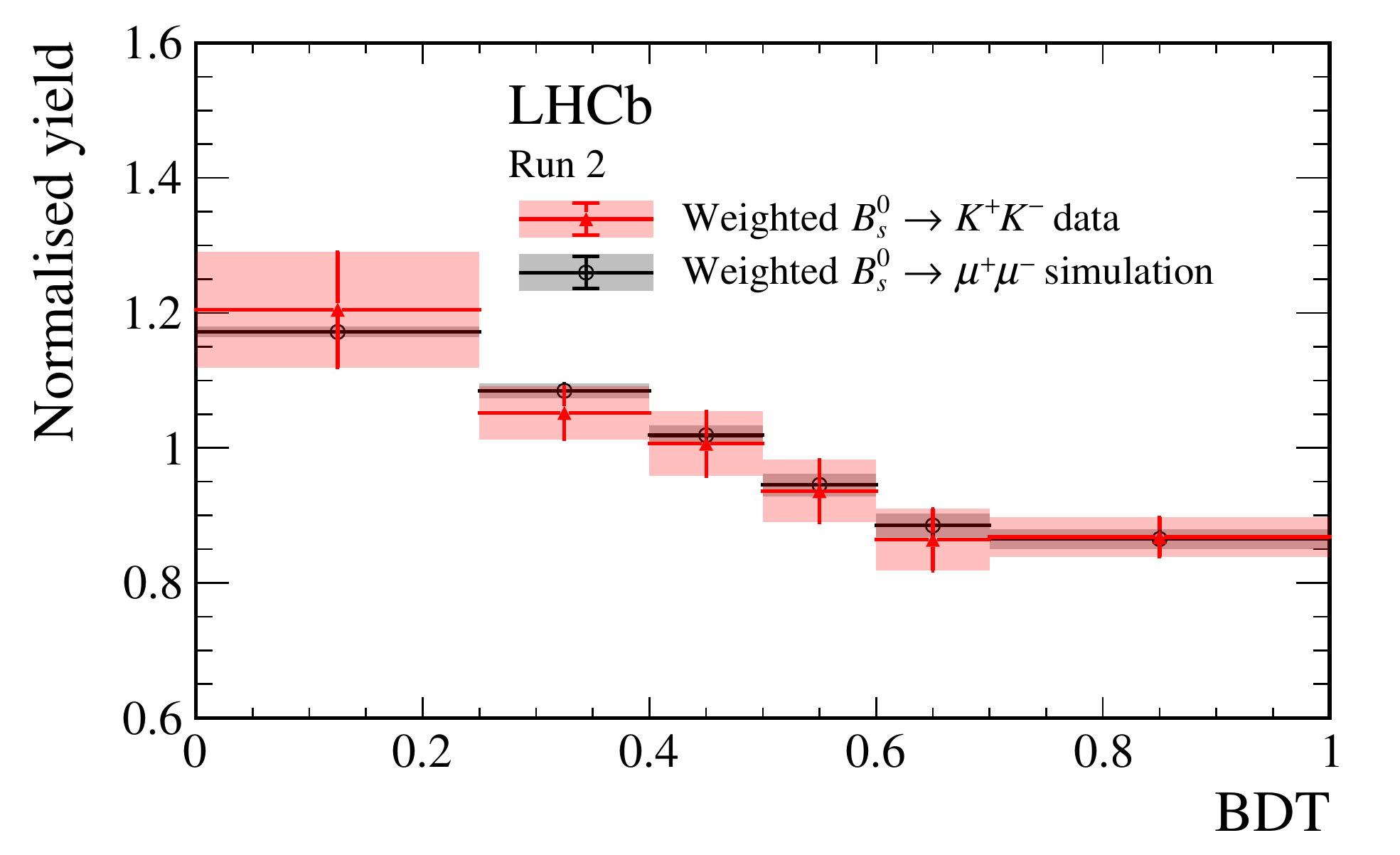}\\
    \includegraphics[width=0.49\textwidth]{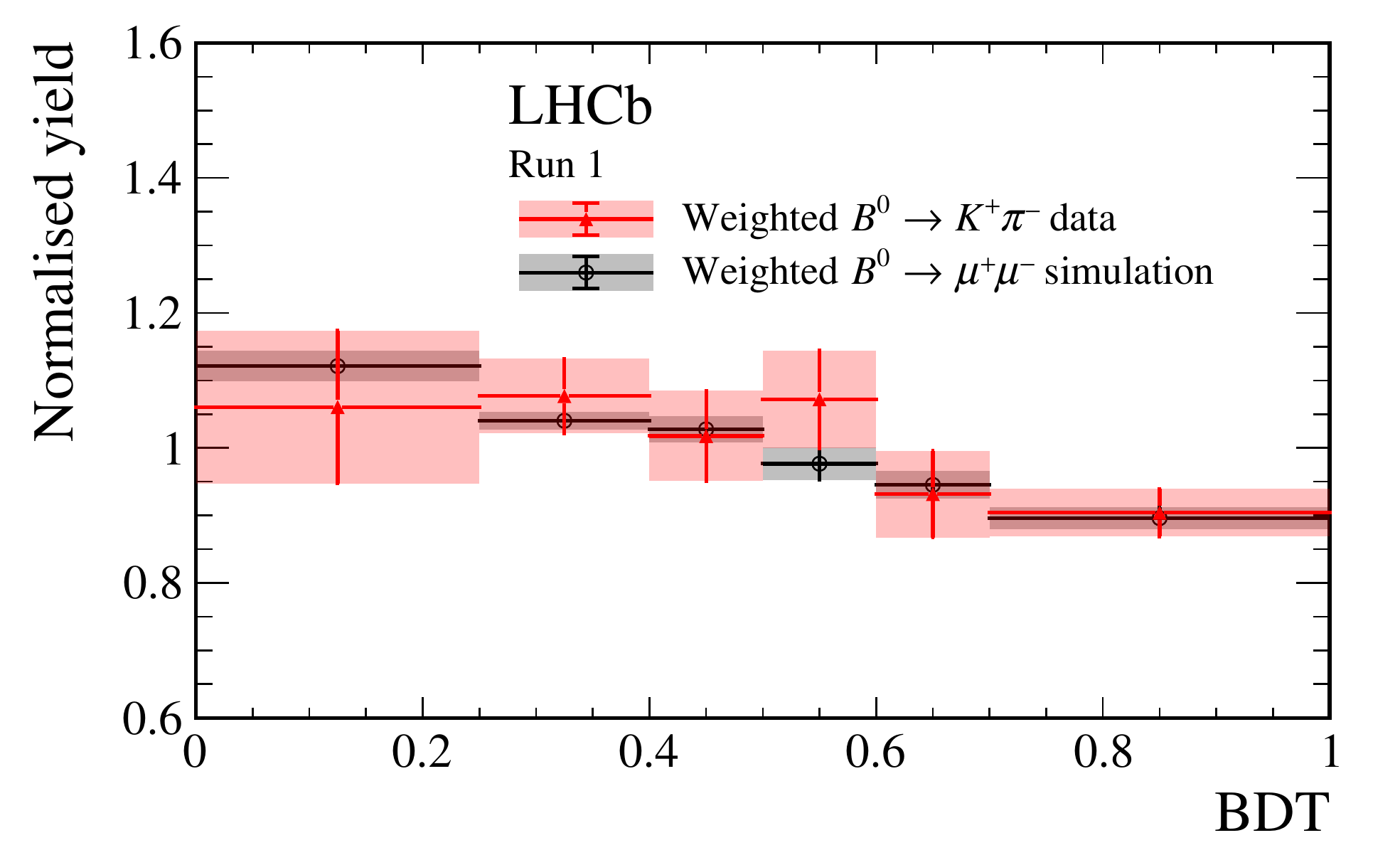}
    \includegraphics[width=0.49\textwidth]{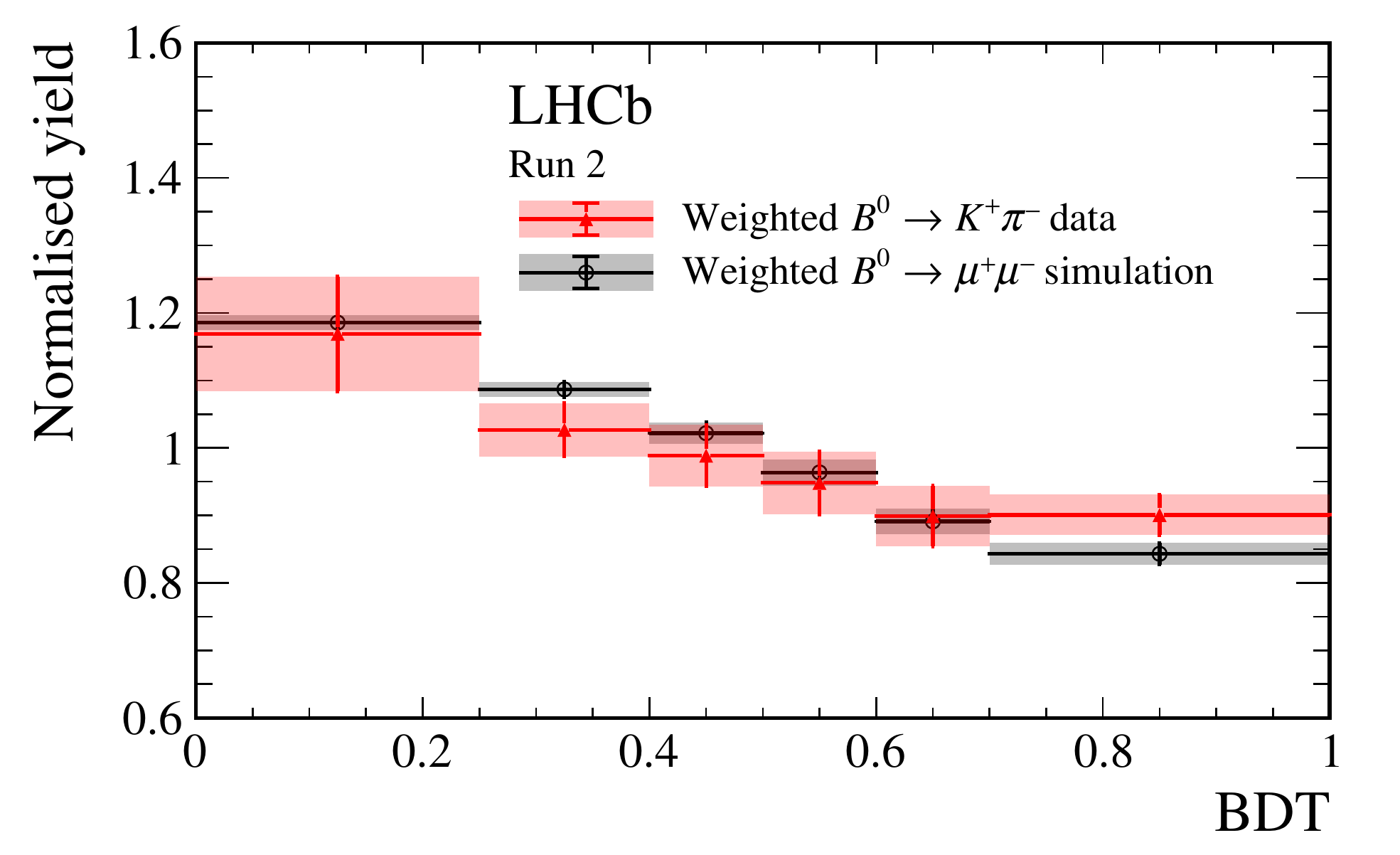}
    \caption{The BDT distributions of (top) \bsmumu and \bskk decays and (bottom) \bdmumu and \BdKpi decays in (left) Run 1 and (right) Run 2 data, including the total uncertainty on the fraction per BDT region. For \bsmumu and \bdmumu decays, the distributions are determined on corrected simulated samples, as described in the text, and are shown in black. The \bskk and \BdKpi distributions are determined using fits to data as described in the text and are shown in red.}
    \label{fig:bdt_calibration_bmumu_bhh}
\end{figure}

Therefore, the fraction of \Bhh decays per BDT region, \fBDThh, can be described as
\begin{equation}
    \fBDThh=\fBDThhData \cdot \fBDThhPID \cdot \fBDThhtrig, 
    \label{eq:bdt_Bhhdata}
\end{equation}
where \fBDThhPID and \fBDThhtrig are the relative PID and trigger efficiencies for \mbox{\bhh} decays versus \Bmm decays. The corrected \BdKpi and \bskk distributions are compared with those determined on corrected simulated \bdmumu and \bsmumu samples, respectively, as shown in Fig.~\ref{fig:bdt_calibration_bmumu_bhh}. The BDT distributions of \bsmumu and \bskk decays are compared for the same effective lifetime, namely for \ADeltaGamma = 1. Because of the good agreement between the two different methods used to calibrate the BDT response, no additional systematic uncertainty is assigned to the BDT distribution. 
 
\section{Normalisation}
\label{sec:Normalisation}

The branching fractions of the signal channels are estimated by comparing their yields in data with those of two normalisation channels with well-known branching fractions, \bujpsik and \BdKpi, according to 
\begin{equation}
\mathcal{B}(\bdsmumu(\gamma)) =  \frac{f_{\rm{norm}}}{f_ {\rm{sig}}} \frac{ \varepsilon_{\rm{norm}}}{\varepsilon_{\rm{sig}}} \frac{N_{\rm sig} }{N_{\rm norm} }\mathcal{B}_{\rm norm}
= \alpha_{\rm{sig}} N_{\rm sig},
\label{eq:norm}
\end{equation}
where $\mathcal{B}$, $\varepsilon$ and $N$ are the branching fraction, efficiency and yield of the corresponding channel and $f_{{\rm sig}({\rm norm})}$ indicates the fragmentation fraction of the relevant \B meson. Signal candidates having $\bdt<0.25$ are not included in the fit to the dimuon mass distribution. The parameter $\alpha_{\rm sig}$ is the single-event sensitivity. In the following, the different elements entering Eq.~\ref{eq:norm} are described. The final single-event sensitivity is obtained for each signal channel as the weighted average of those obtained with the two normalisation channels, taking the correlations between the inputs into account. The branching fractions of the two normalisation channels are taken as $\BRof \bujpsik = (6.02 \pm 0.17)\times 10^{-5}$~\cite{Bdkpi_babar,bdkpi_belle,Bdkpi_cleo}, including the $\decay{\jpsi}{\mup\mun}$ branching fraction, and \mbox{$\mathcal{B}(\BdKpi) = (1.96 \pm 0.05)\times 10^{-5}$}~\cite{PDG2020,bujpsik_belle2019,bujpsik_belle17,bujpsik_babar06,bujpsik_babar05,bujpsik_babar04,bujpsik_belle02,bujpsik_cleo2,bujpsik_cleo,bujpsik_argus}. The normalisation for the \bsmumugamma decay is calculated only in the region \mbox{$m_{\mumu}>4.9\gevcc$} where the branching fraction is measured.

\subsection{Normalisation channel yields}
\label{ssec:NormChannelYields}
\begin{figure}[!t]
    \centering
    \includegraphics[width=0.49\linewidth]{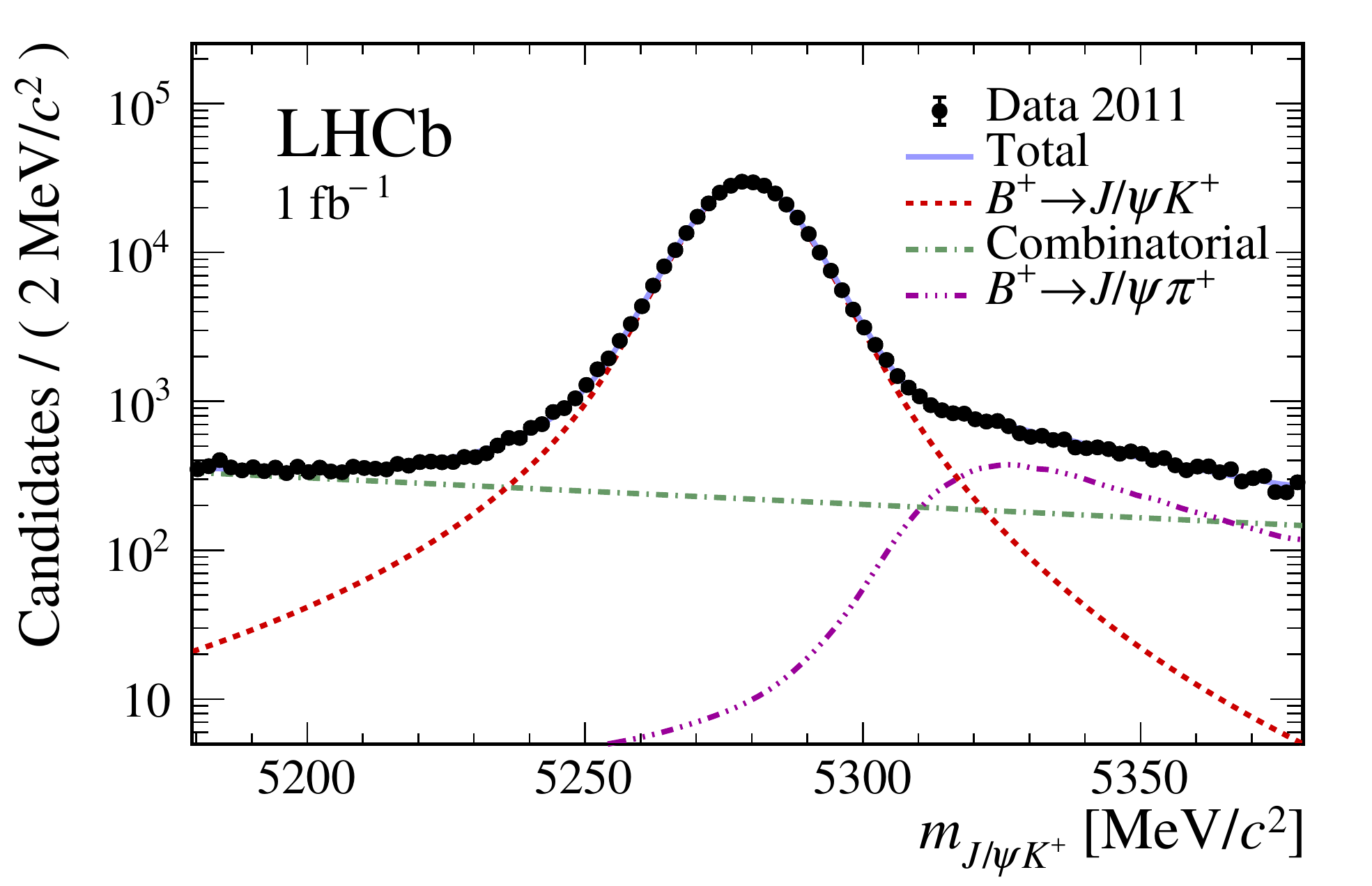}      
    \includegraphics[width=0.49\linewidth]{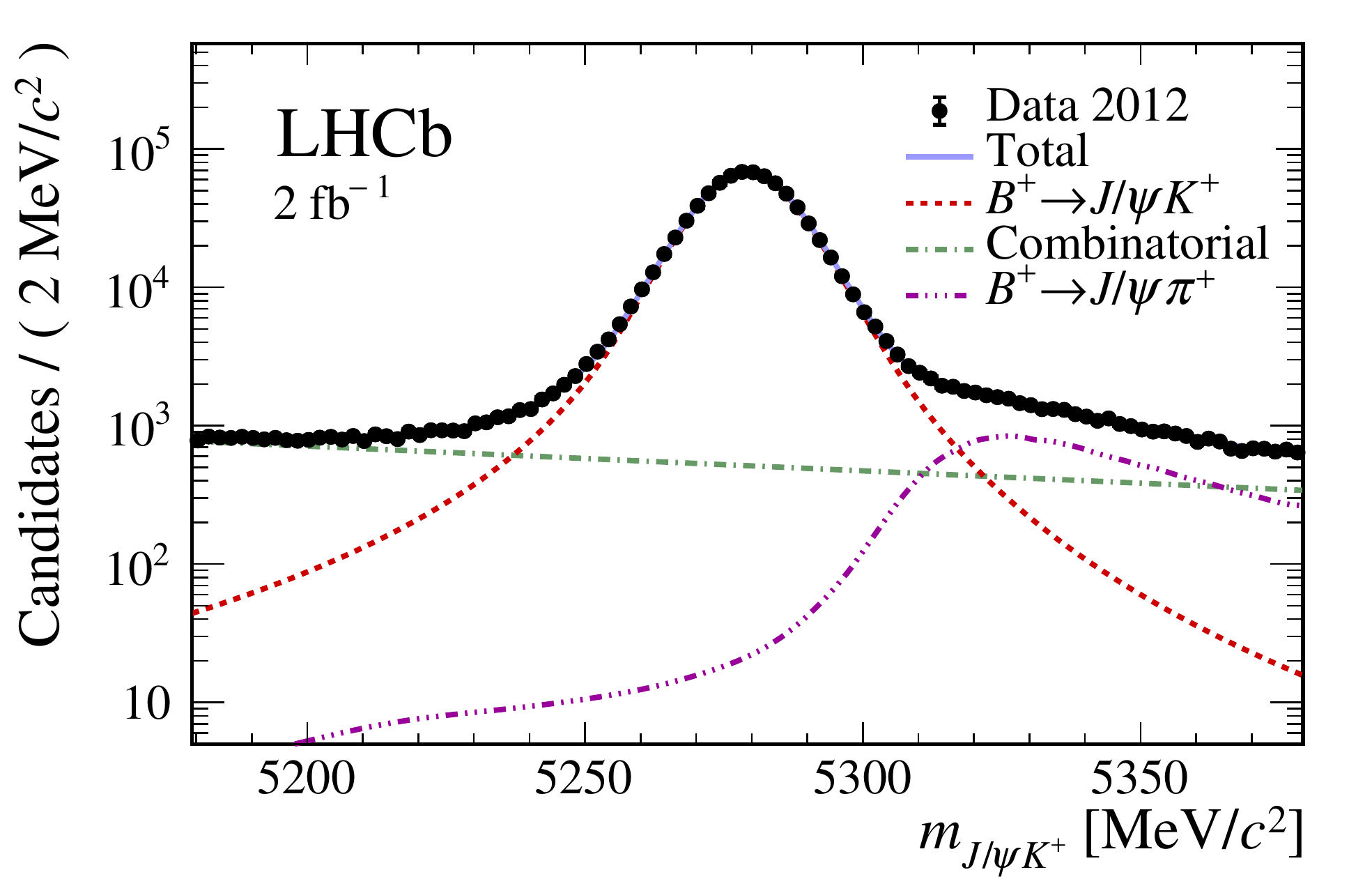}\\   
    \includegraphics[width=0.49\linewidth]{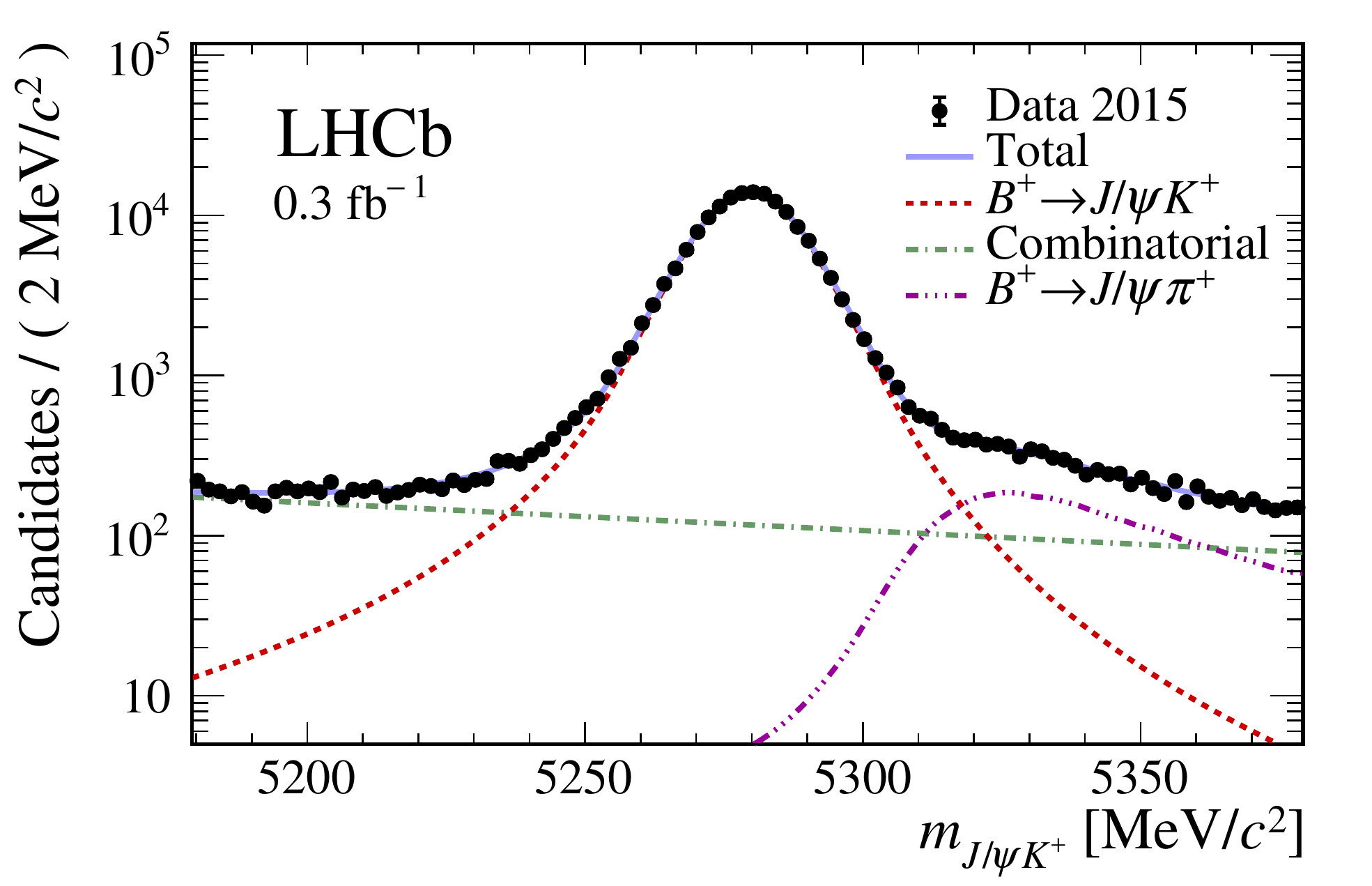}   
    \includegraphics[width=0.49\linewidth]{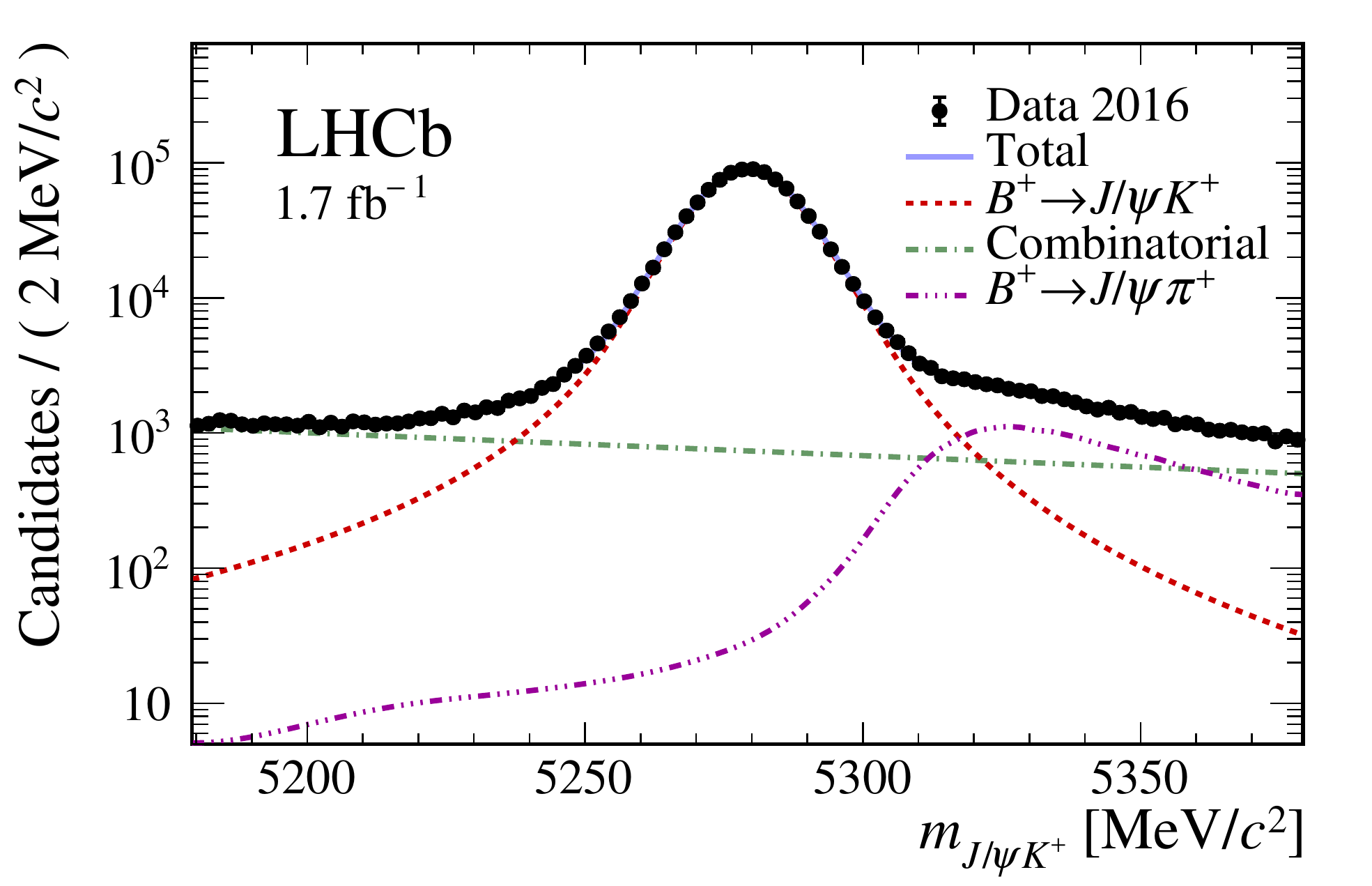}\\
    \includegraphics[width=0.49\linewidth]{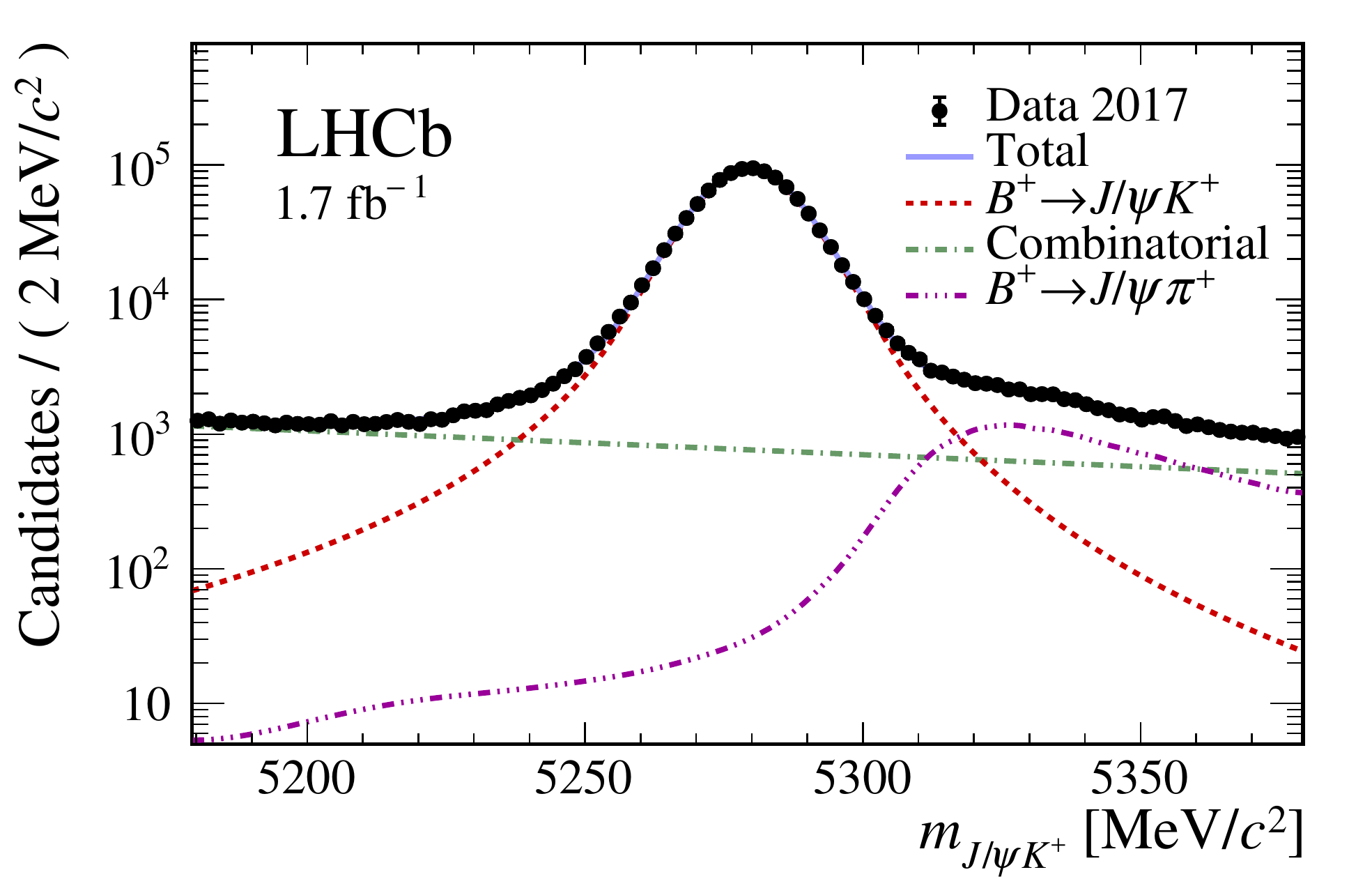}   
    \includegraphics[width=0.49\linewidth]{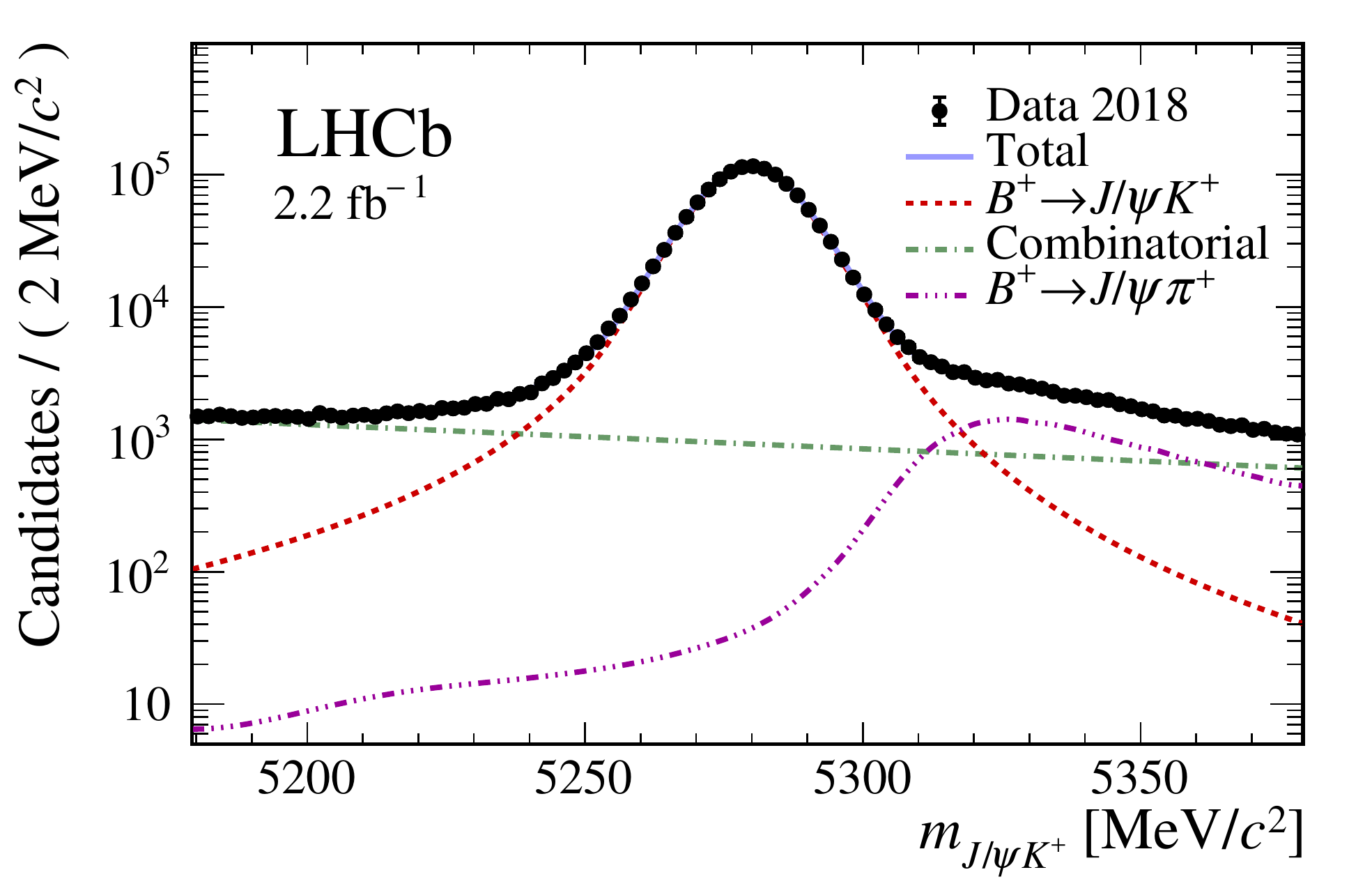}  
    \caption{Mass distribution of \bujpsik candidates in data for different data-taking years. 
    Superimposed is a fit to the distribution: the blue line shows the total fit, the red dashed line is the \bujpsik component, 
    the green dash-dotted line is the combinatorial background, the purple dash-three-dotted line is the \bujpsipi misidentified background.}
    \label{fig:bujpsik_mass}
\end{figure}

The yields of the normalisation channels are obtained through unbinned extended maximum-likelihood fits to the mass distributions of the candidates for each data-taking year separately, after the corresponding selection described in Sec.~\ref{sec:Selection}. To improve the mass resolution of \bujpsik candidates, the \jpsi mass is constrained to its known value~\cite{PDG2020,jpsi_mass_kedr,jpsi_mass_e760,jpsi_mass_olya,jpsi_mass_baglin}. The mass distributions for selected \bujpsik candidates are shown in Fig.~\ref{fig:bujpsik_mass} for the different data taking periods. The mass distribution of signal \bujpsik decays is described by a Hypatia function~\cite{Santos:2013gra} with parameters Gaussian-constrained to the values derived from simulation within their uncertainties, except for the mean and width, which are free to vary in the fit. In addition to signal, the selected candidates contain a contribution of combinatorial background and a small contamination from the \bujpsipi decay, where a pion is misidentified as a kaon. The mass distribution of the combinatorial background is described by an exponential function with the slope left free to vary in the fit. The \bujpsipi decay, which is expected to peak at higher mass values than the \Bu mass due to assigning the kaon mass to a pion track, is described by an analytical function developed in Ref.~\cite{vanVeghel:2020pmk}. This function is obtained by transforming a Gaussian mass distribution under the pion hypothesis to one under the incorrectly-assigned kaon mass hypothesis, using an analytical description of the candidate kinematics. As an alternative model a non-parametric function tuned on a simulated \bujpsipi sample, where the events were reconstructed under the \bujpsik hypothesis as in Ref.~\cite{LHCb-PAPER-2019-020}, has been used to cross check the results. The fits with the two different descriptions for \bujpsipi decays give compatible results for the \bujpsik yields.

Additional possible background from $\decay{\Bc}{\jpsi \Kp\Km\pip}$ and $\decay{\Lb}{\jpsi \Pp \Km}$ decays has been investigated and found to be small and evenly distributed in the considered mass range, and hence is considered negligible. 

\begin{figure}[tb]
    \centering
    \includegraphics[width=0.49\linewidth]{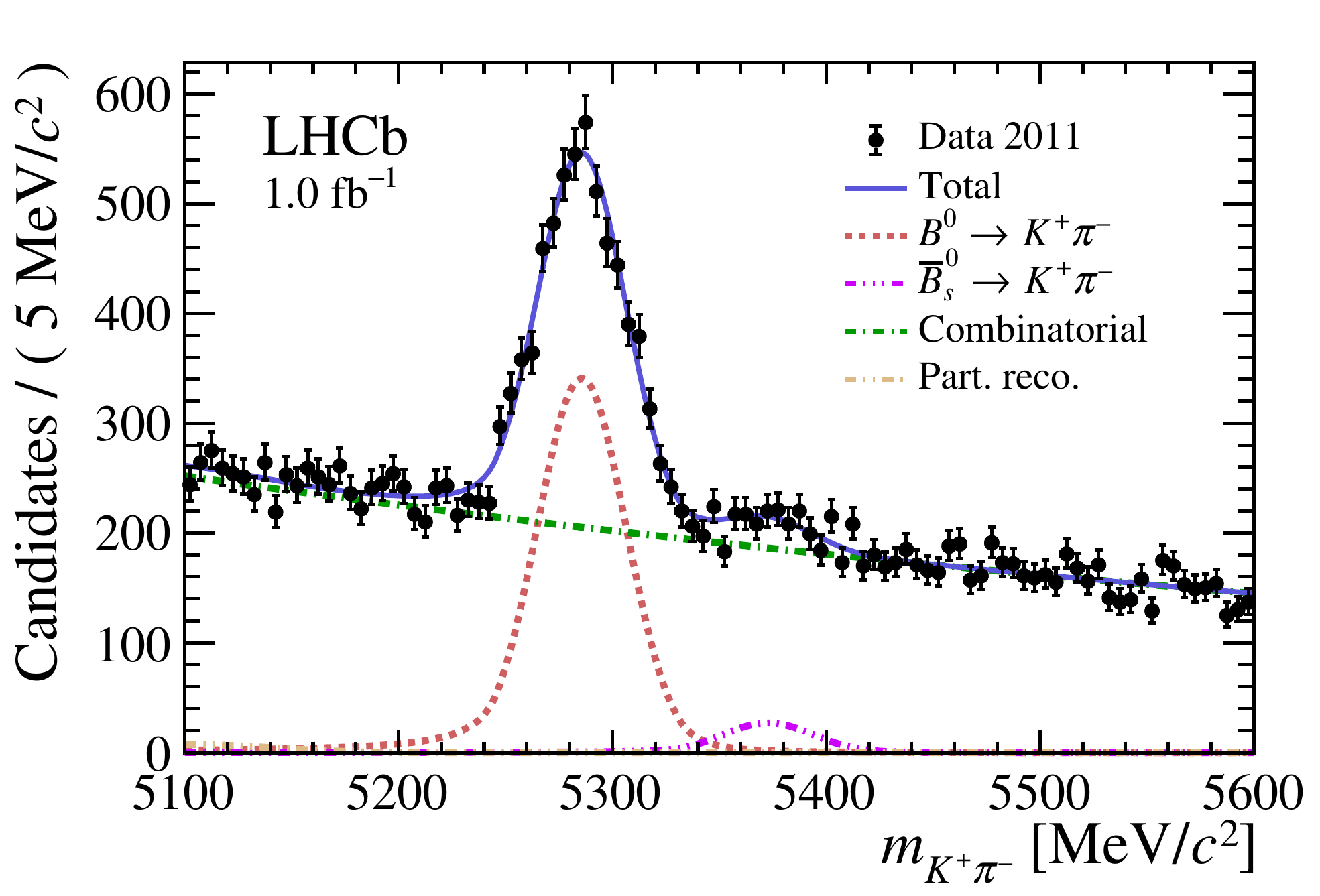}      
    \includegraphics[width=0.49\linewidth]{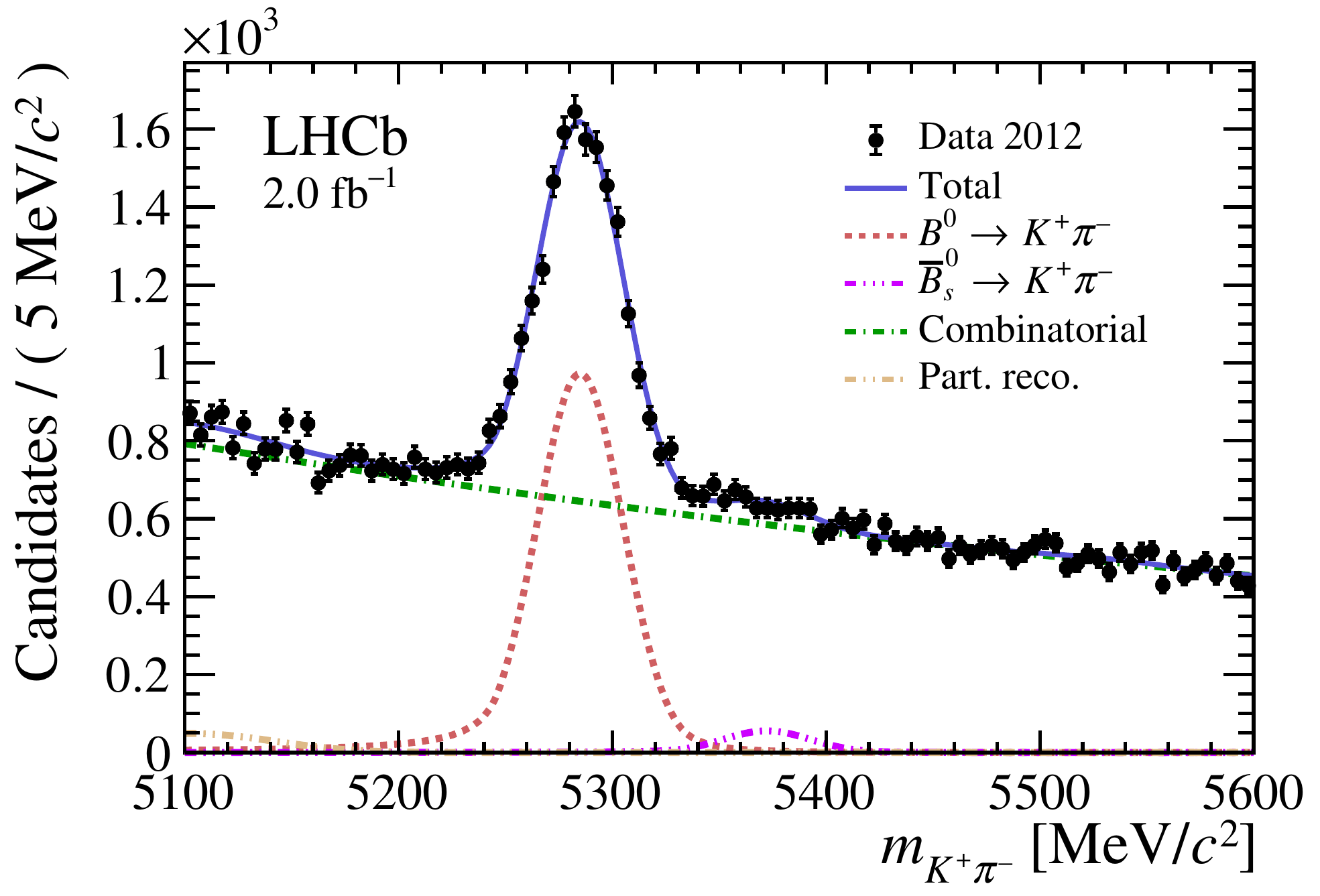}\\   
    \includegraphics[width=0.49\linewidth]{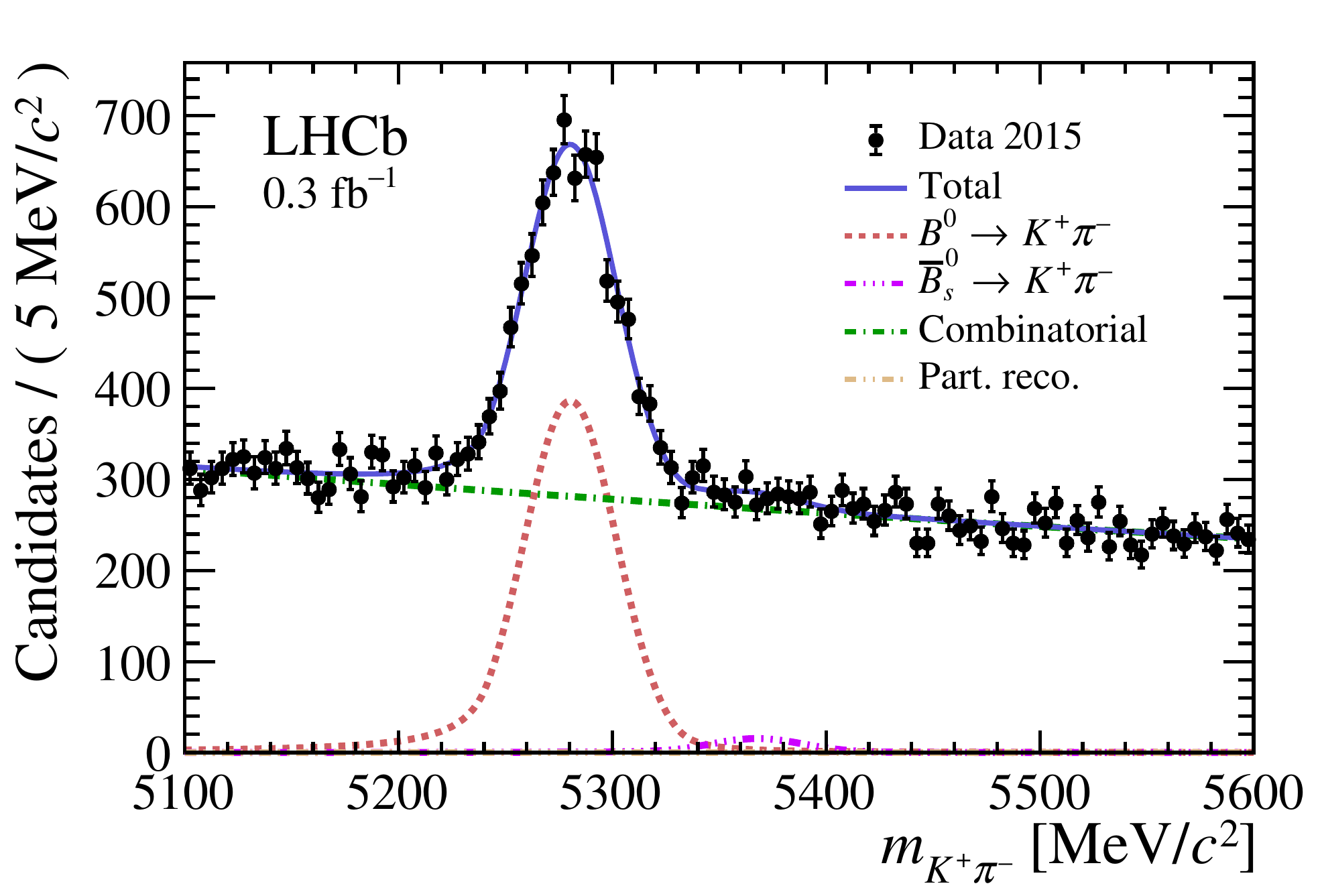}   
    \includegraphics[width=0.49\linewidth]{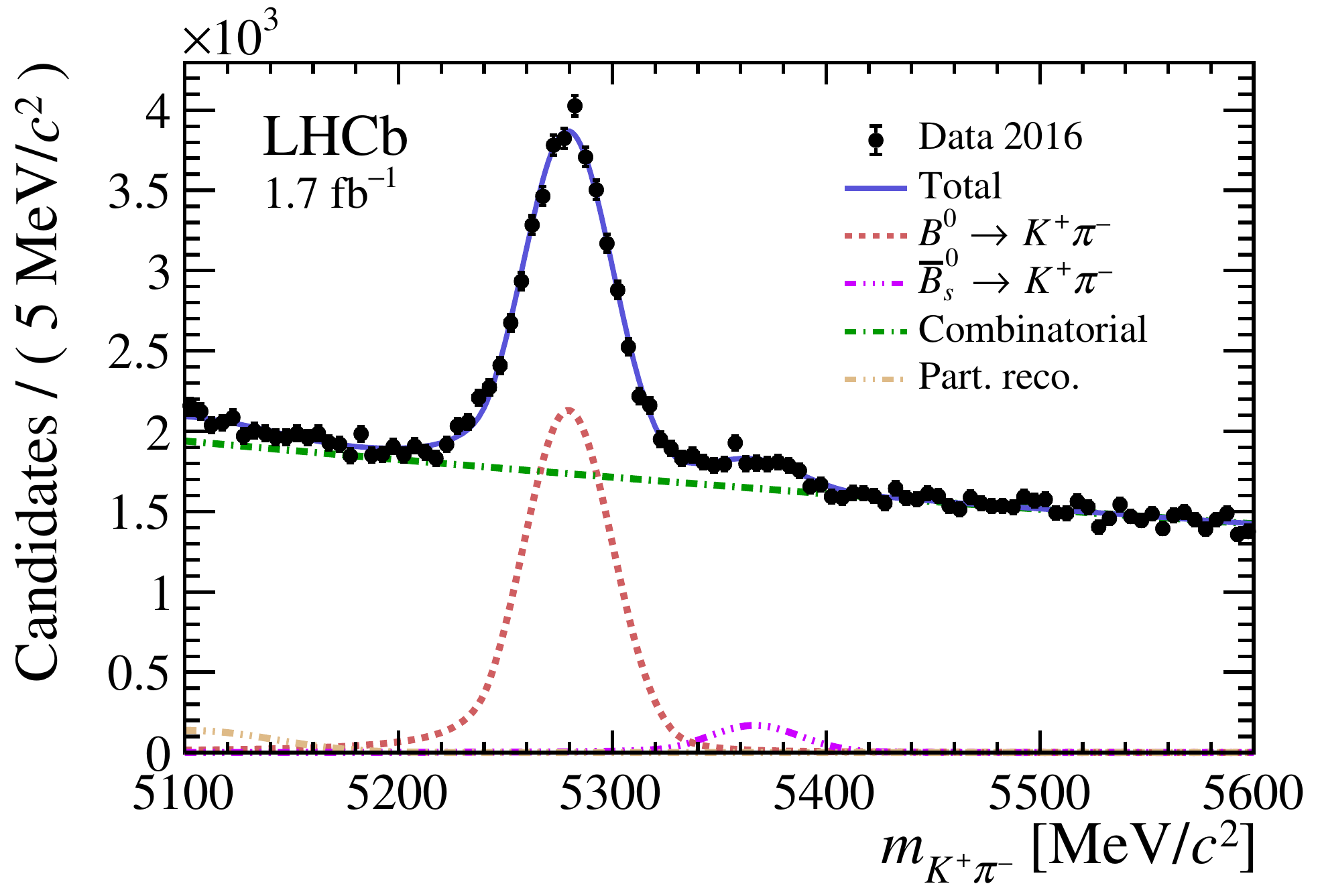}\\
    \includegraphics[width=0.49\linewidth]{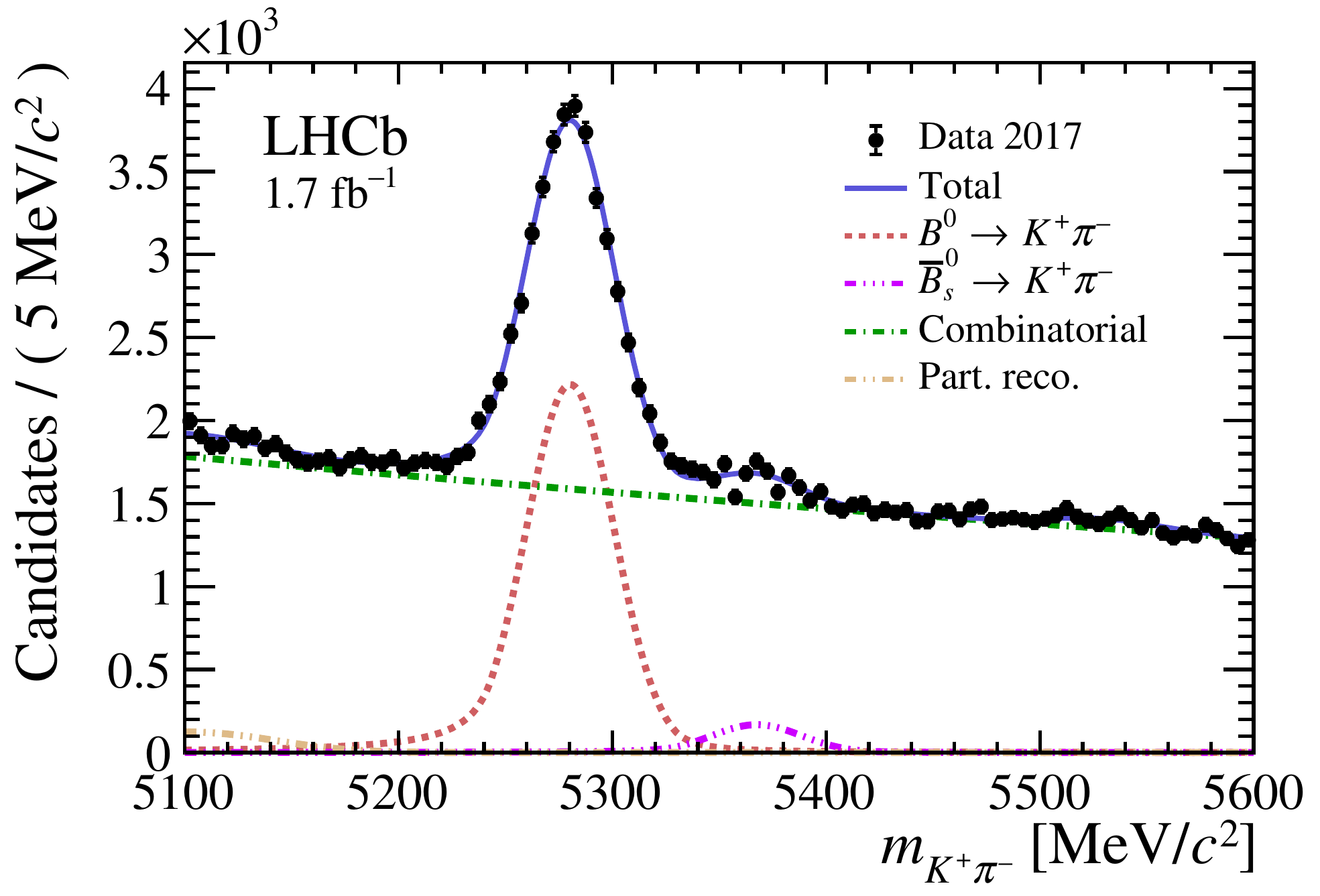}   
    \includegraphics[width=0.49\linewidth]{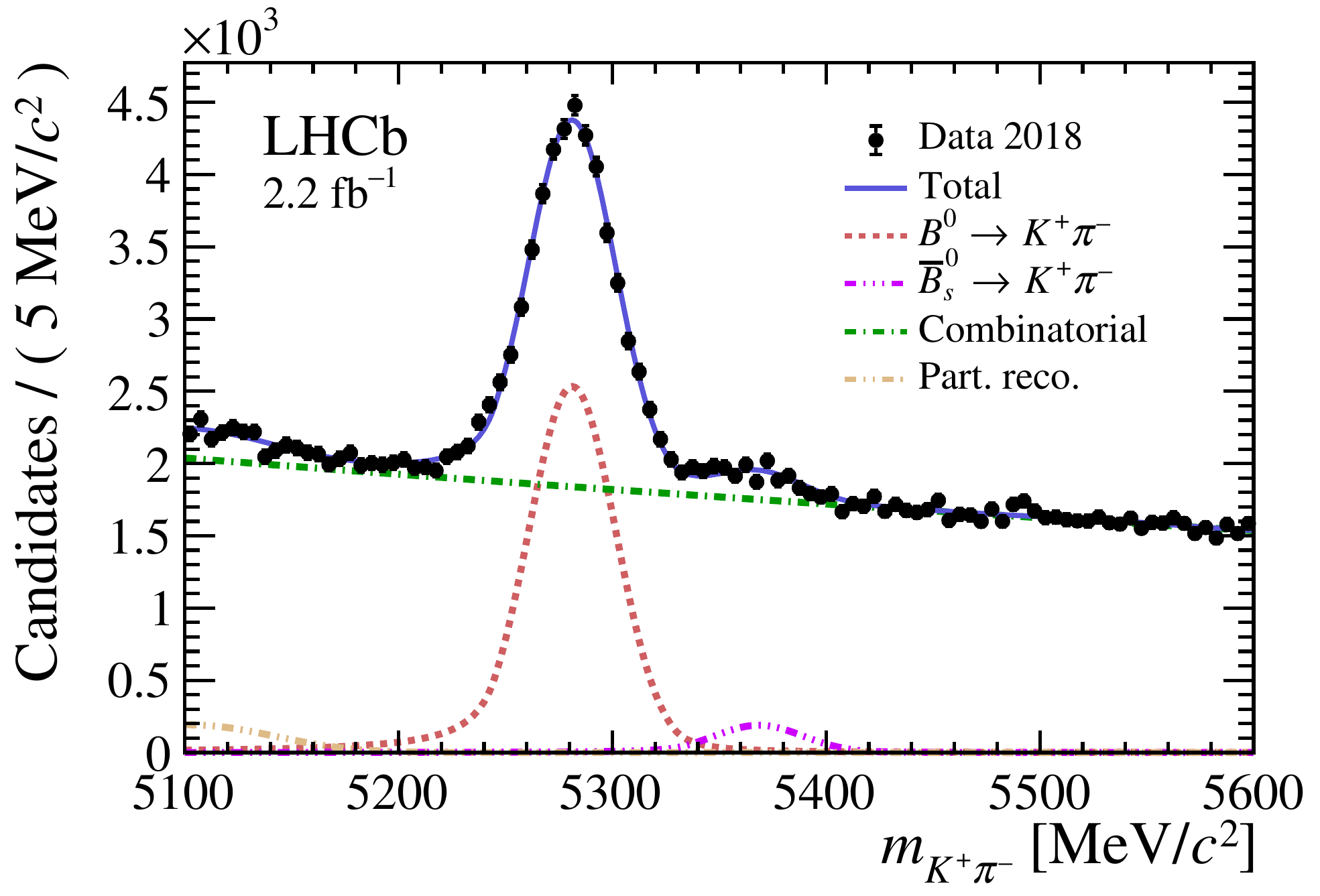}
    \caption{Mass distribution of \BdKpi candidates in data for different data-taking years, triggered independently of the signal. 
    Superimposed is a fit to the distribution: the blue line shows the total fit, the red dashed line is the \BdKpi component, 
    the magenta dashed line is the \BsbarKpi component, 
    the green dashed line is the combinatorial background, 
    and the brown dashed line is the partially reconstructed background component.}
    \label{fig:bdkpi_mass_norm}
\end{figure}

The yield of \BdKpi candidates is determined with a binned maximum-likelihood fit to the data, using the fit model for \BdKpi candidates described in Sec.~\ref{sec:mass_calib}. In contrast to the mass calibration, events are required to be triggered independently of the \BdKpi signal, such that the trigger efficiency for this hadronic channel can be determined on \bujpsik decays with the same TISTOS method that is used to determine the signal efficiency. The mass distribution of \BdKpi candidates is shown in Fig.~\ref{fig:bdkpi_mass_norm} together with the result of the fit, for the different data-taking years. 
The yields of the two normalisation channels are reported in Table~\ref{tab:yields}, for the different data-taking years and for the two data-taking periods combined. 

\begin{table}[tbp]
    \begin{center}
        \caption{Yields of the two normalisation channels with their combined statistical and systematic errors. }
        \label{tab:yields}
                \begin{tabular}{lrr}
                \toprule 
                   Period & \bujpsik  & \BdKpi \\ 
                    \midrule
                    2011 	 & $(3.479 \pm 0.008)\times 10^{5}$ 	 & $(3.73 \pm 0.13)\times 10^{3}$ \\ 
                    2012 	 & $(7.780 \pm 0.012)\times 10^{5}$ 	 & $(10.32 \pm 0.23)\times 10^{3}$ \\ 
                    2015 	 & $(1.676 \pm 0.005)\times 10^{5}$ 	 & $(4.43 \pm 0.14)\times 10^{3}$ \\ 
                    2016 	 & $(10.369 \pm 0.015)\times 10^{5}$ 	 &$(2.37 \pm 0.06)\times 10^{4} $ \\  
                    2017 	 & $(10.820 \pm 0.014)\times 10^{5}$ 	 & $(2.43 \pm 0.06)\times 10^{4}$ \\ 
                    2018 	 & $(13.208 \pm 0.015)\times 10^{5}$ 	 & $(2.75 \pm 0.06)\times 10^{4}$ \\ 
                    \midrule
                    Run~1 	 & $(11.259 \pm 0.015)\times 10^{5}$ 	 & $(14.05 \pm 0.26)\times 10^{3}$ \\ 
                    Run~2 	 & $(36.072 \pm 0.026)\times 10^{5}$ 	 & $(7.99 \pm 0.10)\times 10^{4}$ \\ 
                    \bottomrule 
                \end{tabular}
        \end{center}
\end{table}
                         
\subsection{Fragmentation fractions } 
\label{ssec:frag}

The fragmentation fractions, denoted as $f_u$, $f_d$, $f_s$, and $f_{\rm baryon}$, are the probabilities for a \bquark quark to hadronise into a \Bu, \Bd, \Bs meson or a \bquark baryon, respectively. These fractions include contributions from intermediate states decaying to the aforementioned hadrons via the strong or electromagnetic interactions. The ratio of fragmentation fractions, \fsfd, used in this analysis has been measured at LHCb using several \B decay modes: semileptonic $B\to D \mu X$ decays at $7\tev$~\cite{LHCb-PAPER-2011-018} and at $13\tev$~\cite{LHCb-PAPER-2018-050}; hadronic  $B\to D h$ decays, where $h = \pi, K$, at 7, 8 and $13\tev$~\cite{LHCb-PAPER-2012-037, LHCb-PAPER-2020-021}; $B\to J/\psi X$ decays at 7, 8 and $13\tev$~\cite{LHCb-PAPER-2019-020}. These measurements have been combined in Ref.~\cite{LHCb-PAPER-2020-046}. The value of \fsfd is found to be dependent on \B transverse momentum and $pp$ collision centre-of-mass energy, while it is found not to be dependent on pseudorapidity. Here only the integrated values at different energies are used since the average \pt of \bquark-hadrons in this analysis is found to be compatible with those used in the determination of the fragmentation fractions~\cite{LHCb-PAPER-2020-046}.

Since the reported values at $7$, $8$ and $13\,\tev$ are strongly correlated because their uncertainties are dominated by external measurements, only the $13\,\tev$ value of
\begin{align*}
\fsfd\ (13\tev) = 0.254 \pm 0.008
\end{align*}
is used, while the Run 1 value is normalised with respect to  \fsfd(13\tev) by the ratio~\cite{LHCb-PAPER-2020-046}
\begin{eqnarray*}
 \frac{\fsfd\ (13\tev)}{\fsfd\ (\rm{Run\ 1})} = 1.064 \pm 0.007\,.
\end{eqnarray*}
Following the approach adopted in Ref.~\cite{LHCb-PAPER-2020-046}, isospin symmetry is assumed to hold in $b$ quark hadronisation at the LHC such that $f_u = f_d$, and hence the same \fsfd values are used in relation to the \bujpsik normalisation channel.

\subsection{Efficiencies}\label{sec:efficiencies}

The efficiencies to detect the signal and normalisation channels can be factorised as 
\begin{equation}
    \varepsilon_{\rm norm(sig)} =  \varepsilon_{\text{RecSel}}~\cdot ~\varepsilon_{\text{PID}}~\cdot  ~\varepsilon_{\text{Trig}}~(\cdot~\varepsilon_{\text{BDT}})\,\, 
    \label{eq:normalisation_efficiency}
\end{equation}
into reconstruction within the LHCb detector and selection (RecSel), PID, trigger (Trig) efficiencies and exclusion of the first BDT region ($\bdt>0.25$) on signal candidates. These are evaluated separately on top of each preceding stage. 

The acceptance, reconstruction and selection efficiencies are evaluated using simulation with corrections applied to improve the agreement with data. The efficiency to detect and reconstruct tracks is evaluated on a sample of $\decay{\jpsi}{\mumu}$ decays in data~\cite{LHCb-DP-2013-002}, using a tag-and-probe method. These samples are used to determine efficiency correction factors as a function of the particle kinematics, which are convolved with the simulated samples to calculate the total efficiency correction. The corrections are at the level of 1\% for all channels and data-taking years. When considering the ratio of signal and normalisation channels in the normalisation formulae~(Eq.~\ref{eq:norm}), uncertainties on these corrections are treated as 100\% correlated. The total efficiencies for the reconstruction within the LHCb detector and selection are listed for the relevant channels in Table~\ref{tab:accrecsel}, where the efficiency with which the muon system detects muons is included. Correction factors for the imperfect modelling of the muon system efficiency simulation are estimated using a sample of \bujpsik decays in data, which are selected without particle identification criteria~\cite{LHCb-PUB-2016-021,LHCb-DP-2018-001}.  These corrections are applied to the signal, \bujpsik and \bsjpsiphi channels, modifying the efficiencies by 1-3\%.

\begin{table}[tbp]
    \begin{center}
        \caption{Efficiencies of reconstruction within the LHCb detector and selection for the signal and normalisation channels, 
        averaged for the two running periods. The uncertainties include the statistical uncertainty from the simulated samples and the uncertainty of the tracking efficiency corrections.}
        \label{tab:accrecsel}
                \begin{tabular}{lccc}
                \toprule 
                    &\multicolumn{2}{c}{$\varepsilon_{\text{RecSel}}$}\\
                    \midrule
                    & Run 1 & Run 2 \\ 
                    \midrule
                    \bsmumu  	 & $0.0602 \pm 0.0003$ 	 & $0.0640 \pm 0.0004$ \\ 
                    \bdmumu  	 & $0.0594 \pm 0.0003$ 	 & $0.0635 \pm 0.0004$ \\ 
                    \bsmumugamma& $0.0508 \pm 0.0003$ 	 & $0.0546 \pm 0.0004$ \\ 
                    \bdkpi  	 & $0.0462 \pm 0.0007$ 	 & $0.0500 \pm 0.0006$ \\ 
                    \bujpsik  	 & $0.0290 \pm 0.0003$ 	 & $0.0305 \pm 0.0003$ \\ 
                    \bottomrule 
                \end{tabular}
        \end{center}
\end{table}
                         
The efficiency of the PID requirements described in Sec.~\ref{sec:Selection} is measured using high-purity control samples of each particle species obtained from data~\cite{LHCb-PUB-2016-021,LHCb-DP-2018-001}. These control samples are obtained by means of kinematic requirements only, with muons obtained from $\decay{\jpsi}{\mup \mun}$ and \bujpsik decays, pions and kaons from $\decay{\Dz}{\Km \pip}$ decays selected via $\decay{\Dstarp}{\Dz\pip}$, and protons from $\decay{\PLambda}{p \pim}$ and $\decay{\Lc}{p \Km \pip}$ decays. The muon PID efficiencies are evaluated as a function of the muon momentum and pseudorapidity, as well as the track multiplicity of the event using a dedicated procedure~\cite{LHCb-PUB-2016-021}. The resulting efficiency maps are then applied to simulated samples to determine the integrated efficiency for a specific channel. The efficiency measurements for the different hadronic species are described in Sec.~\ref{ssec:misid}. The results for the signal and normalisation channels are shown in Table~\ref{tab:pid} and include for the channels with muons the data-simulation correction of the muon system identification. For \bujpsik candidates, only the correction to the muon system identification efficiency is computed, as no further requirements on the multivariate PID classifier are applied when selecting these decays. The systematic uncertainties arise from modelling the dependencies of the PID efficiency maps.

\begin{table}[tbp]
    \begin{center}
        \caption{Particle identification efficiencies for the signal and normalisation channels, 
        averaged for the two running periods, where  the first uncertainty is statistical and the second systematic. A data-simulation correction of the muon-system identification is included for channels with muons. 
        For the \bujpsik channel only the data-simulation correction part of the muon identification is reported, as no multivariate PID requirement is applied to this channel. 
        }
        \label{tab:pid}
                \begin{tabular}{lccc}
                \toprule 
                    &\multicolumn{2}{c}{$\varepsilon_{\text{PID}}$}\\
                    \midrule
                    & Run 1 & Run 2 \\ 
                    \midrule
                    \bsmumu  	 & $0.8580 \pm 0.0006 \pm 0.0053$ 	 & $0.8822 \pm 0.0003 \pm 0.0039$ \\ 
                    \bdmumu  	 & $0.8518 \pm 0.0007 \pm 0.0063$ 	 & $0.8759 \pm 0.0004 \pm 0.0046$ \\ 
                    \bsmumugamma  	 & $0.8487 \pm 0.0006 \pm 0.0088$ 	 & $0.8785 \pm 0.0003 \pm 0.0064$ \\ 
                    \bdkpi  	 & $0.4741 \pm 0.0049 \pm 0.0010$ 	 & $0.5004 \pm 0.0027 \pm 0.0012$ \\ 
                    \midrule
                    \bujpsik  	 & $1.0096 \pm 0.0005$ 	 & $1.00260 \pm 0.00018$ \\ 
                    \bottomrule 
                \end{tabular}
        \end{center}
\end{table}

The trigger efficiencies are determined from data with the TISTOS method~\cite{LHCb-PUB-2014-039}. Trigger information is associated to the reconstructed candidates during the offline processing. The event of a reconstructed signal candidate can be classified into three categories: events triggered on signal (TOS), triggered on part of the underlying event that is independent of the tracks forming the signal candidate (TIS), or triggered on both elements of the signal candidate and the underlying event. 

The trigger efficiency can be estimated by exploiting the overlap between the TIS and TOS categories (TIS\&TOS) and assuming signal decays uncorrelated with the rest of the event. The trigger efficiency, $\varepsilon_{\rm{trig}}$, of a given decay channel, with respect to a total of $N_{\rm{Tot}}$ events, can be computed as
\begin{eqnarray}
 \varepsilon_{\rm{trig}} =\frac{N_{\rm{trig}}}{N_{\rm{Tot}}} =  \frac{N_{\rm{trig}}}{N_{\rm{TIS}}} \cdot \varepsilon_{\rm{TIS}}
 =
 \frac{N_{\rm{trig}}}{N_{\rm{TIS}}}\frac{N_{\rm{TIS\& TOS}}}{N_{\rm{TOS}}}\,,
\label{eq:tistos}
\end{eqnarray}
where $N_X$ is the number of background-subtracted candidates triggered within the category $X$ and the efficiency $\varepsilon_{\rm TIS} = N_{\rm TIS\& TOS} / N_{\rm TOS}$ is estimated under the already mentioned independence assumption, which is verified in Ref.~\cite{LHCb-PUB-2014-039}.

The trigger efficiency estimation is done in two steps: the hardware and first software-level trigger efficiencies are estimated from data as described in the following; the second software-level efficiency, being aligned to the offline selection, is estimated from simulation and included in the full trigger efficiency presented in this paragraph. The trigger efficiencies for signal and normalisation channels with muons are calibrated using the \bujpsik channel. In order to reduce residual kinematic correlations between the decay in question and the rest of the event, the calibration is performed in intervals of kinematic quantities. Yields of \bujpsik decays for each trigger category and different kinematic ranges are obtained by performing a mass fit as described in Sec.~\ref{ssec:NormChannelYields}. Efficiency tables are obtained as a function of the maximum \pt of the two muons and of the product of the \pt of the two muons, as these are the variables used in the muon hardware trigger. These efficiency distributions are then convolved with the simulated samples of the relevant channels. The trigger efficiencies for the \BdKpi channel are also obtained in data by measuring the TIS trigger efficiency in Eq.~\ref{eq:tistos} through the more abundant \bujpsik channel, since  the TIS efficiencies do not depend on the control channel used to evaluate it.

The trigger efficiencies for the signal and normalisation channels in each data-taking period are presented in Table~\ref{tab:trig}. The systematic uncertainty on the trigger efficiency is comprised of a number of sources. A systematic effect is associated with the choice of the mass model used for the \bujpsik channel. This is estimated by fitting the \bujpsik data using a double Crystal Ball function to model the signal and taking the difference with the default fit as systematic uncertainty. A second systematic uncertainty stems from the precision of the TISTOS method and is obtained by comparing the efficiency determined by applying the TISTOS method to simulation. A third source of systematic effect is due to the difference in the phase space between the \bujpsik decay and the decay channel for which the trigger efficiency is evaluated. The corresponding systematic uncertainty is estimated by comparing the results of the method applied to simulated events using the \bujpsik channel or the considered channel. The last source of systematic effect arises from the choice of the kinematic ranges for which the efficiencies are evaluated, and its uncertainty is determined from the change of the efficiency when these ranges are varied. The resulting shifts in trigger efficiency from each source are added in quadrature, and assigned as the total systematic uncertainty in Table~\ref{tab:trig}.

\begin{table}[tbp]
    \begin{center}
        \caption{Trigger efficiencies for the signal and normalisation channels, 
        averaged for the two data taking periods. The first uncertainty is statistical and the second systematic.  }
        \label{tab:trig}
                \begin{tabular}{lccc}
                \toprule                     &\multicolumn{2}{c}{$\varepsilon_{\text{Trig}}$}\\
                    \midrule
                    & Run 1 & Run 2 \\ 
                    \midrule
                    \bsmumu  	 & $0.9579 \pm 0.0033 \pm 0.0164$ 	 & $0.9712 \pm 0.0014 \pm 0.0093$ \\ 
                    \bdmumu  	 & $0.9570 \pm 0.0032 \pm 0.0176$ 	 & $0.9708 \pm 0.0014 \pm 0.0097$ \\ 
                    \bsmumugamma  	 & $0.9538 \pm 0.0032 \pm 0.0195$ 	 & $0.9694 \pm 0.0013 \pm 0.0111$ \\ 
                    \bdkpi  	 & $0.0433 \pm 0.0002 \pm 0.0016$ 	 & $0.0727 \pm 0.0002 \pm 0.0020$ \\ 
                    \bujpsik  	 & $0.8810 \pm 0.0040 \pm 0.0080$ 	 & $0.9033 \pm 0.0016 \pm 0.0089$ \\ 
                    \bottomrule 
                \end{tabular}
        \end{center}
\end{table}

The efficiencies of the exclusion of the first BDT region on the signal decays are evaluated using the calibrated BDT response described in Sec.~\ref{sec:bdt_calibration} and are listed in Table~\ref{tab:BDTeff}, combining statistical and systematic uncertainties.
\begin{table}[tbp]
    \begin{center}
        \caption{Efficiency on the signal channels of excluding the BDT region $\text{BDT}<0.25$, averaged for the two data taking periods. The uncertainties combine statistical and the second systematic uncertainties.  }
        \label{tab:BDTeff}
                \begin{tabular}{lccc}
                \toprule                     &\multicolumn{2}{c}{$\varepsilon_{\text{BDT}}$}\\
                    \midrule
                    & Run 1 & Run 2 \\ 
                    \midrule
                    \bsmumu  	 & $0.723 \pm 0.006$ 	 & $0.7071 \pm 0.0026$ \\ 
                    \bdmumu  	 & $0.720 \pm 0.006$ 	 & $0.7036 \pm 0.0027$ \\ 
                    \bsmumugamma  	 & $0.656 \pm 0.007$ & $0.6531 \pm 0.0035$ \\ 
                    \bottomrule 
                \end{tabular}
        \end{center}
\end{table}

\subsection{Single-event sensitivities}

The single-event sensitivities (defined in Eq.~\ref{eq:norm}) for the three signal channels in \runone, \runtwo and the full data sample, are reported in Table~\ref{tab:norm_full}. Single-event sensitivities for $\bdt>0.25$ are obtained using the two normalisation channels \BdKpi and \bujpsik separately, which are combined into a normalisation using a weighted average, taking into account the relevant correlations. In the same table, the expected number of signal candidates for $\bdt>0.25$ is reported, assuming the SM branching fraction. 

\begin{table}[htbp]
    \begin{center}
        \caption{Single-event sensitivities, $\alpha(\Bu)$, $\alpha(\Bd)$ and $\alpha(\rm{Comb})$ for the three signal channels obtained for $\bdt>0.25$ with the two normalisation channels, \BdKpi and \bujpsik, and combined, 
        for \runone, \runtwo and the full data set. The first uncertainty is statistical and the second systematic. The expected yields assuming SM branching fractions, $N_{\rm{exp}}$, are also reported. The \bsmumugamma expected number does not include an uncertainty on the branching fraction.}
        \label{tab:norm_full}
        \scalebox{0.8}{
        \begin{tabular}{lcccccc}
            \toprule
            & \bdmumu & \bsmumu & \bsmumugamma \\
            \midrule
             \multicolumn{4}{c}{Run 1} \\
             $\alpha(\Bu)$  	 & $(3.96 \pm 0.13 \pm 0.09)\times 10^{-11}$  	 & $(1.57 \pm 0.07 \pm 0.03)\times 10^{-10}$  	 & $(2.11 \pm 0.10 \pm 0.05)\times 10^{-10}$ \\
             $\alpha(\Bd)$  	 & $(3.79 \pm 0.14 \pm 0.16)\times 10^{-11}$  	 & $(1.50 \pm 0.07 \pm 0.06)\times 10^{-10}$  	 & $(2.01 \pm 0.10 \pm 0.09)\times 10^{-10}$ \\
             $\alpha(\rm{Comb})$  	 & $(3.93 \pm 0.10 \pm 0.08)\times 10^{-11}$  	 & $(1.56 \pm 0.06 \pm 0.03)\times 10^{-10}$  	 & $(2.09 \pm 0.09 \pm 0.05)\times 10^{-10}$ \\
             $N_{\rm{exp}}$  	 & $2.62 \pm 0.14 \pm 0.05$ 	 &  $23.5 \pm 1.3 \pm 0.4$ 	 &  $0.479 \pm 0.020 \pm 0.011$\\
             \midrule
             \multicolumn{4}{c}{Run 2} \\
             $\alpha(\Bu)$  	 & $(1.214 \pm 0.037 \pm 0.018)\times 10^{-11}$  	 & $(4.54 \pm 0.20 \pm 0.07)\times 10^{-11}$  	 & $(5.86 \pm 0.26 \pm 0.10)\times 10^{-11}$ \\
             $\alpha(\Bd)$  	 & $(1.176 \pm 0.035 \pm 0.037)\times 10^{-11}$  	 & $(4.40 \pm 0.19 \pm 0.14)\times 10^{-11}$  	 & $(5.67 \pm 0.25 \pm 0.18)\times 10^{-11}$ \\
             $\alpha(\rm{Comb})$  	 & $(1.204 \pm 0.023 \pm 0.014)\times 10^{-11}$  	 & $(4.50 \pm 0.16 \pm 0.05)\times 10^{-11}$  	 & $(5.81 \pm 0.21 \pm 0.08)\times 10^{-11}$ \\
             $N_{\rm{exp}}$  	 & $8.55 \pm 0.45 \pm 0.10$ 	 &  $81.3 \pm 4.3 \pm 0.9$ 	 &  $1.721 \pm 0.063 \pm 0.023$\\
             \midrule
             \multicolumn{4}{c}{All} \\
             $\alpha(\Bu)$  	 & $(9.27 \pm 0.28 \pm 0.12)\times 10^{-12}$  	 & $(3.53 \pm 0.15 \pm 0.04)\times 10^{-11}$  	 & $(4.61 \pm 0.20 \pm 0.07)\times 10^{-11}$ \\
             $\alpha(\Bd)$  	 & $(8.95 \pm 0.26 \pm 0.23)\times 10^{-12}$  	 & $(3.41 \pm 0.15 \pm 0.09)\times 10^{-11}$  	 & $(4.45 \pm 0.19 \pm 0.12)\times 10^{-11}$ \\
             $\alpha(\rm{Comb})$  	 & $(9.20 \pm 0.14 \pm 0.09)\times 10^{-12}$  	 & $(3.51 \pm 0.12 \pm 0.03)\times 10^{-11}$  	 & $(4.57 \pm 0.16 \pm 0.05)\times 10^{-11}$ \\
             $N_{\rm{exp}}$  	 & $11.20 \pm 0.57 \pm 0.11$ 	 &  $104.4 \pm 5.4 \pm 1.0$ 	 &  $2.186 \pm 0.077 \pm 0.026$\\
            \bottomrule
        \end{tabular}}
    \end{center}
\end{table}

In order to cross-check the normalisation of the signal channels, the ratio of the efficiency-corrected yields of the two normalisation channels
 \begin{equation*}
    \frac{\BRof\BdKpi}{\BRof\bujpsik}=\frac{N_{\BdKpi}}{N_{\bujpsik}} \frac{\varepsilon_{\bujpsik}}{\varepsilon_{\BdKpi}}\frac{f_u}{f_d}
    \label{eq:bd_bu_us}
\end{equation*}
is measured using data, where $f_u=f_d$ is assumed. This ratio is found to be \mbox{$0.340\pm 0.016 \stat$} and \mbox{$0.336\pm 0.012 \stat$} in \runone and \runtwo, respectively, in agreement with the ratio of the world averages of these branching fractions, \mbox{$0.326\pm 0.012$}~\cite{PDG2020}.

To cross-check the ratio of the \Bs and \Bu fragmentation fractions and its stability over the data taking, the ratio of \bujpsik and \bsjpsiphi efficiency-corrected yields,
\begin{equation}
    \mathcal{R}=\frac{N_{\bsjpsiphi}}{N_{\bujpsik}}\frac{\varepsilon_{\bujpsik}}{\varepsilon_{\bsjpsiphi}}=\frac{f_s}{f_u}\frac{\BRof\bsjpsiphi}{\BRof\bujpsik},
\end{equation}
is also measured, following a similar approach to Ref.~\cite{LHCb-PAPER-2019-020}. The ratios are found to be similar to those measured in Ref.~\cite{LHCb-PAPER-2019-020}, although the two methods explore different kinematic regions. A dependence on the centre-of-mass energy is seen and found to be consistent with  Ref.~\cite{LHCb-PAPER-2019-020} and the combined analysis of Ref.~\cite{LHCb-PAPER-2020-046}, justifying the use of different \fsfd values for the \runone and \runtwo data samples.

\section{Background}
\label{sec:Backgrounds}

Three classes of background events are considered in the analysis: combinatorial background; \Bhh decays; semileptonic $\bquark$-hadron decays. The combinatorial background, mainly composed of real muons originating from two different \B decays, is modelled using an exponential function with the slope left free to float in the mass fit, as described in Sec.~\ref{sec:BFfit}. The other background sources are included as separate components in the fit, with mass shapes evaluated on simulated events and with yields that are Gaussian-constrained to their estimated values, as explained in the following sections.

\subsection{Hadron misidentification rates}
\label{ssec:misid}
To estimate the yield of physical background where one or two final-state particles are misidentified as a muon, it is crucial to perform unbiased measurements of the probability for protons, pions and kaons to pass the muon identification requirements. These measurements are carried out as a function of the track momentum and transverse momentum, using the  data control samples listed in Sec.~\ref{sec:efficiencies}. The dedicated procedure from Ref.~\cite{LHCb-PUB-2016-021} is used for protons, while a different method is developed to determine the pion and kaon misidentification rates, using \decay{\Dz}{\Km\pip} from \decay{\Dstarp}{\Dz\pip} decays. For these particles, especially at low momenta, a sizeable contribution to the misidentification rate originates from hadrons decaying to muons. When the hadron decays in flight, the momentum resolution of the reconstructed track degrades by an amount that depends on the distance the hadron has travelled before decaying and on the fraction of energy inherited by the daughter muon. As a consequence, the mass distribution of the \Dz candidates broadens significantly and the efficiency to select \decay{\Dz}{\Km\pip} decays in the \Dz mass selection window, $1825 \leq m_{K\pi}\leq 1910\,\mevcc$, decreases. If this effect is not taken into account, a significant underestimation of the misidentified hadron yield would occur. The misidentification efficiencies for pions and kaons are determined by measuring the \Dz yield from a two-dimensional fit to the $m_{K\pi\pi}-m_{K\pi}$, $m_{K\pi}$ distribution with or without the muon requirement applied to the particle in question. The shape of the signal \Dz mass distribution when the PID selection is applied includes the tail arising from hadron decays-in-flight, estimated from simulated events. The resulting misidentification probability is then corrected for the fraction of \decay{\Dz}{\Km\pip} decays with the \Km\pip mass falling outside the \Dz selection window, estimated from simulated events.

\subsection{\boldmath \Bhh decays}
The \Bhh decays can appear as background when both final state hadrons are misidentified as muons. These candidates have a broad mass distribution centred close to the \Bd mass, as determined from simulation where each of the four modes is weighted according to its expected yield.

The expected yield of doubly misidentified \Bhh events is estimated by normalising to the \BdKpi channel as
\begin{equation}
    N_{\Bhh\to\mu^+\mu^-}=\varepsilon^{\rm trig}_{\bsmumu}\cdot\frac{N^{\rm TIS}_{hh}}{\varepsilon^{\rm TIS}}\cdot\varepsilon_{hh\to\mu\mu},
    \label{eq:bhh}
\end{equation}
where $\varepsilon^{\rm trig}_{\bsmumu}$ is the signal trigger efficiency, $N^{\rm TIS}_{hh}$ is the number of \Bhh TIS events evaluated by correcting the \BdKpi TIS yield by the expected fraction of this mode, $\varepsilon^{\rm TIS}$ is the TIS efficiency (Sec.~\ref{sec:Calibration}), and $\varepsilon_{hh\to\mu\mu}$ represents the double misidentification rate, which is estimated using data control samples (Sec.~\ref{ssec:misid}) and found to be in the range $10^{-6}-10^{-5}$, depending on the data set and BDT region. An independent estimate of the \Bhh background is performed on $\pi\mu$ and $K\mu$ combinations, selected from data samples of \B candidates with two tracks in the final states applying strong muon and hadron identification requirements on the tracks. Their mass spectra are fitted and the resulting yields are scaled by the $\pi\to\mu$ and $K\to\mu$ misidentification rates. The ratio between this result and the default estimate is assigned as a correction factor to the misidentification efficiency. The estimated \Bhh background yields in each BDT region with $\bdt > 0.25$ are summarised in Table~\ref{tab:bkg_summary_run1} for Run 1 and Table~\ref{tab:bkg_summary_run2} for Run 2 data.

\subsection{Semileptonic decays}

Several semileptonic $\bquark$-hadron decays, with branching fractions ranging from $10^{-8}$ to $10^{-4}$, are considered in the fit:~\bcjpsimunu, with $\jpsi \to \mumu$, and \Bpimumu decays have two real muons in the final state, while \bdpimunu, \bskmunu and \lbpmunu decays represent non-negligible background when the final-state hadron is misidentified as a muon. When reconstructed as dimuon candidates, these decays are partially reconstructed and therefore populate the lower \Bd/\Bs sideband, but can have tails reaching into the signal region.

For each of the above channels, the number of expected candidates is estimated by normalising to the yield of \bujpsik decays, according to
\begin{equation}
    N_x=\frac{f_x}{f_u}\cdot\frac{N_{\bujpsik}}{\mathcal{B}_{\bujpsik}\cdot\varepsilon^\text{tot}_{\bujpsik}}\cdot\;\mathcal{B}_x\cdot\varepsilon^\text{tot}_x,
\end{equation}
where $f_x$ is the hadronisation fraction of the initial-state hadron for the decay mode $x$, $\mathcal{B}_x$ is the branching fraction and $\varepsilon^\text{tot}_x$ its total selection and trigger efficiency. The efficiencies are estimated from simulation, except for the PID, which is estimated from data control samples as described in Secs.~\ref{sec:efficiencies} and~\ref{ssec:misid}. The branching fraction, including the hadronisation fraction of the \bcjpsimunu channel, is taken from Ref.~\cite{LHCb-PAPER-2019-033}, while those of \Bpimumu and \bdpimunu channels are obtained from Refs.~\cite{PDG2020,Wang:2012ab}, assuming $f_u = f_d$. LHCb measurements for the \lbpmunu and \bskmunu branching fractions~\cite{LHCb-PAPER-2015-013, LHCb-PAPER-2020-038} and hadronisation fractions~\cite{LHCb-PAPER-2018-050, LHCb-PAPER-2020-046} are used. The estimated yields in each BDT region with $\bdt >0.25$ are shown in Table~\ref{tab:bkg_summary_run1} for Run 1 and Table~\ref{tab:bkg_summary_run2} for Run 2 data.

\begin{landscape}
    \begin{table}[p]%[tb]
        \begin{center}
            \caption{Expected background yields per BDT region and for $\bdt >0.25$ with their total estimated uncertainties for Run 1 data.}
	        \begin{tabular}{c|ccccccc}
                \toprule
                BDT region & \bhh & \bdpimunu & \bskmunu & \bupimumu & \bdpimumu & \lbpmunu & \bcjpsimunu \\
                \midrule
                {[0.25-0.4]} & $3.8 \pm0.9 $ & $ 8.8\pm0.7$ & $0.94\pm0.20$ & $ 2.72\pm0.35$ & $0.9 \pm0.4 $ & $0.29\pm0.32$ & $ 2.84\pm0.21$ \\
                {[0.4-0.5]}  & $1.57\pm0.20$ & $ 6.0\pm0.4$ & $0.76\pm0.16$ & $ 1.47\pm0.19$ & $0.65\pm0.29$ & $0.18\pm0.20$ & $ 1.38\pm0.13$ \\
                {[0.5-0.6]}  & $1.61\pm0.21$ & $ 5.8\pm0.4$ & $0.90\pm0.19$ & $ 1.22\pm0.16$ & $0.65\pm0.28$ & $0.20\pm0.22$ & $ 0.80\pm0.09$ \\
                {[0.6-0.7]}  & $1.65\pm0.21$ & $ 5.8\pm0.4$ & $1.03\pm0.22$ & $ 0.99\pm0.13$ & $0.65\pm0.29$ & $0.19\pm0.21$ & $ 0.58\pm0.08$ \\
                {[0.7-1.0]}  & $5.3 \pm0.7 $ & $11.6\pm0.8\hphantom{0}$ & $2.5 \pm0.5 $ & $ 1.46\pm0.19$ & $1.4 \pm0.6 $ & $0.32\pm0.35$ & $ 0.41\pm0.06$ \\
                {[0.25-1.0]} &$ 13.9\pm1.2\hphantom{0}$ & $30.8\pm1.8\hphantom{0}$ & $9.0 \pm1.1 $ & $ 7.6\pm0.5$ & $4.2 \pm0.9 $ & $1.2\pm0.6$ & $ 6.01\pm0.28$ \\
                \bottomrule 
            \end{tabular}
            \label{tab:bkg_summary_run1}
        \end{center}
    \end{table}
    \begin{table}[p]%[tb]
        \begin{center}
            \caption{Expected background yields per BDT region and for $\bdt >0.25$ with their total estimated uncertainties for Run 2 data.}
	        \begin{tabular}{c|ccccccc}
                \toprule
                BDT region & \bhh & \bdpimunu & \bskmunu & \bupimumu & \bdpimumu & \lbpmunu & \bcjpsimunu \\
                \midrule
                {[0.25-0.4]} & $6.2 \pm1.0 $ & $ 40.1\pm2.6$ & $ 3.3\pm0.7$ & $ 9.9\pm1.3$ & $ 3.5\pm1.5$ & $1.1\pm1.2$ & $ 11.7 \pm0.6 $ \\
                {[0.4-0.5]}  & $2.90\pm0.26$ & $ 24.6\pm1.6$ & $ 2.5\pm0.5$ & $ 5.2\pm0.7$ & $ 2.3\pm1.0$ & $0.8\pm0.8$ & $  4.59\pm0.27$ \\
                {[0.5-0.6]}  & $2.82\pm0.26$ & $ 24.4\pm1.6$ & $ 2.9\pm0.6$ & $ 4.2\pm0.5$ & $ 2.3\pm1.0$ & $0.9\pm0.9$ & $  2.85\pm0.19$ \\
                {[0.6-0.7]}  & $2.68\pm0.24$ & $ 23.5\pm1.5$ & $ 3.3\pm0.7$ & $ 3.3\pm0.4$ & $ 2.3\pm1.0$ & $0.9\pm1.0$ & $  1.56\pm0.12$ \\
                {[0.7-1.0]}  & $8.1 \pm0.7 $ & $ 44.6\pm2.9$ & $ 7.7\pm1.6$ & $ 4.8\pm0.6$ & $ 4.7\pm2.1$ & $1.6\pm1.8$ & $  0.85\pm0.08$ \\
                {[0.25-1.0]}  & $22.7 \pm1.3 \hphantom{0}$ & $ 130\pm5\hphantom{0}$ & $ 21.5\pm2.5\hphantom{0}$ & $ 26.5\pm1.7\hphantom{0}$ & $ 14.7\pm3.0\hphantom{0}$ & $5.3\pm2.7$ & $  21.6\pm0.7$ \\
                \bottomrule 
            \end{tabular}
            \label{tab:bkg_summary_run2}
        \end{center}
    \end{table}
\end{landscape}

\section{Measurement of signal branching fractions}
\label{sec:BFfit}
\begin{figure}[tbp]
    \centering
    \includegraphics[width=0.4\textwidth]{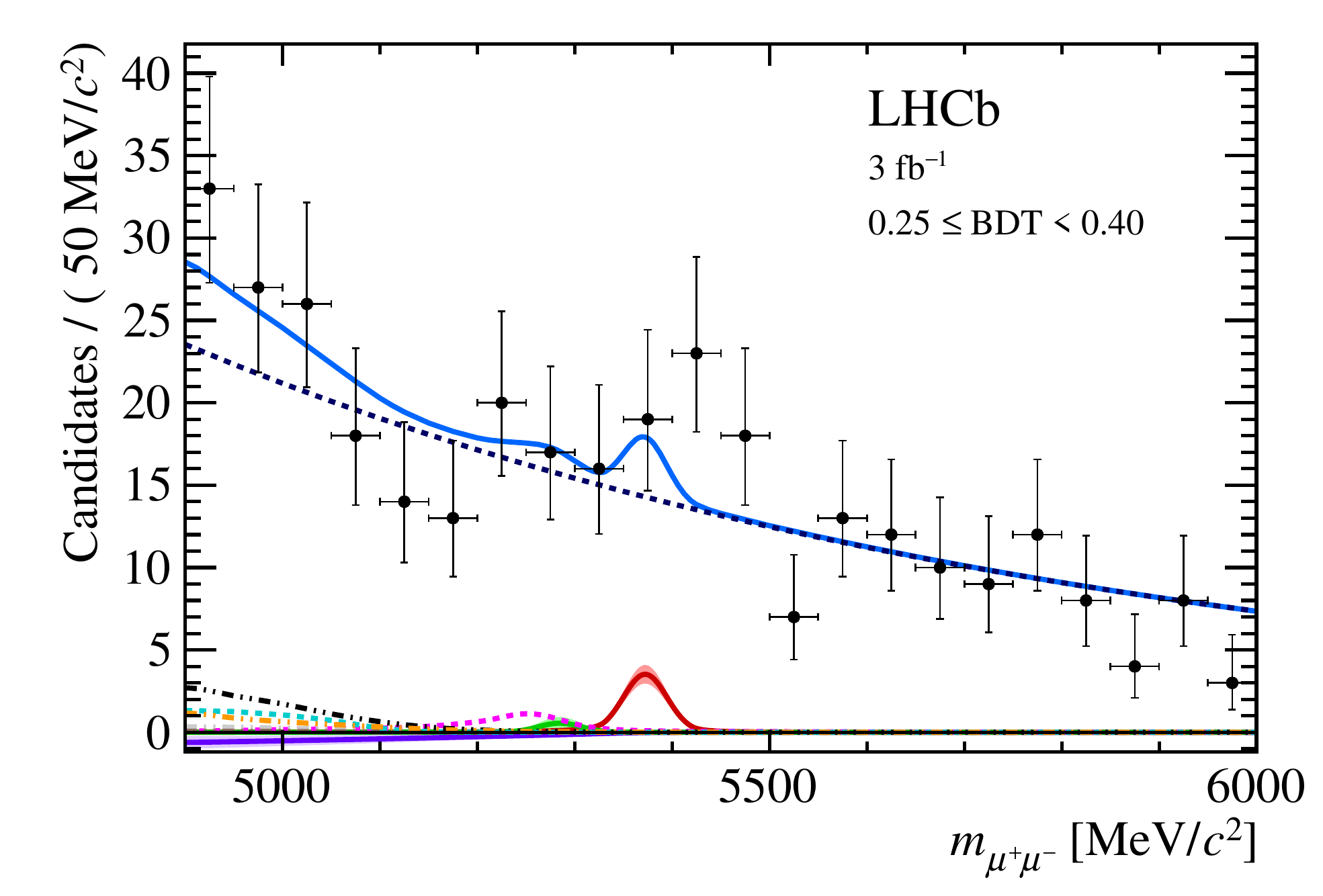}
    \includegraphics[width=0.4\textwidth]{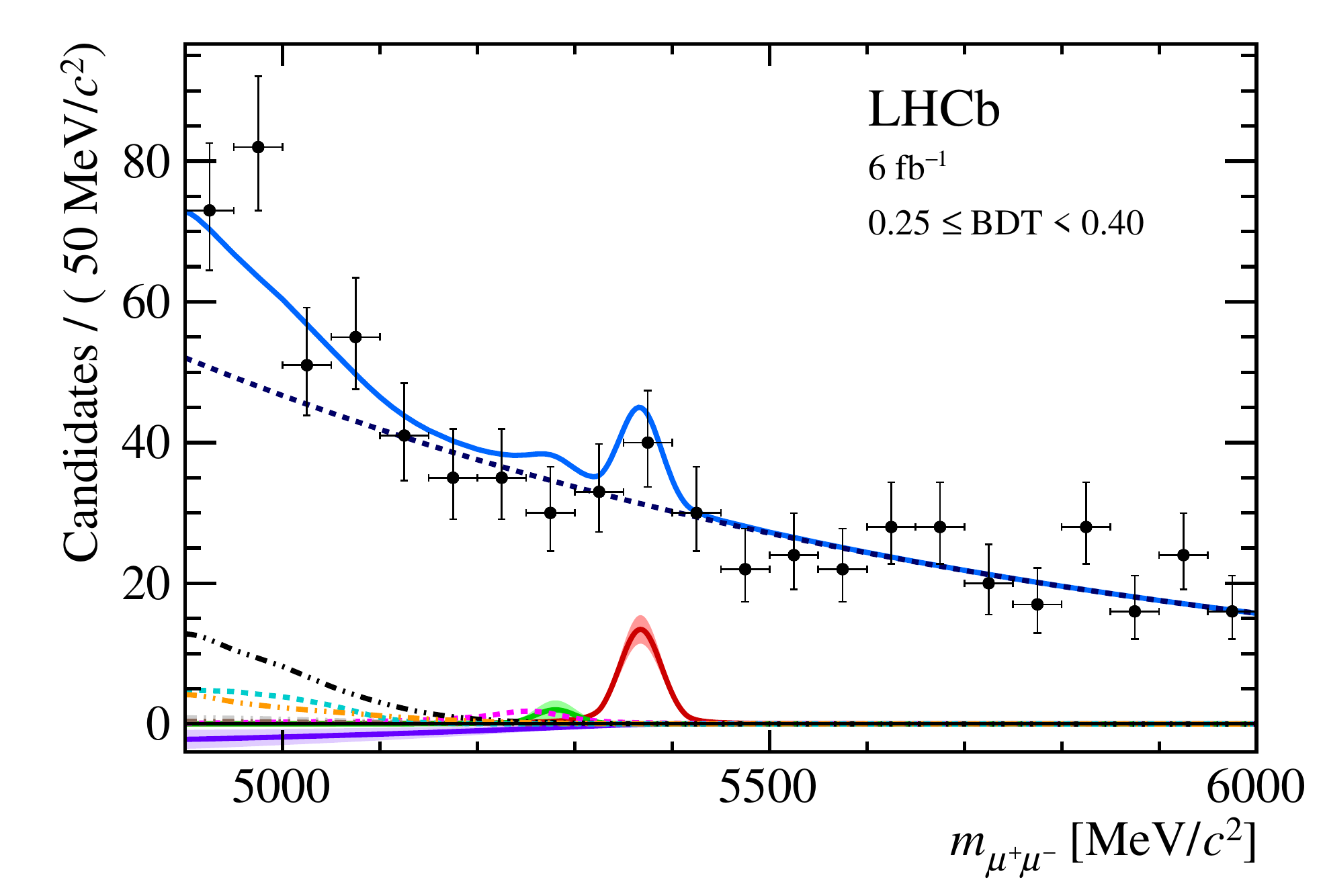}\\
    \includegraphics[width=0.4\textwidth]{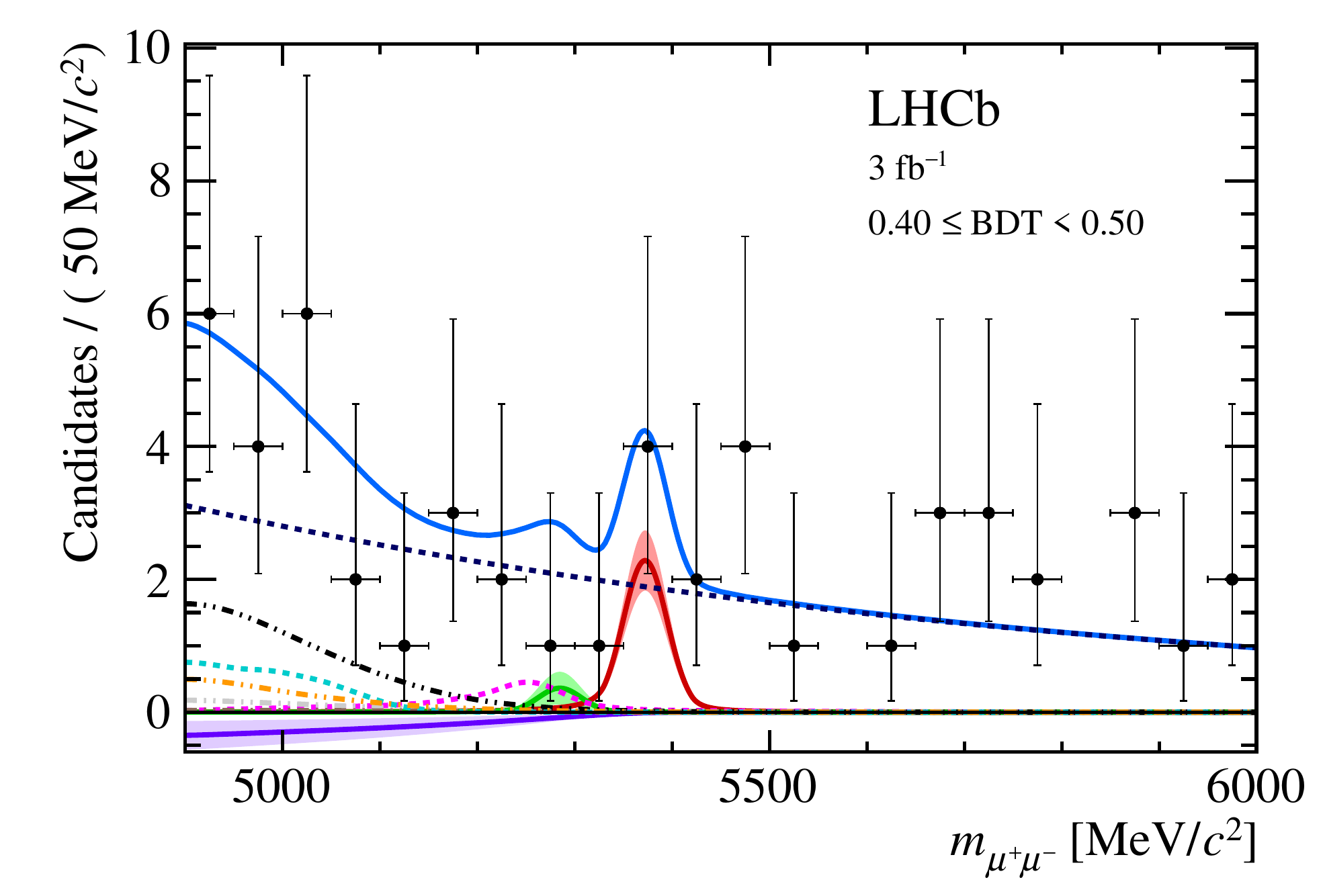}
    \includegraphics[width=0.4\textwidth]{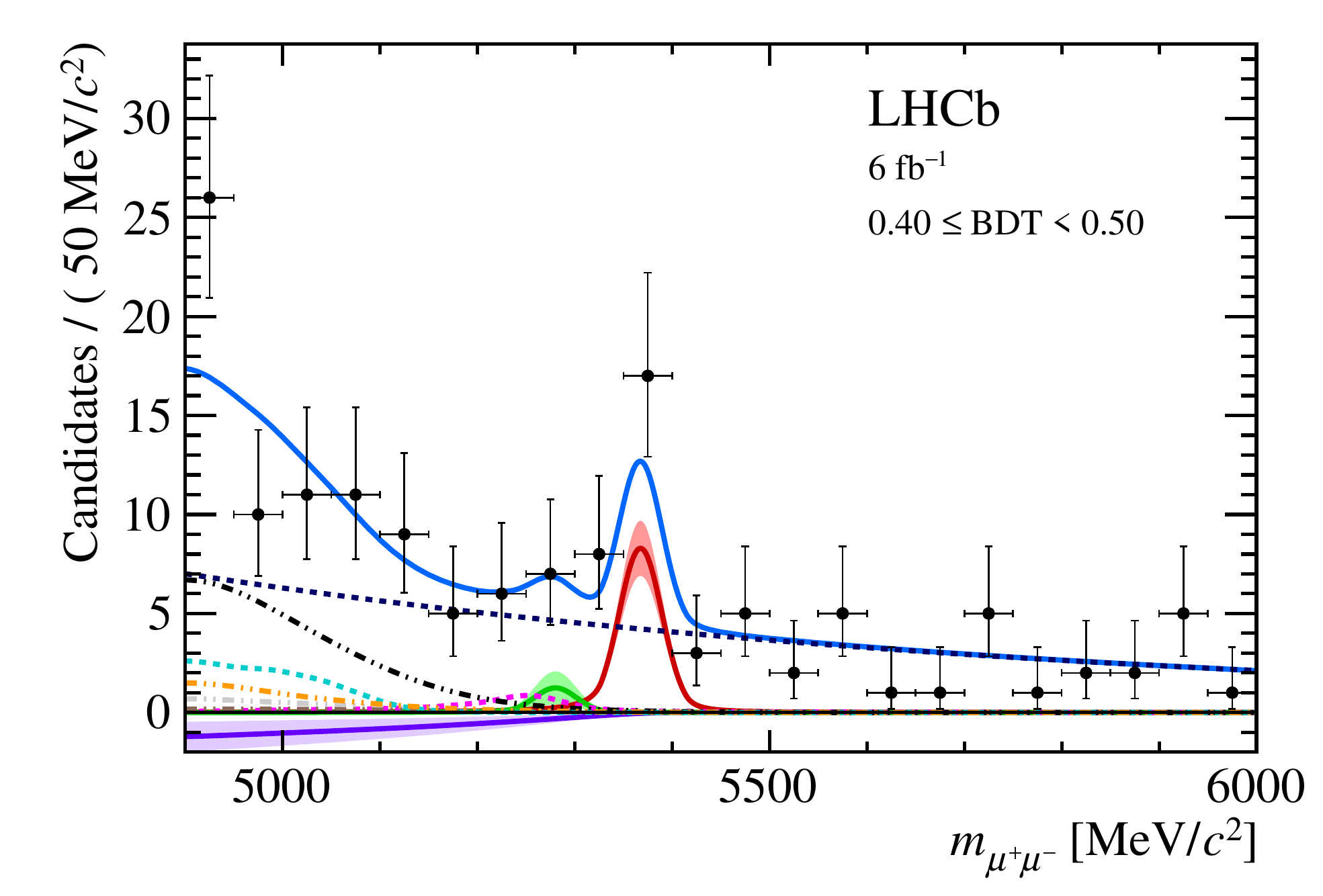}\\
    \includegraphics[width=0.4\textwidth]{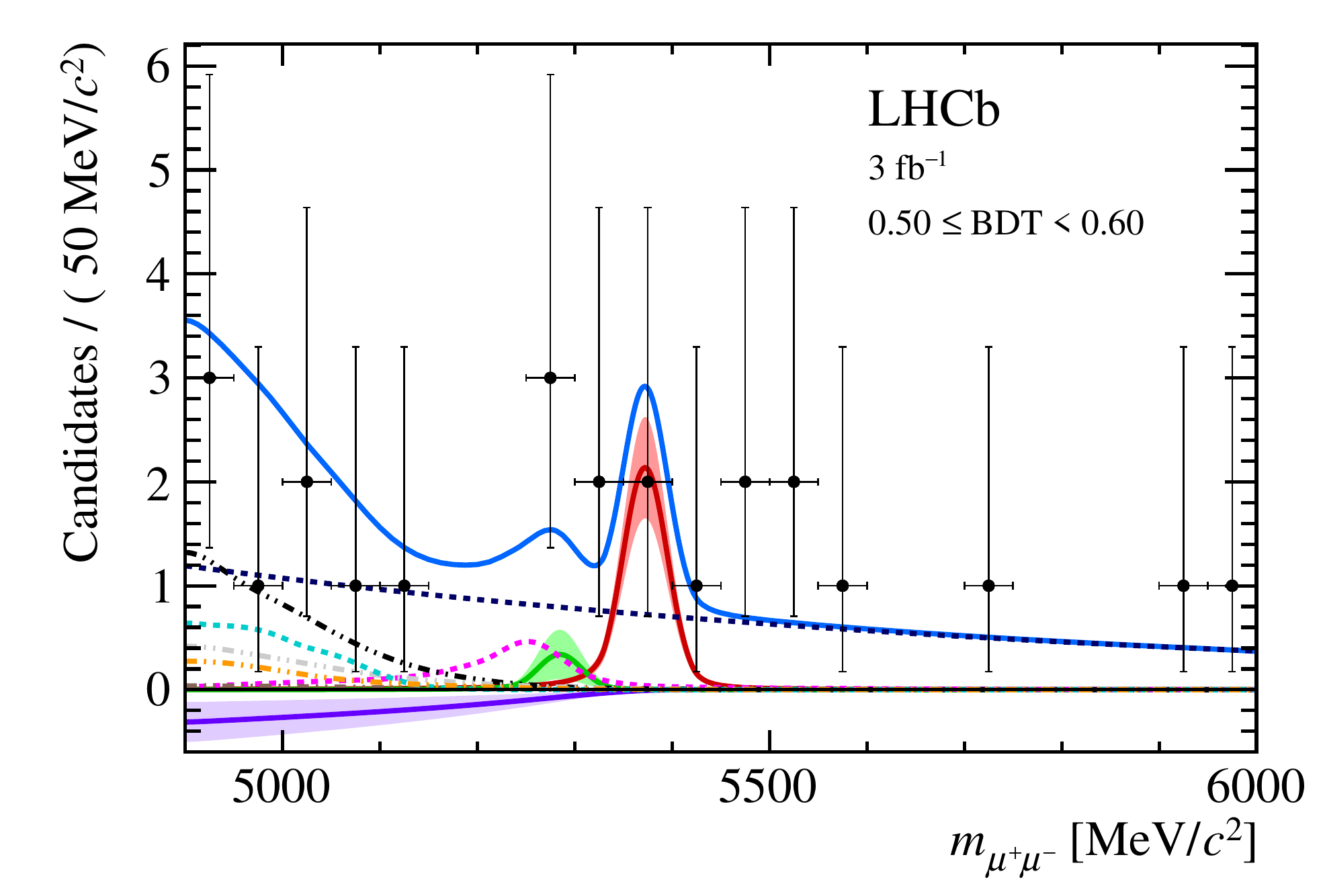}
    \includegraphics[width=0.4\textwidth]{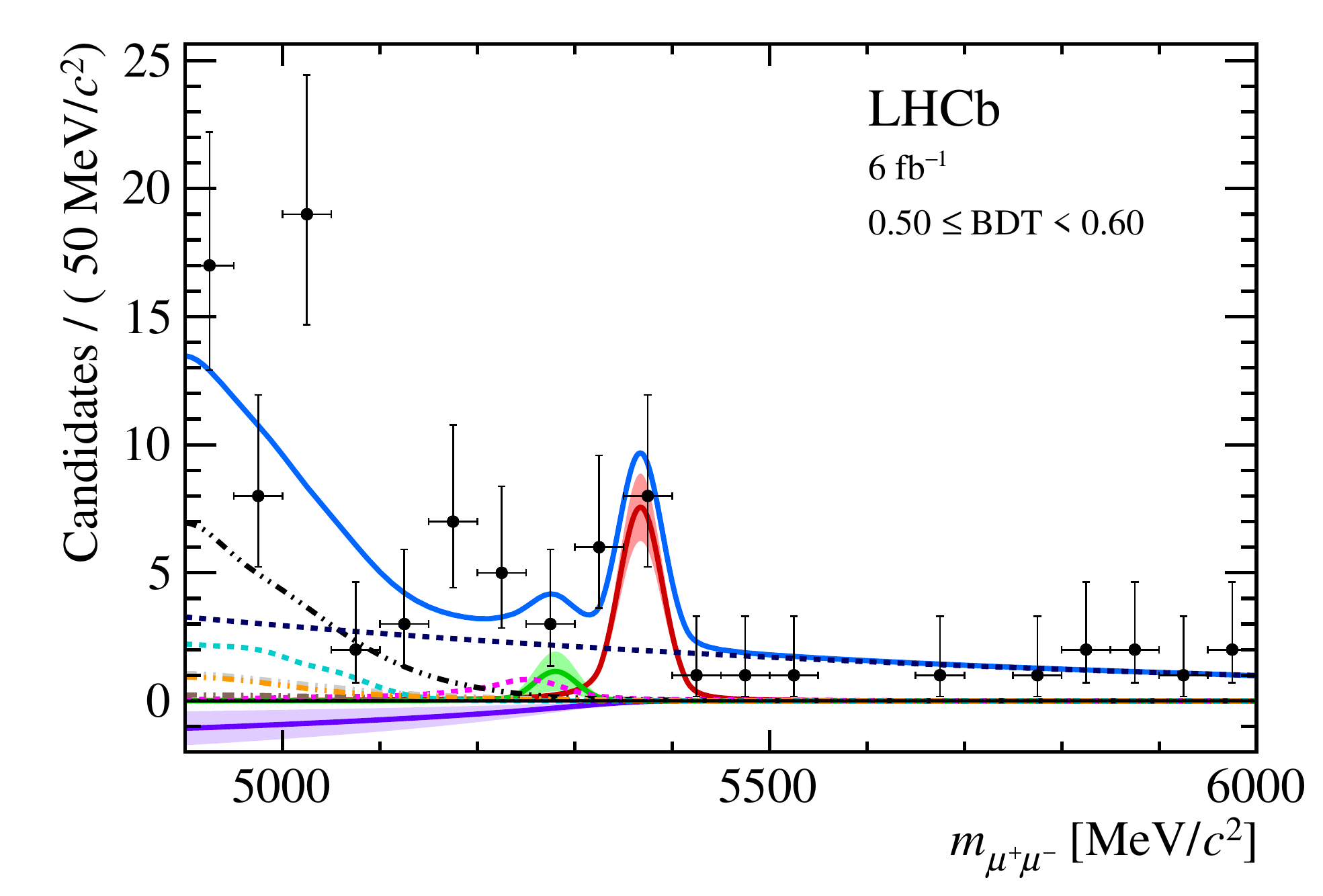}\\
    \includegraphics[width=0.4\textwidth]{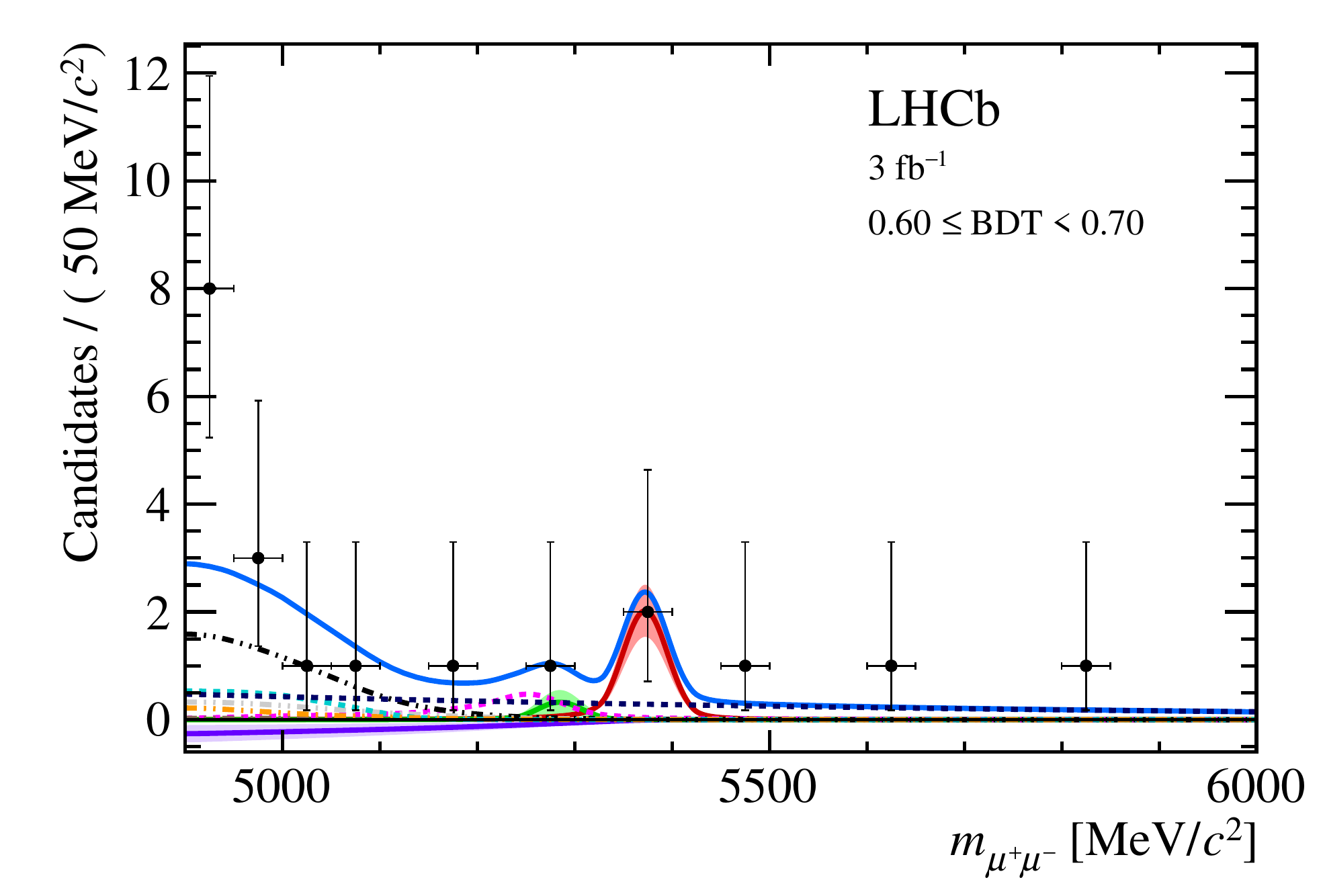}
    \includegraphics[width=0.4\textwidth]{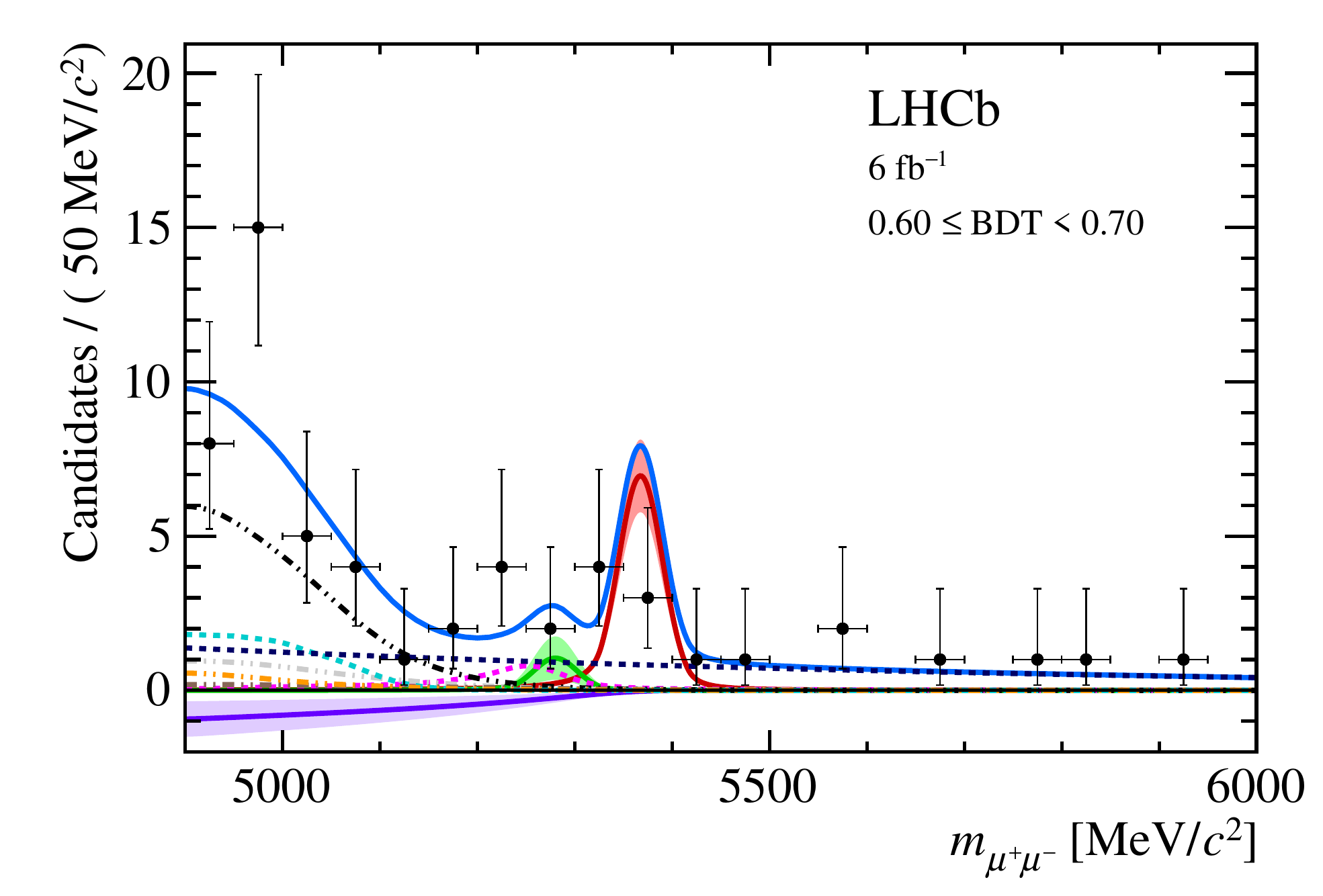}\\
    \includegraphics[width=0.4\textwidth]{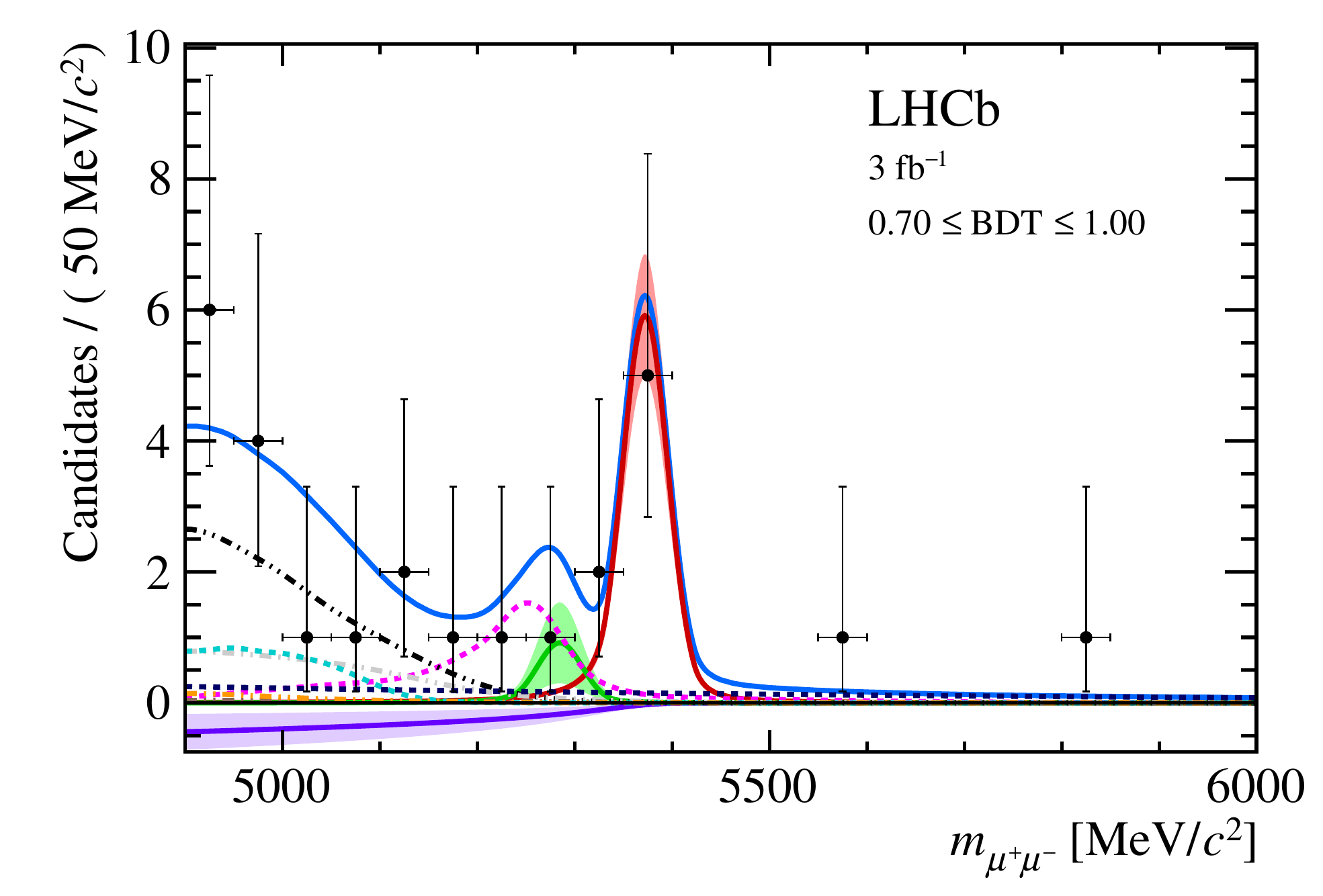}
    \includegraphics[width=0.4\textwidth]{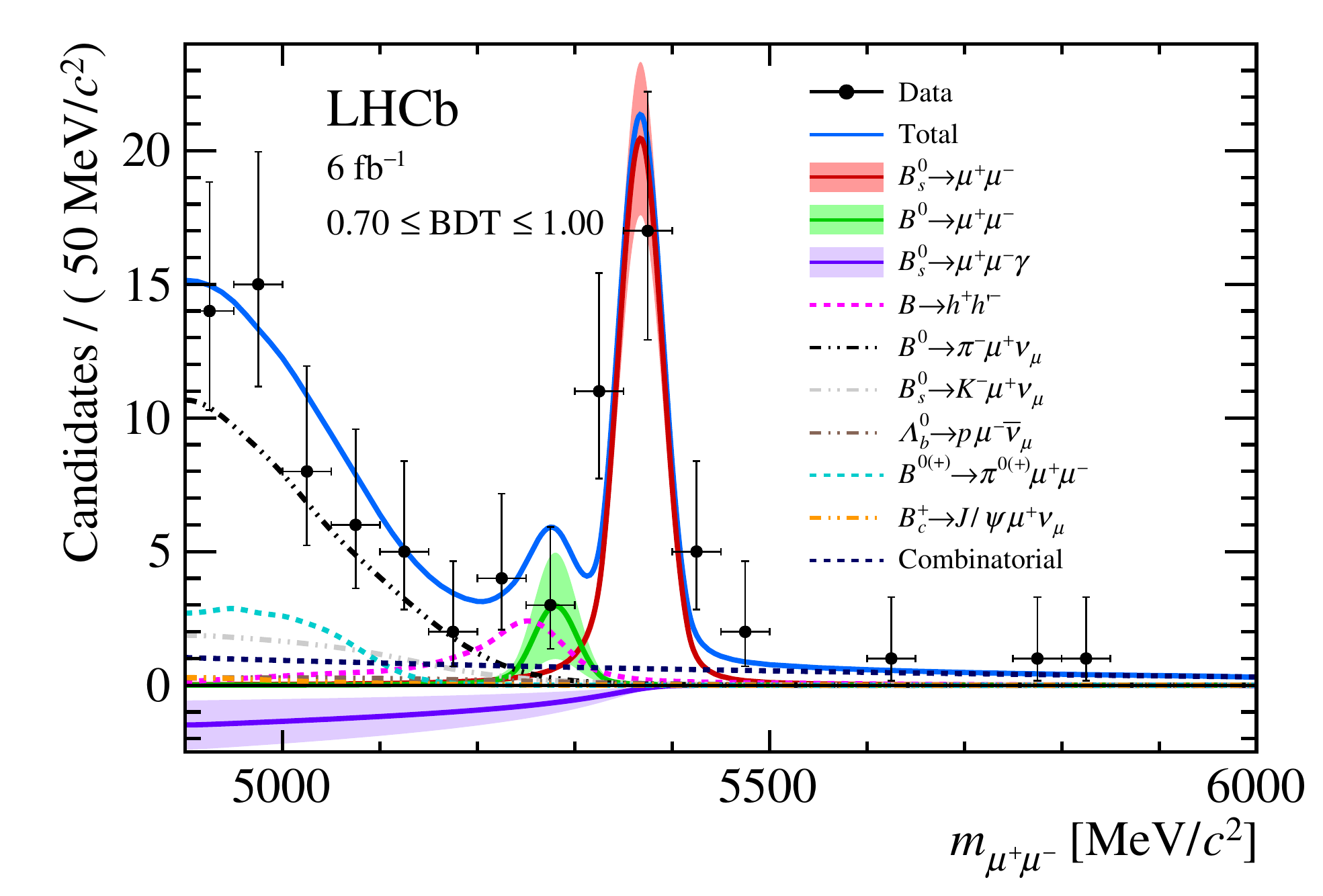}

    \caption{Mass distribution of signal candidates (black dots) for (left) \runone and (right) \runtwo samples in regions of BDT. 
    The result of the fit is overlaid (blue line) and the different components detailed in the legend. The solid bands represent the variation of the signal branching fractions within their total uncertainty.} 
    \label{fig:fit_nominal}
\end{figure}

The data sample containing \Bdsmumugamma candidates is divided into the two data-taking periods, \runone and \runtwo, which are further divided into six subsets based on the BDT response, using the intervals defined in Sec.~\ref{sec:Selection}. The branching fractions of the signal decays are determined using an unbinned extended maximum-likelihood fit to the dimuon mass distributions, performed simultaneously on all the subsets. Due to substantial contamination from combinatorial background, the lowest BDT region, \mbox{$0 \leq \bdt<0.25$}, is excluded from the data set but its fraction is taken into account in the total normalisation of the BDT shape. The fit is performed in a mass window of $4900\leq m_{\mu\mu}\leq6000\,\mevcc$. The dimuon mass distribution is shown in Fig.~\ref{fig:fit_nominal} for the \runone and \runtwo samples in all BDT intervals. The low mass region is populated by the partially reconstructed background and \bsmumugamma decays, while the higher mass region is dominated by combinatorial background. 

The probability density functions (PDFs) of the \bsmumu and \bdmumu decays are described by DSCB functions, defined in Eq.~\ref{eq:dscb}, with their parameters Gaussian-constrained to the values measured in Sec.~\ref{sec:Calibration}. The mass distribution of the \Bsmumugamma decay is described using an empirical threshold function
\begin{equation}
\label{eq:mumugammashape}
 f(m_{\mu\mu}) \propto \left( 1 -\frac{ m_{\mu\mu}}{M_{B_s}}\right)^b - a \sqrt{1 -e^{\frac{m_{\mu\mu} - M_{B_s}}{s}}},
\end{equation}
where the parameters $a$, $b$ and $s$, are determined from simulation, which is based on the theoretical predictions and form factors of Ref.~\cite{Melikhov:2004mk}. The parameter $b$ is found to be close to 0.5, while the other parameters vary across the BDT regions. This threshold function is convolved with a Gaussian resolution function which models the effect of the detector resolution. The parameter values are estimated from kinematically weighted simulated events, and fixed in the fit. While the branching fraction prediction for the \Bsmumugamma decay is dependent on the exact form-factor parametrisation used~\cite{Kozachuk:2017mdk}, the distribution of the dimuon mass at high $q^2$ is found to not depend significantly on the choice of the form-factor, and so the same threshold function can be used for a range of scenarios. Moreover, varying the detector resolution parameter within the known uncertainties has a negligible effect on the yield of \bsmumugamma decays. The combinatorial background is modelled with a single exponential function with an independent yield in each BDT region but with common slope parameters for each data-taking period. Both the yields and the parameters are free to float in the fit. The \Bhh and semileptonic $b$-hadron contributions are described using kernel estimation techniques~\cite{Cranmer:2000du} applied to simulated events in each BDT region. Their expected yields in each BDT region are Gaussian-constrained according to the values reported in Sec.~\ref{sec:Backgrounds}. Moreover, common parameters, such as the yields of the normalisation channels, efficiencies and branching fractions are shared across all BDT regions and their values are Gaussian-constrained to their estimated values and uncertainties.

The result of the fit in each subset is shown in Fig.~\ref{fig:fit_nominal}.
The resulting branching fractions of the \Bsmm, \Bdmm and \Bsmumugamma decays are
\begin{eqnarray*}
\BRof \Bsmumu &=&\Bsbr,\\ 
\BRof \Bdmumu &=&\Bdbr,  \\
\BRof \Bsmumugamma &=&\Bsmmgbr\mbox{ with }m_{\mu\mu}>4.9\gevcc . 
\end{eqnarray*}
The statistical uncertainties are evaluated by repeating the fit with all nuisance parameters fixed to the value obtained in the standard fit, where all nuisance parameters are free to float within their constraints. The systematic uncertainties are then computed by subtracting in quadrature the statistical uncertainties from the total ones. The main contribution to the systematic uncertainty of $\BRof \Bsmumu$ originates from the knowledge of $f_s/f_d$ (3\%), while that of $\BRof\Bdmumu$ is dominated by the knowledge of the \Bhh and semileptonic $b$-hadron background (9\%). The correlation between the \Bdmm and \Bsmm branching fractions is found to be $-11\%$. The correlation between the \Bsmumugamma and \Bdmm branching fractions is $-25\%$, while that with \Bsmm is below 10\%. 

\begin{figure}[tb]
    \centering
     \includegraphics[width=0.90\linewidth]{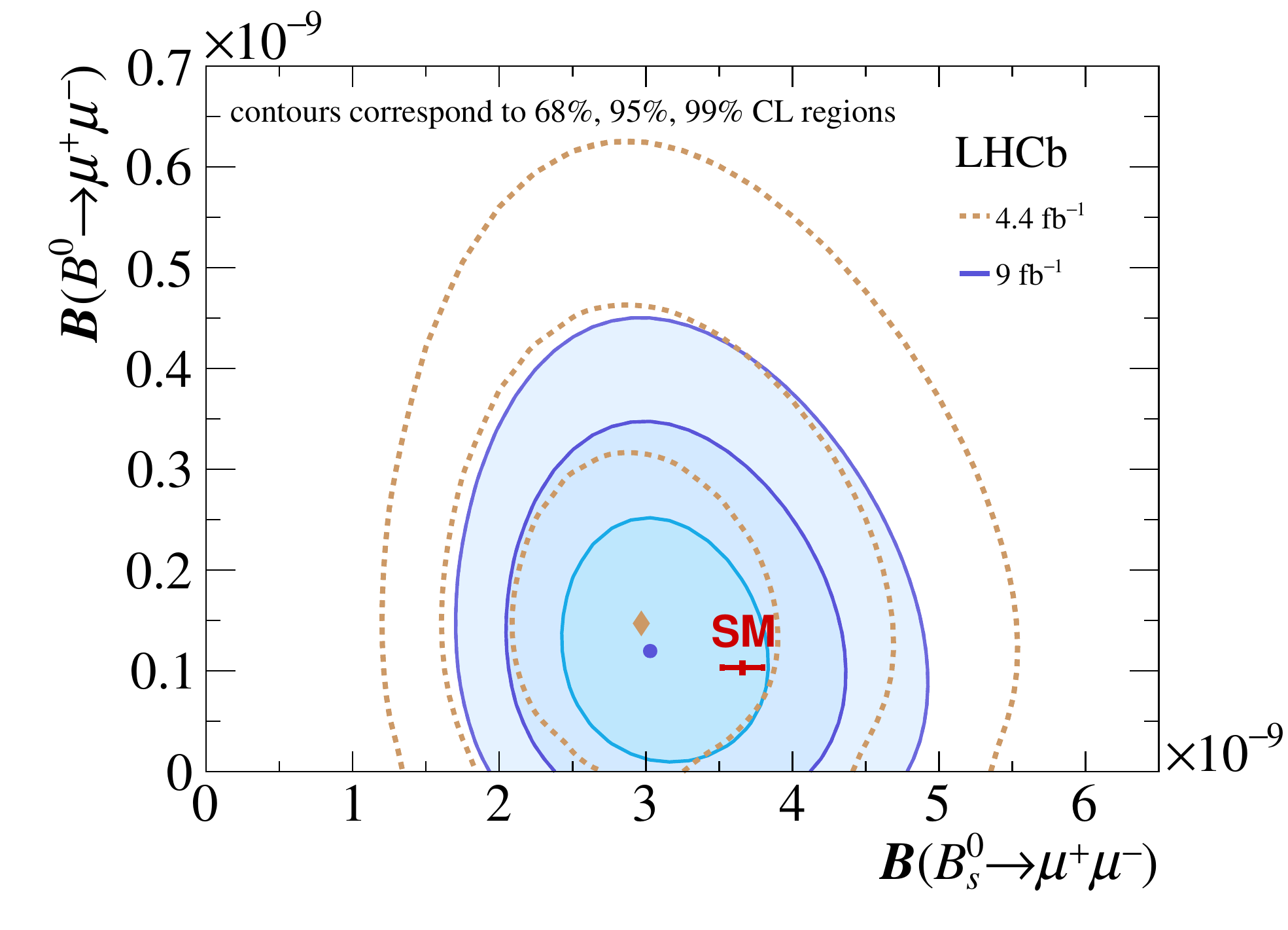}\\     
     \includegraphics[width=0.49\linewidth]{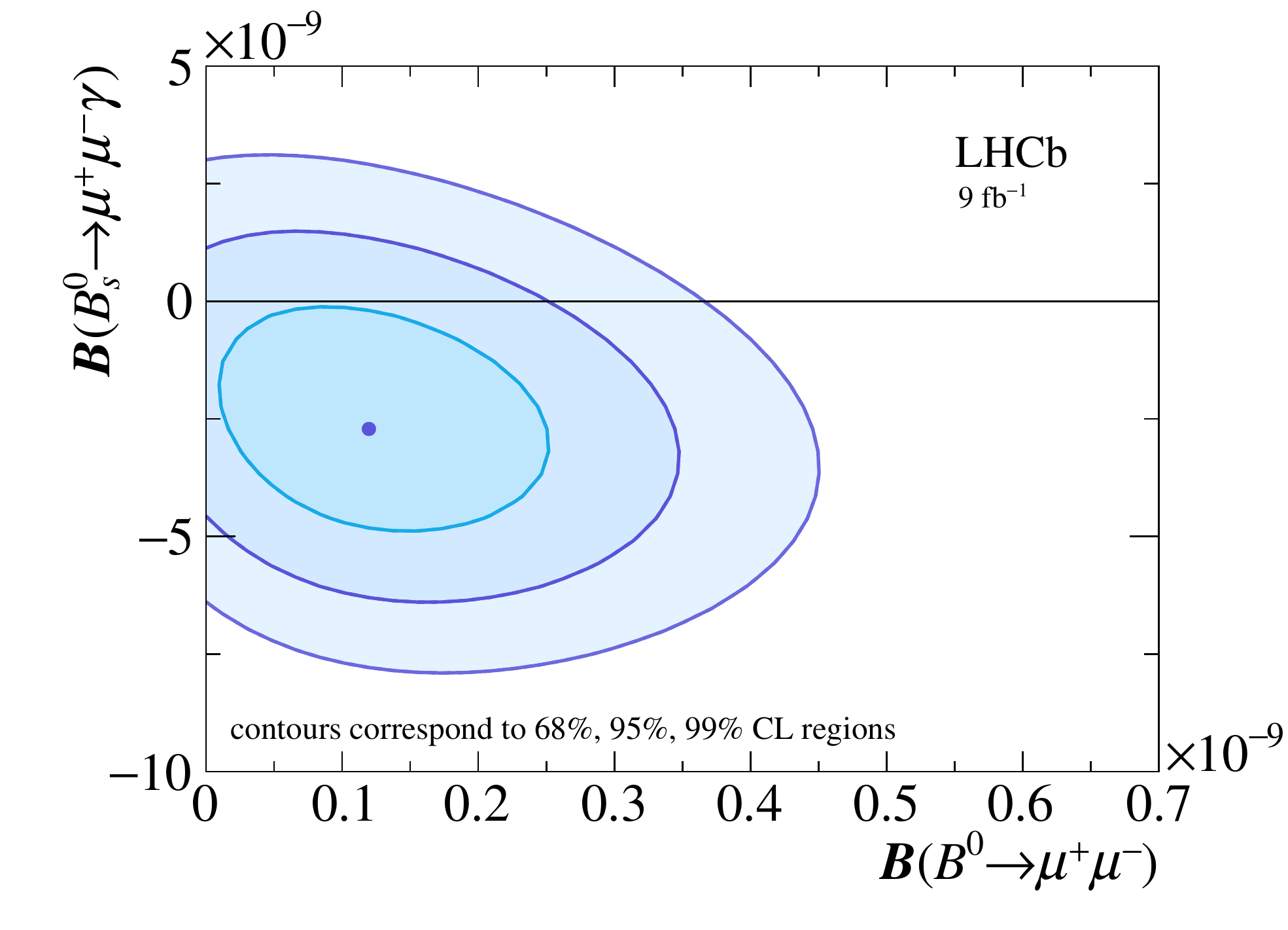}
     \includegraphics[width=0.49\linewidth]{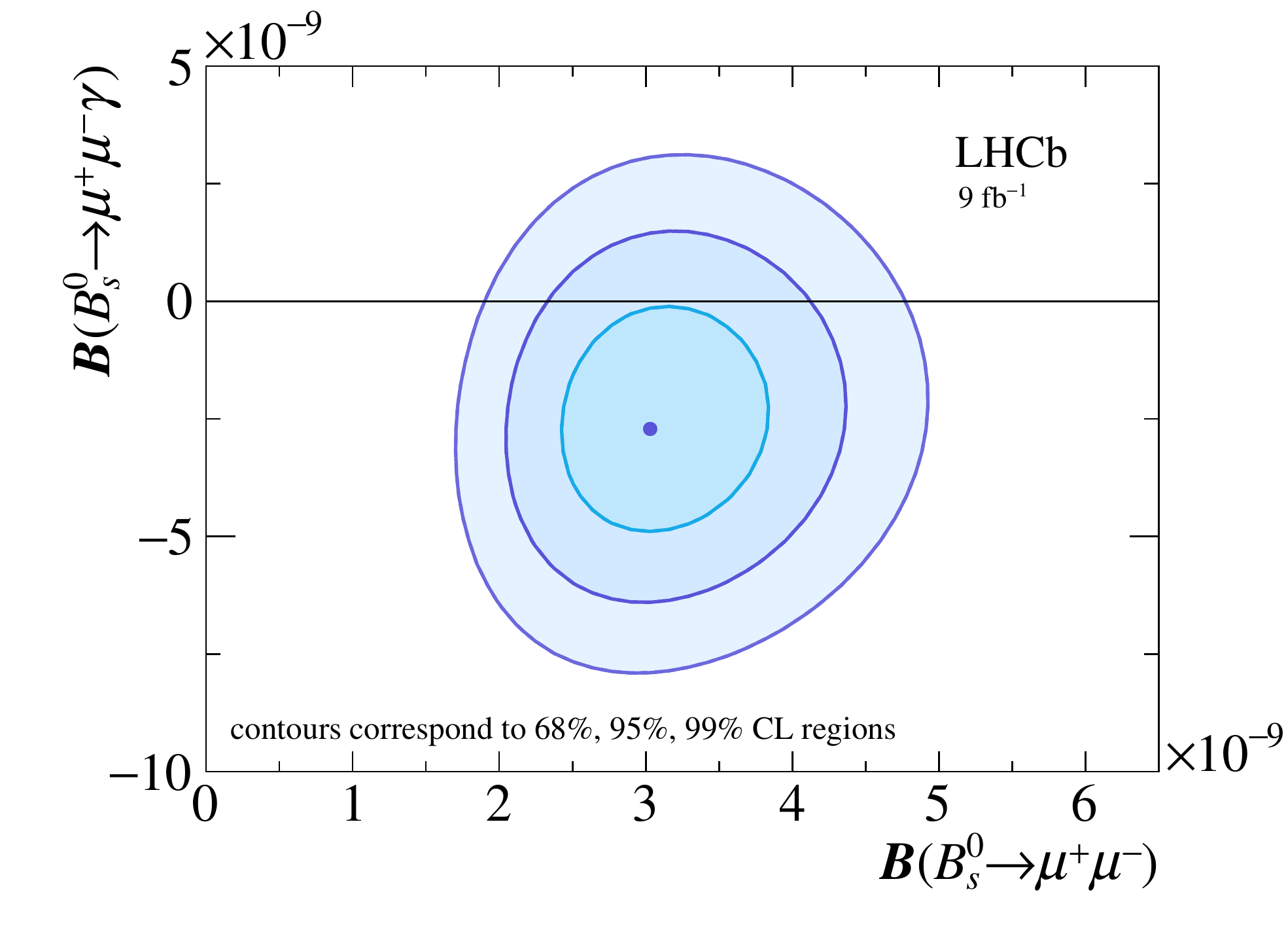}
\caption{Two-dimensional representations of the branching fraction measurements for the decays (top) \mbox{\Bsmm} and \Bdmm, (bottom left) \Bdmm and \Bsmumugamma and (bottom right) \Bsmm and \Bsmumugamma. The \Bsmumugamma branching fraction is limited to the range $m_{\mu\mu}>4.9\gevcc$. The measured central values of the branching fractions are indicated with a blue dot. The profile likelihood contours for 68\%, 95\% and 99\% CL regions of the result presented in this paper are shown as blue contours, while in the top plot the brown contours indicate the previous measurement~\cite{LHCb-PAPER-2017-001} and the red cross shows the SM prediction. Figure reproduced from Ref.~\cite{LHCb-PAPER-2021-007}.}
\label{fig:2Dprof}
\end{figure}

Two-dimensional profile likelihoods are evaluated in the plane of the possible combinations of two branching fractions. These are obtained by taking the ratio of the likelihood value of a fit where the parameters of interest are fixed and the likelihood value of the standard fit. The results are shown in Fig.~\ref{fig:2Dprof}.

\begin{figure}[tb]
    \centering
    \includegraphics[width=0.49\textwidth]{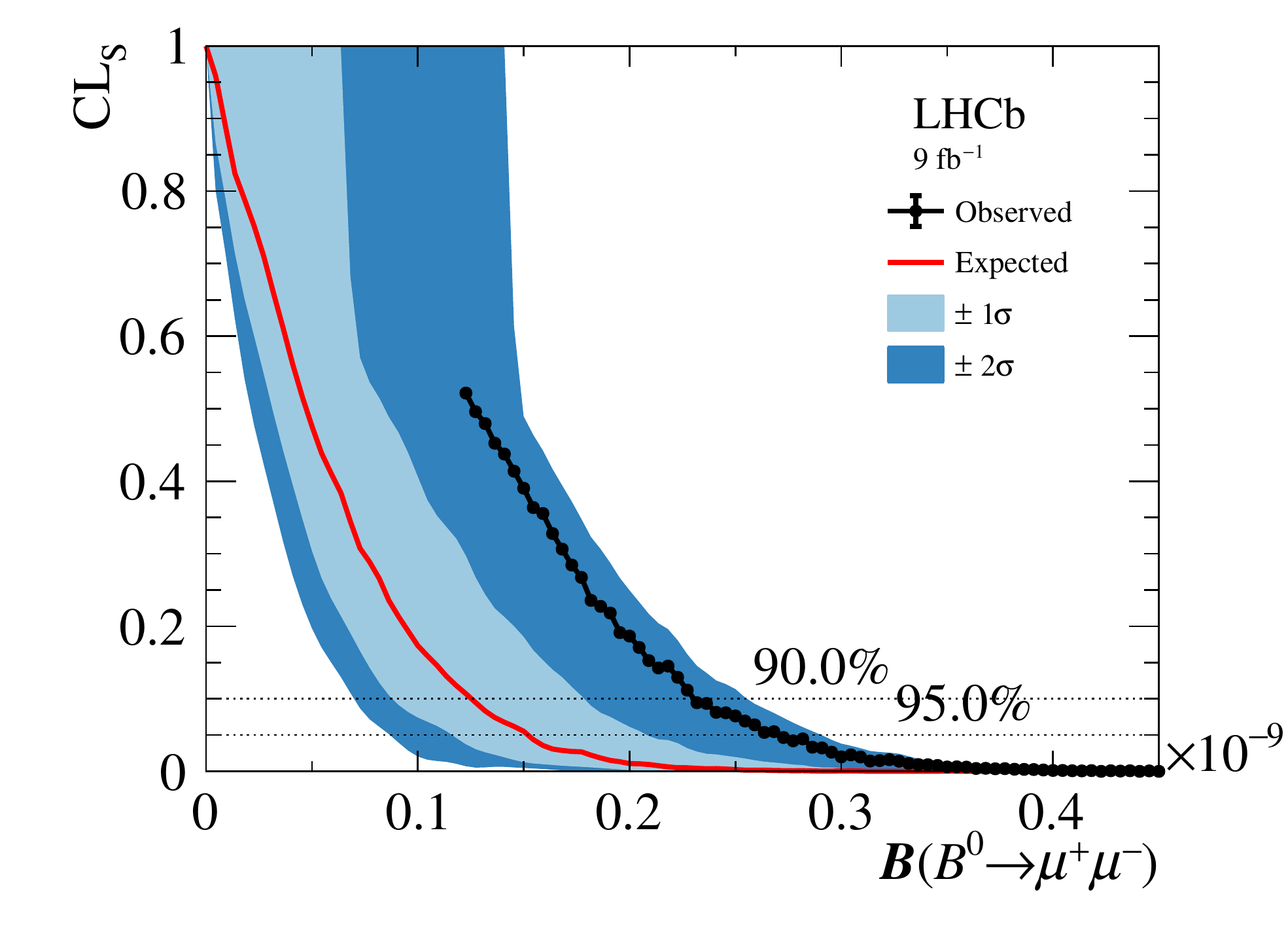}
    \includegraphics[width=0.49\textwidth]{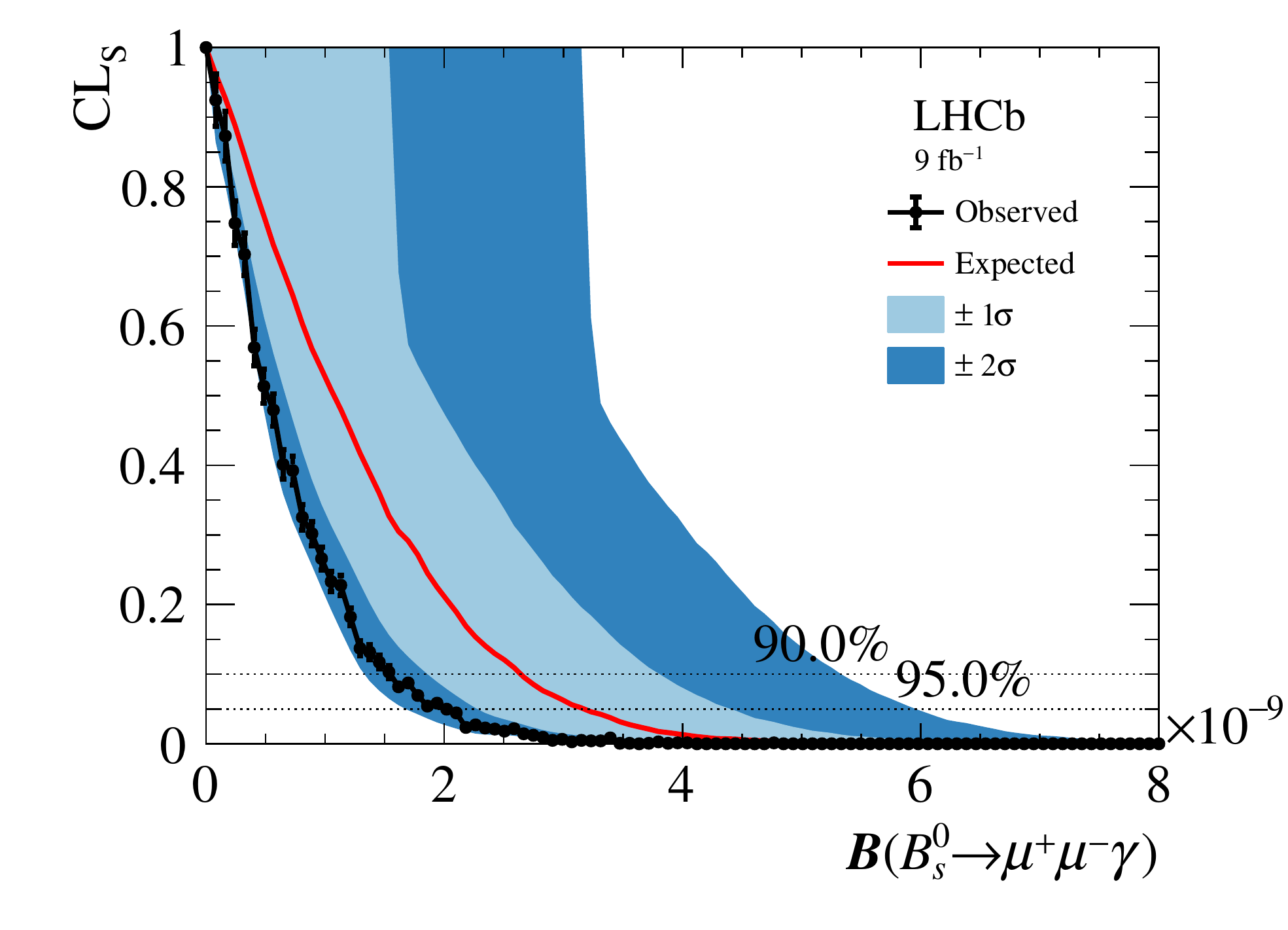}

    \caption{Results from the CL$_{\text{s}}$ scan used to obtain the limit on (left) $\BF(\Bdmm)$ and (right) $\BF(\Bsmumugamma)$. The background-only expectation is shown by the red line and the 1$\sigma$ and 2$\sigma$ bands are shown as light blue and blue bands respectively. The observation is shown as the solid black line. The two dashed lines intersecting with the observation indicate the limits at 90\% and 95\% CL for the upper and lower line, respectively.}
    \label{fig:obslimits}
\end{figure}

The difference between the logarithm of the likelihood values under the presence or the absence of a specific signal component is used to evaluate the significance with Wilks' theorem~\cite{Wilks:1938dza}. The \Bsmm signal exceeds the background-only hypothesis more than \Bssigma standard deviations ($\sigma$), while the statistical significance of the \Bdmm signal is $\Bdsigma\,\sigma$ and the \bsmumugamma signal is compatible with the background-only hypothesis within $\Bsmumugammasigma\,\sigma$. Since no evidence for \Bdmm and \bsmumugamma decays is found, upper limits on their branching fractions are evaluated using the \CLs method~\cite{Read:2002hq} with a one-sided test statistic~\cite{Cowan:2010js} as implemented in Refs.~\cite{GammaCombo,LHCb-PAPER-2016-032}. The one-sided test statistic for a given branching fraction value is defined as twice the negative logarithm of the profile likelihood ratio if it is larger than the measured branching fraction and zero otherwise. Its distribution is determined from pseudoexperiments, where nuisance parameters are set to their best fit values for toy generation while central values of the Gaussian-constraints are independently fluctuated within their uncertainty for each pseudoexperiment as described in Ref.~\cite{Bodhisattva:2009uba}. The \CLs curves are shown in Fig.~\ref{fig:obslimits} from which the limit on the \Bdmumu and \Bsmumugamma branching fractions are found to be 
\begin{eqnarray*}
\BRof \Bdmumu  &<& \Bdobslimit, \\  
\BRof \bsmumugamma &<& \Bsmmgobslimit\mbox{ with }m_{\mu\mu}>4.9\gevcc,
\end{eqnarray*}
at $90\% \,(95\%)$ CL.
The measured upper limits are shown in Fig.~\ref{fig:obslimits}, together with the expected ones. 

To quantify the impact of the \Bsmumugamma component on the other signal modes, the fit is repeated by fixing its branching fraction to zero. Using this configuration, \BRof \Bsmumu increases by 2\% while the limit on \BRof \bdmumu decreases by 12\%. 

As described in Sec.~\ref{sec:bdt_calibration}, the BDT calibration of \Bsmm decays depends on the effective lifetime which introduces a model dependence in the measured time-integrated branching fraction. In the fit, the SM value of $\ADeltaGamma=1$ is assumed for \Bsmm and \bsmumugamma decays. The model dependence is evaluated by repeating the fit under the assumptions $\ADeltaGamma=0$ and $-1$, finding an increase of the \Bsmumu branching fraction with respect to the SM hypothesis of 4.7\% and 10.9\%, respectively. On the contrary, the \Bsmumugamma branching fraction decreases with respect to the $\ADeltaGamma=1$ hypothesis by 2\% and 5\% with a neglible impact on its limit. The dependence for \Bsmumu and \Bsmumugamma is approximately linear in the physically allowed \ADeltaGamma range. To evaluate the ratio of the branching fractions $\mathcal{R}_{\mu^+\mu^-}$ defined in Eq.~\ref{eq:ratio}, the fit is modified such that $\mathcal{R}_{\mu^+\mu^-}$ and \BRof \Bsmumu are floating observables, which allows for the cancellation of common uncertainties, while \BRof \Bsmumugamma is kept as a floating observable. The ratio is found to be
\begin{eqnarray*}
\mathcal{R}_{\mu^+\mu^-} &=&\ratio\ , 
\end{eqnarray*}
where the first uncertainty is statistical and the second systematic. Using the \CLs method described above, the upper limit is evaluated to be
\begin{equation*}
\mathcal{R}_{\mu^+\mu^-}<\ratiolimit
\end{equation*}
at $90\% \,(95\%)$ CL.

\section{\boldmath Measurement of \Bsmumu effective lifetime}
\label{sec:Lifetime}
The effective lifetime of the \Bsmumu decay is measured using the same data sample as for the branching fraction measurement but with a slightly different selection, described in Sec.~\ref{sec:Selection}. The data are divided into two regions of the BDT classifier response and unbinned extended maximum-likelihood fits are performed to the dimuon mass distribution in each region in order to calculate weights using the \sPlot method \cite{Pivk:2004ty}. These weights are then used to extract the \Bsmumu signal decay-time distributions. Finally, the effective lifetime is measured using an unbinned maximum-likelihood fit to the weighted decay-time distributions, performed simultaneously to both BDT regions.

\begin{figure}[tb]
    \centering
    \includegraphics[width=0.48\linewidth]{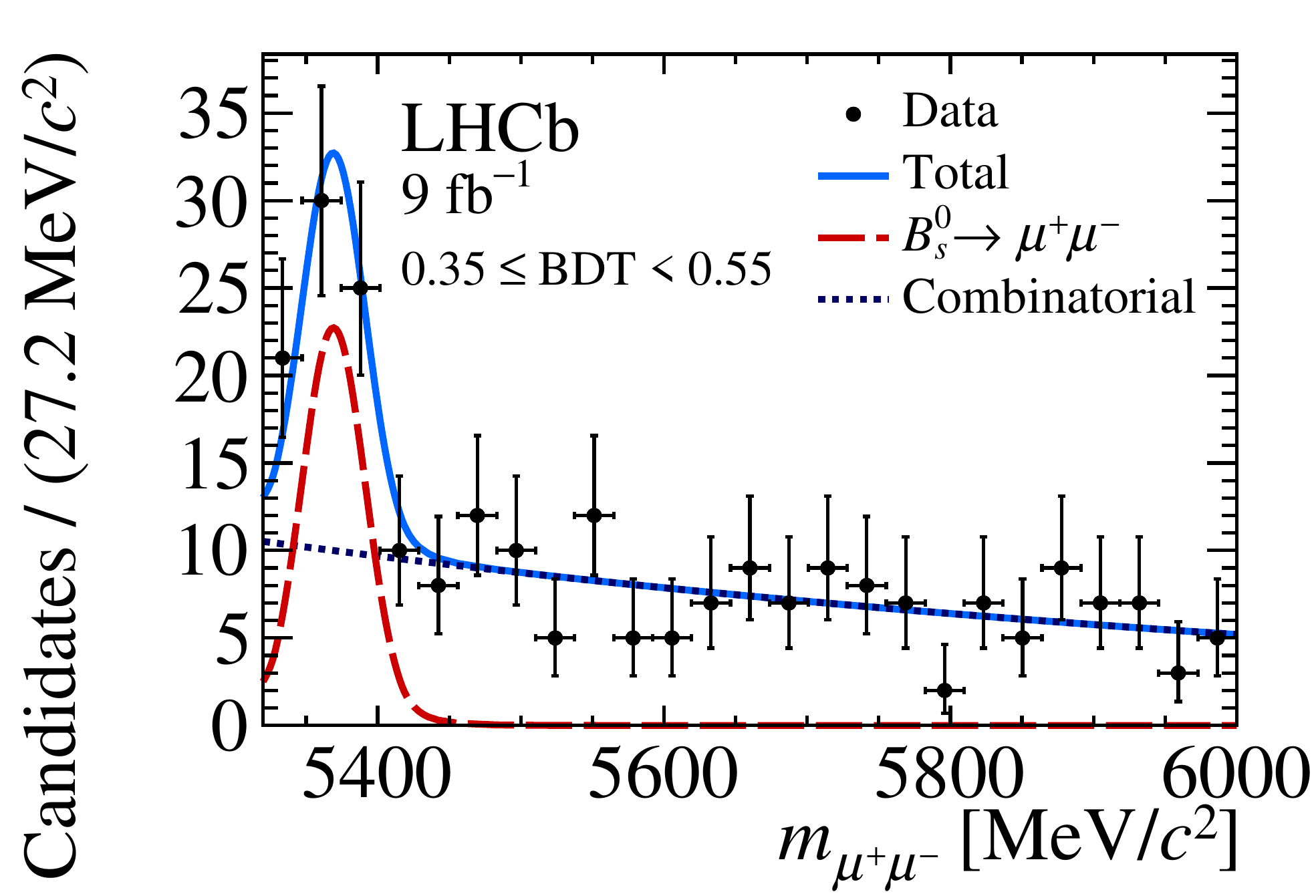}
    \includegraphics[width=0.48\linewidth]{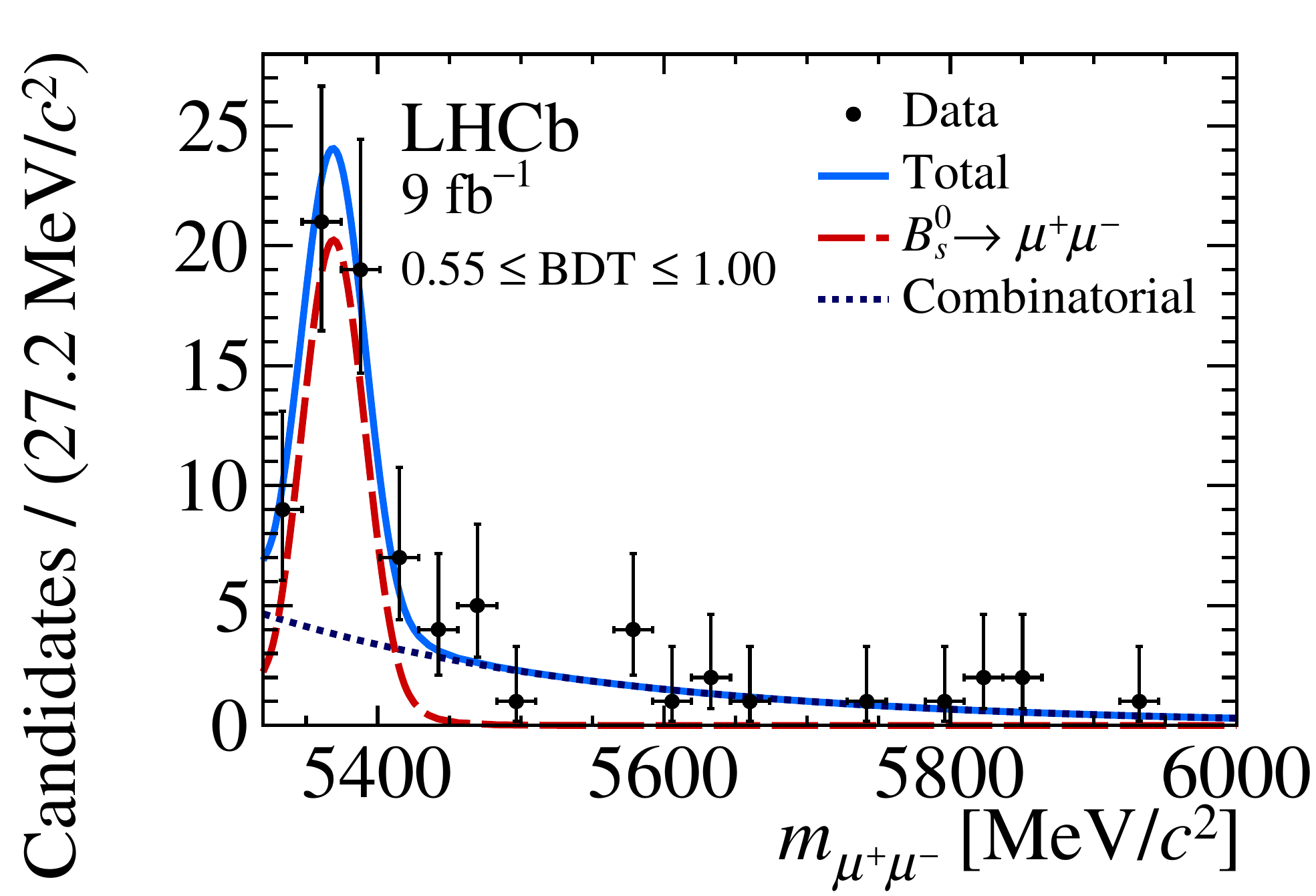}
\caption{Dimuon mass distributions of \bsmumu candidates with the fit model used to perform the background subtraction for the measurement of the \Bsmumu effective lifetime superimposed in the (left) low and (right) high BDT regions.}\label{fig:LT_massfits}
\end{figure}

The fits to the dimuon mass used to extract the weights are performed in the range $5320\leq m_{\mu\mu} \leq 6000\,\mevcc$. The lower limit of 5320\,\mevcc removes the low mass region containing most of the physical background, including \Bdmumu and \Bhh decays, so only \Bsmumu decay and combinatorial background components are included in the fit. Residual contamination from physical background in the fit region is low and is treated as a source of systematic uncertainty. The \Bsmumu signal is modelled using a DSCB PDF and the background with an exponential PDF. The parameters of the signal PDF are determined using the same method as for the branching fraction (Sec.~\ref{sec:BFfit}) and are fixed in the fit, while the signal and background yields and the decay constant of the combinatorial background exponential are allowed to float freely. The distributions of the dimuon mass in the two BDT regions are shown in Fig.~\ref{fig:LT_massfits}.

\begin{figure}[!t]
    \centering
      \includegraphics[width=0.48\linewidth]{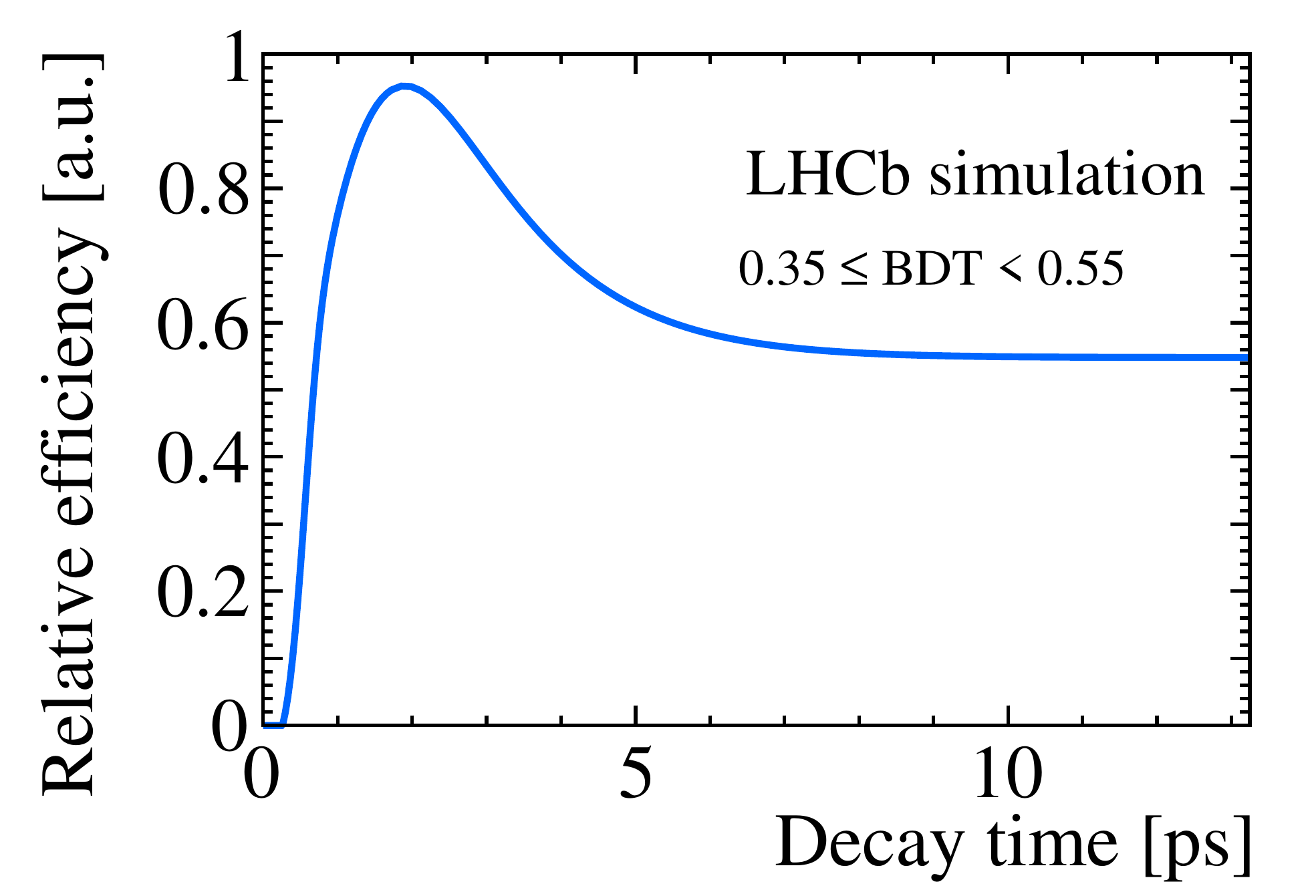}
      \includegraphics[width=0.48\linewidth]{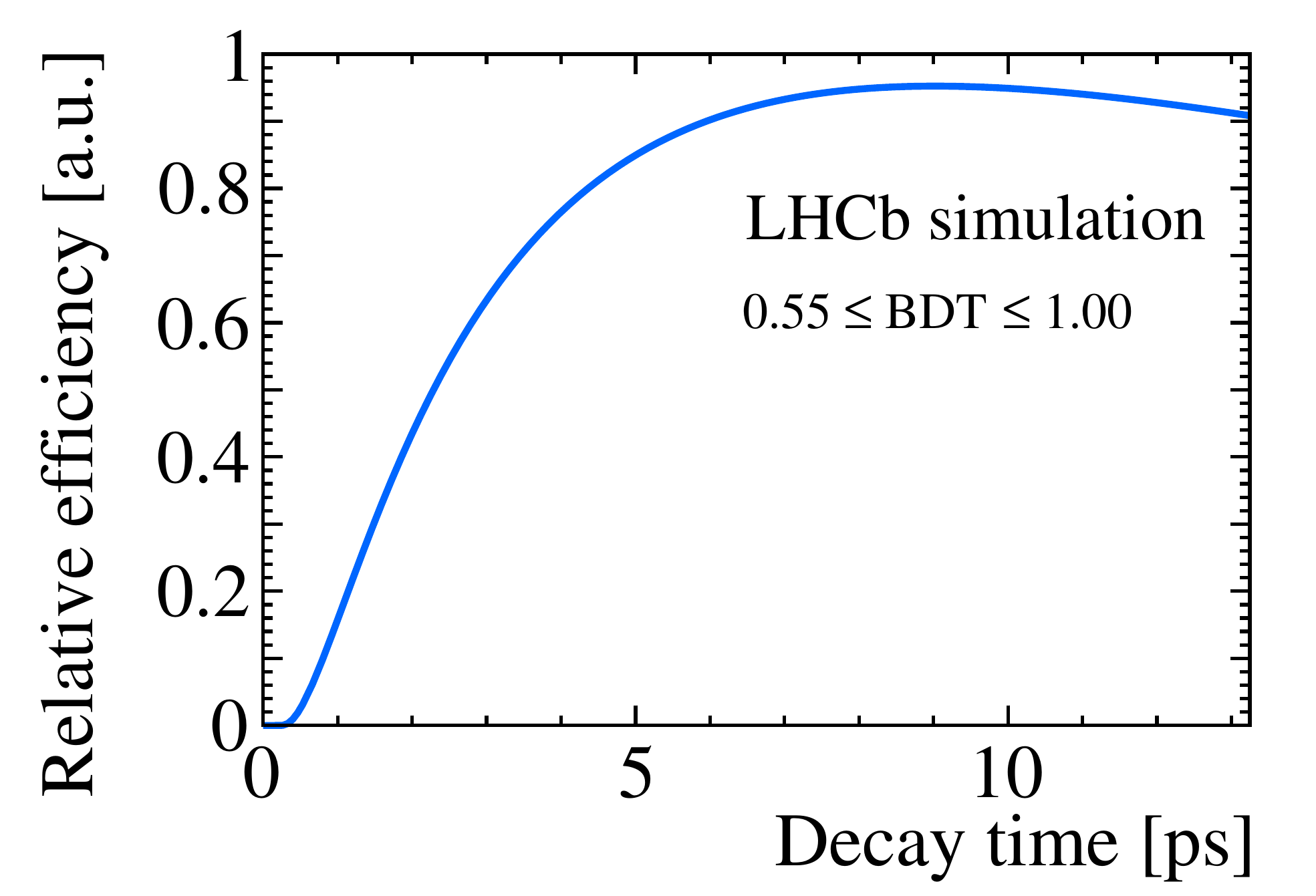}
\caption{The functions used to model the decay-time efficiency in the (left) low and (right) high BDT regions in the fit for the \bsmumu effective lifetime.}\label{fig:acceptance}
\end{figure}

To make an unbiased measurement of the effective lifetime, the decay-time dependence of the combined trigger, reconstruction and selection efficiency must be accurately estimated. This decay-time acceptance is calculated using simulated \Bsmumu candidates, which have been weighted in order to improve agreement with data. The parameters of the functions used to model the decay-time efficiency are extracted using unbinned maximum-likelihood fits to the decay-time distributions of simulated \Bsmumu candidates, with the effective lifetime fixed to its true value. Since the decay-time efficiency has a different form in the two BDT regions, two different empirical functions are used. In the low BDT region the efficiency is modelled as
\begin{equation}\label{eq:accbin1}
\varepsilon(t) = a \times \text{Erf}\left(t \sqrt{b\times\text{tanh}(c\,t^{3})}\right) + \text{exp}\left(-d\,t^e\right) - 1,
\end{equation}
where Erf is the error function, $t$ is the reconstructed decay time, $a$, $b$, $c$, $d$ and $e$ are free parameters and $\varepsilon(t) = 0$ when $t < 0.26 \ps$.
The acceptance in the high region is modelled using
\begin{equation}\label{eq:accbin2}
\varepsilon(t) =  \text{exp}\left(-\frac{1}{2} \left(\frac{\text{ln}\left(t-t_{0}\right) - f}{g}\right)^{2} \right),
\end{equation}
where $f$\kern-0.1em, $g$ and $t_{0}$ are free parameters and $\varepsilon(t) = 0$ when $t \leq t_{0}$. The forms of these functions with respect to the \Bs meson decay time are shown in Fig.~\ref{fig:acceptance}.  The different behaviour in the two intervals reflect the positive correlation of the \Bs-meson decay time with the BDT response, so that the low (high) BDT region contains more signal decays with small (large) decay times.

Finally, the \Bsmumu effective lifetime is determined using a simultaneous fit to the background-subtracted decay-time distributions in the two BDT regions in data, where the decay-time distributions are modelled by the acceptance functions above multiplied by an exponential function. Only the effective lifetime is allowed to float freely in the fit, while the parameters of the acceptance function are Gaussian constrained to the results of the fits to simulation.

Pseudoexperiments are used to evaluate several systematic effects that have the potential to bias the measurement. The fit procedure is found to return an unbiased estimate of the lifetime to a precision of $0.009 \ps$ and good coverage. The effects of residual contamination from physical background, predominantly \Bhh and \Lbpmunu decays, is found to introduce a bias of around $0.012 \ps$. The effect of the decay-time acceptance on the mixture of the light and heavy mass eigenstates is evaluated and found to be negligible~\cite{LHCb-PAPER-2017-001}. A further source of uncertainty is related to the decay-time distribution of the combinatorial background, which is unknown \textit{a priori} and can bias the lifetime measurement if the background lifetime is significantly longer than that of the signal. The decay-time distribution of combinatorial background in the dimuon sample cannot be determined directly from data due to the very small number of candidates and so instead the decay-time model used in the pseudoexperiments is taken from a fit to the decay-time distributions of candidates in the high mass region of the higher yield di-hadron sample, which includes both a short and long lived component. A systematic uncertainty is evaluated by fluctuating the mean lifetimes of both components upwards by $1\,\sigma$, which results in an overall bias on the \Bsmumu effective lifetime of around $0.003 \ps$. The effect of a production asymmetry between \Bs and \Bsb mesons~\cite{LHCb-PAPER-2016-062} is found to be small, at $0.002 \ps$. The effects of ignoring the detector decay-time resolution and also the choice of signal mass PDF are evaluated and are both found to be negligible. Any correlation between mass and decay-time is found to be negligible, as required by the \sPlot method.

\begin{figure}[t]
    \centering
     \includegraphics[width=0.48\linewidth]{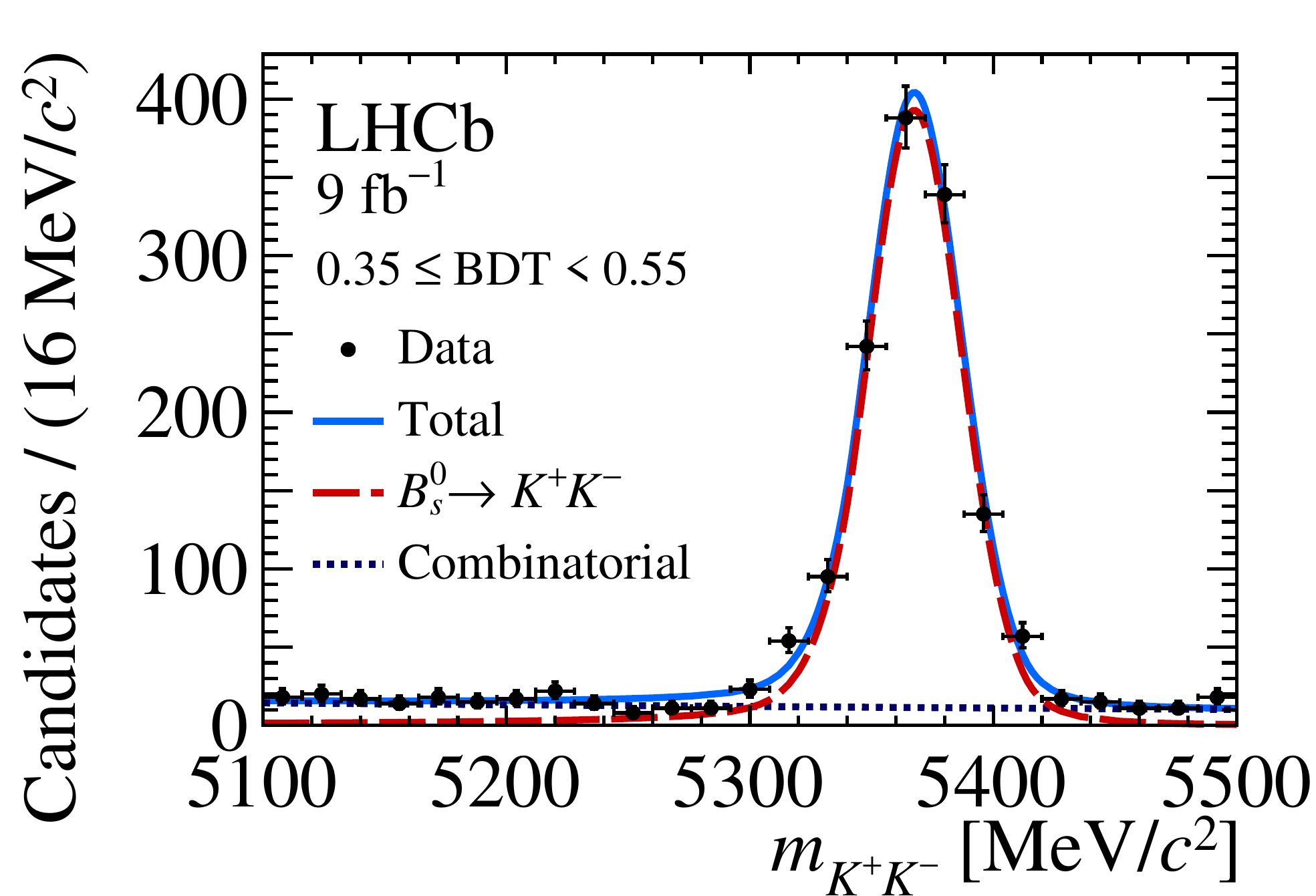}   
     \includegraphics[width=0.48\linewidth]{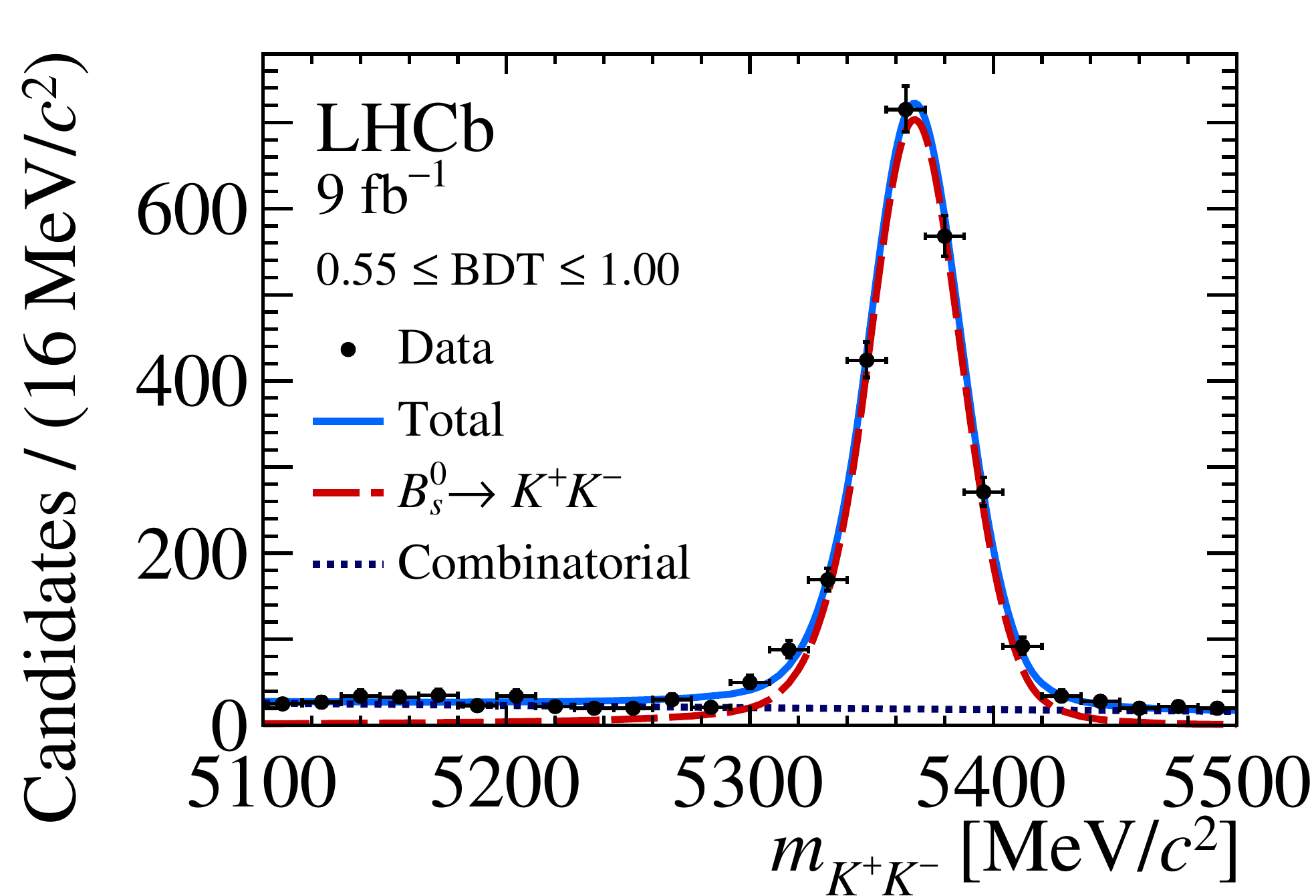}\\    
     \includegraphics[width=0.48\linewidth]{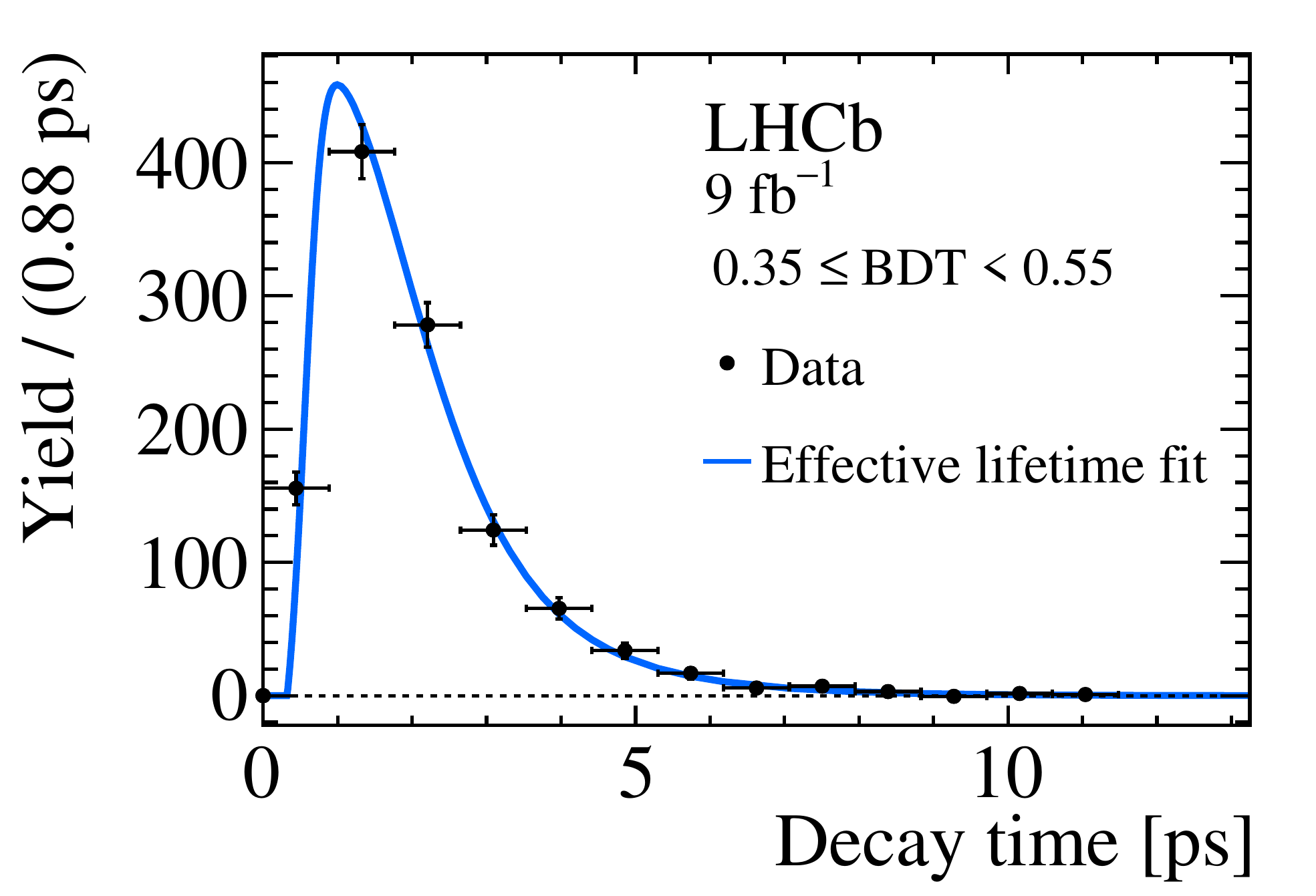}
     \includegraphics[width=0.48\linewidth]{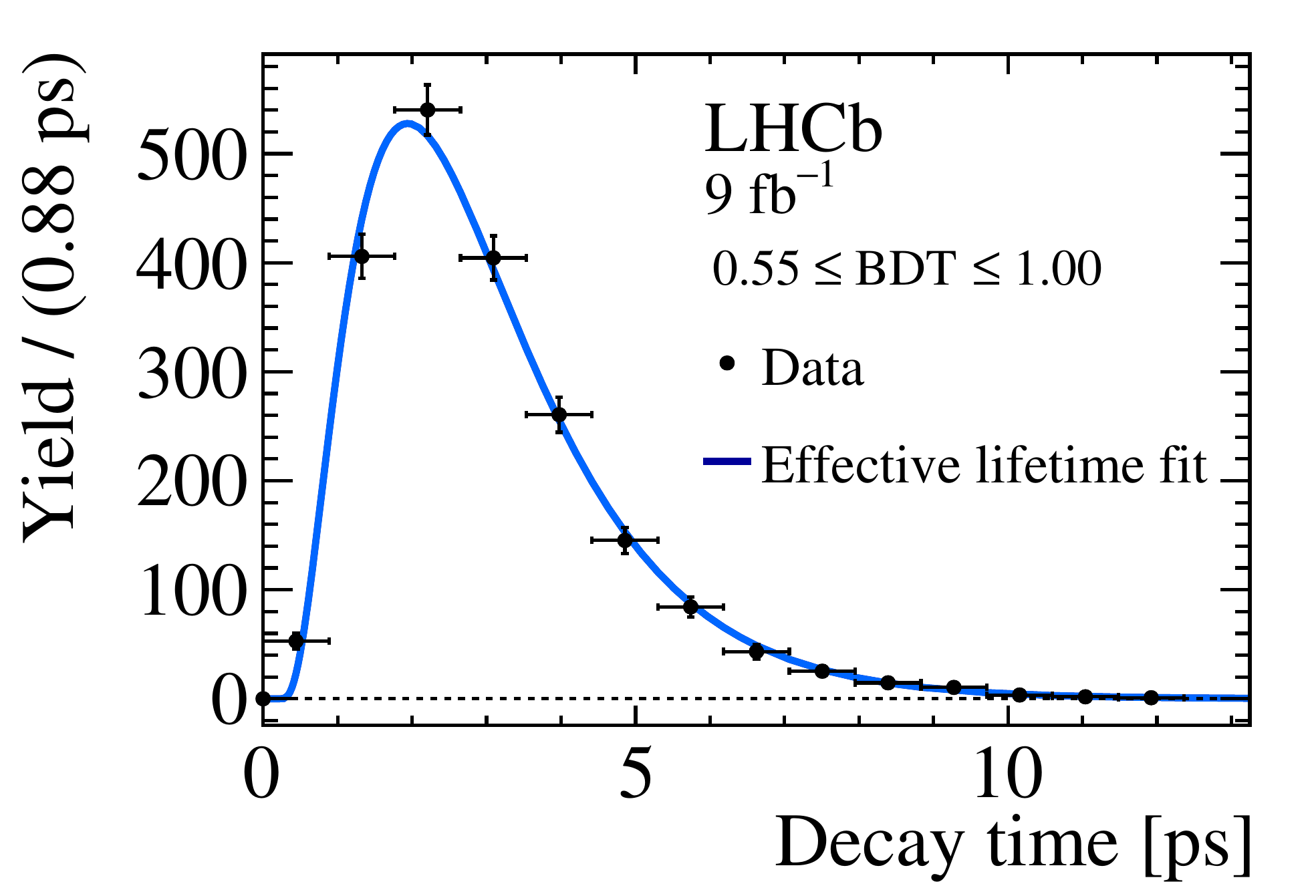} 
\caption{(Top) Distribution of $\Kp\Km$ mass with the fit models used to perform the background subtraction superimposed and (bottom) the background-subtracted decay-time distributions with the fit model used to determine the \BsKK effective lifetime superimposed (bottom row). The distributions in the low and high BDT regions are shown in the left and right columns, respectively.}\label{fig:BsKKfit}
\end{figure}

\begin{figure}[ht]
    \centering
     \includegraphics[width=0.48\linewidth]{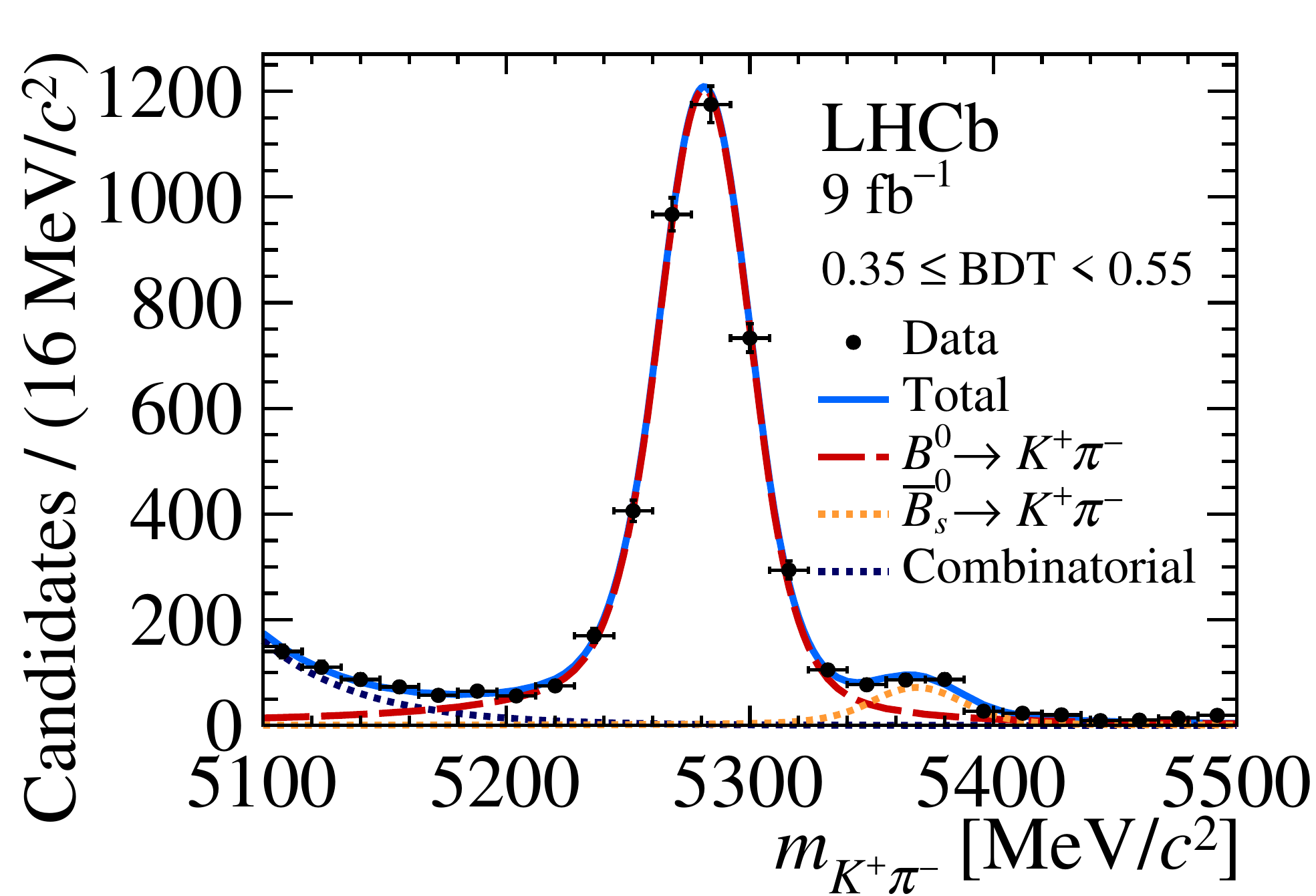}   
     \includegraphics[width=0.48\linewidth]{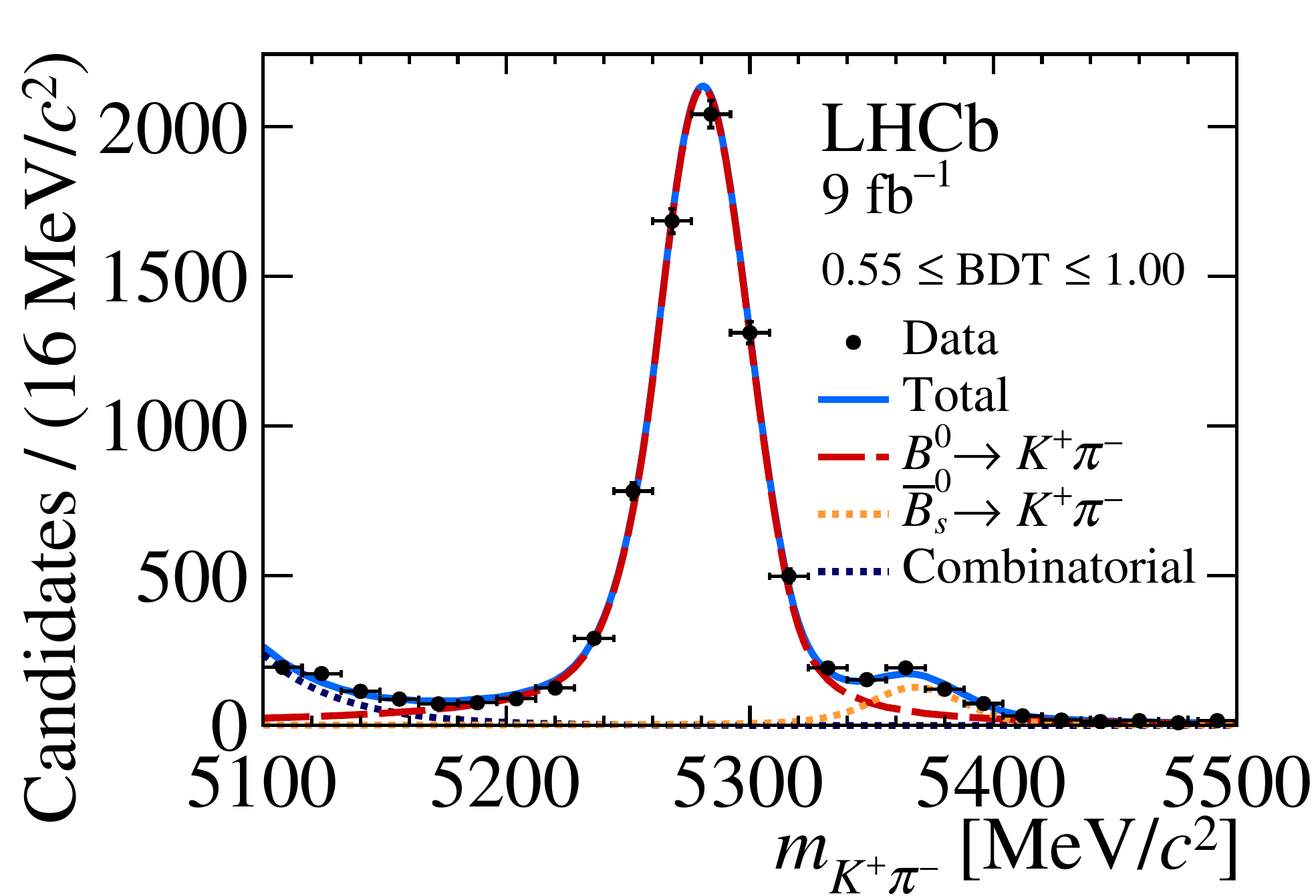}\\    
     \includegraphics[width=0.48\linewidth]{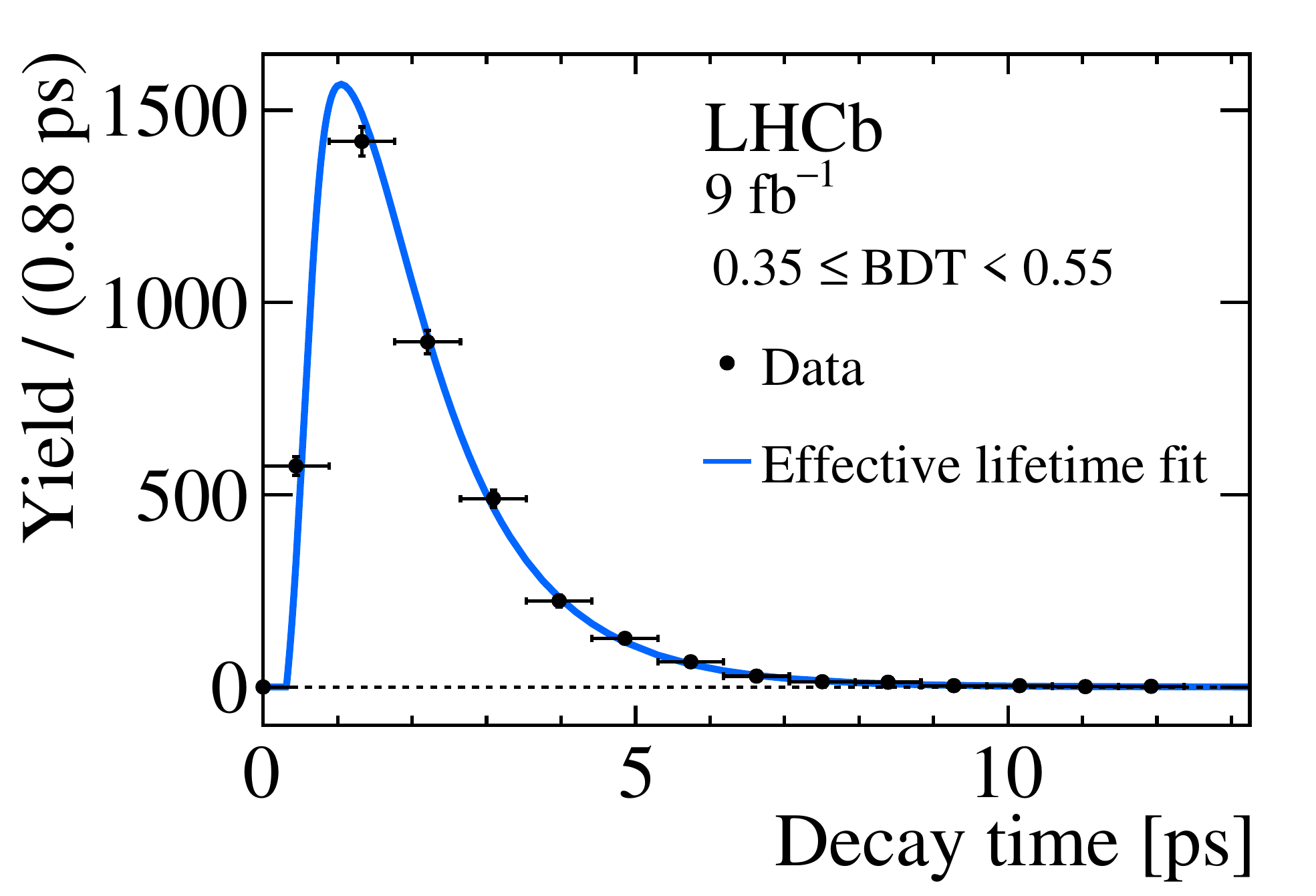}
     \includegraphics[width=0.48\linewidth]{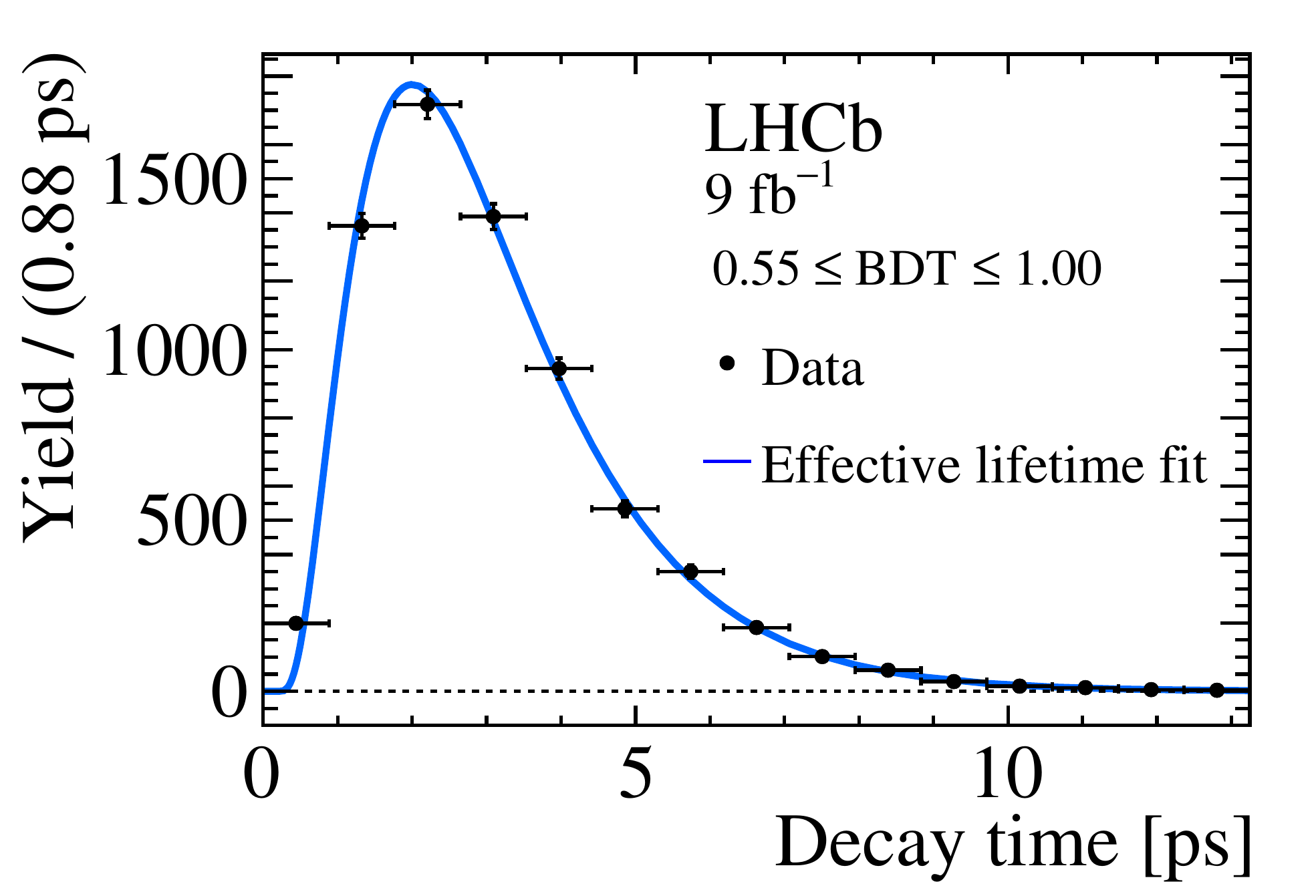} 
\caption{(Top) Distribution of $\Kp\pim$ mass with the fit models used to perform the background subtraction superimposed and (bottom) the background-subtracted decay-time distributions with the fit model used to determine the \BdKpi lifetime superimposed. The distributions in the low and high BDT regions are shown in the left and right columns, respectively.}\label{fig:BdKpifit}
\end{figure}

The entire procedure used to measure the lifetime is cross-checked by performing measurements of the lifetimes of the \BdKpi and \BsKK decays, which have much larger branching fractions and have already been precisely measured. Candidates are selected using similar requirements to those used to select \Bsmumu decays, with a few differences. While the efficiency of the \Bsmumu trigger selection is independent of decay time, this is not true for \Bhh decays where flight distance requirements are imposed on candidates. In order to match the \Bsmumu trigger requirements as closely as possible, \Bhh events are selected with TIS requirements, eliminating any dependence of the trigger efficiency on decay time. Furthermore, different requirements on particle identification information are imposed in order to separate the $\Kp\pim$ and $\Kp\Km$ final states. The decay-time acceptance is evaluated using weighted simulation in the same way as for \Bsmumu and the fit procedure is identical apart from some differences in the fit ranges and the inclusion of a \bskpi component in the \BdKpi fit. The fits to the \BsKK and \BdKpi mass and weighted decay-time distributions are shown in Figs. \ref{fig:BsKKfit} and \ref{fig:BdKpifit}.
The measured values of the \BsKK and \BdKpi lifetimes are found to be
\begin{align*}
\tau_{\BsKK} &= 1.435 \pm 0.026 \ps, \nonumber\\
\tau_{\BdKpi} &= 1.510 \pm 0.015 \ps, \nonumber
\end{align*}
where the uncertainties are statistical only. Systematic uncertainties on these cross-check measurements are not evaluated. The results are in agreement with the values measured previously by the LHCb collaboration of $\tau_{\BsKK} = 1.407 \pm 0.016 \ps$ and $\tau_{\BdKpi} = 1.524 \pm 0.011 \ps$ \cite{LHCb-PAPER-2014-011}. The measurements presented here are performed on data samples with very little overlap with those used to make the measurements published in Ref.~\cite{LHCb-PAPER-2014-011} due to different data-taking periods and trigger requirements. The statistical uncertainty on the measured \BsKK lifetime is taken as the systematic uncertainty associated with the use of simulated events to determine the \Bsmm acceptance function.

\begin{figure}[t]
    \centering
     \includegraphics[width=0.48\linewidth]{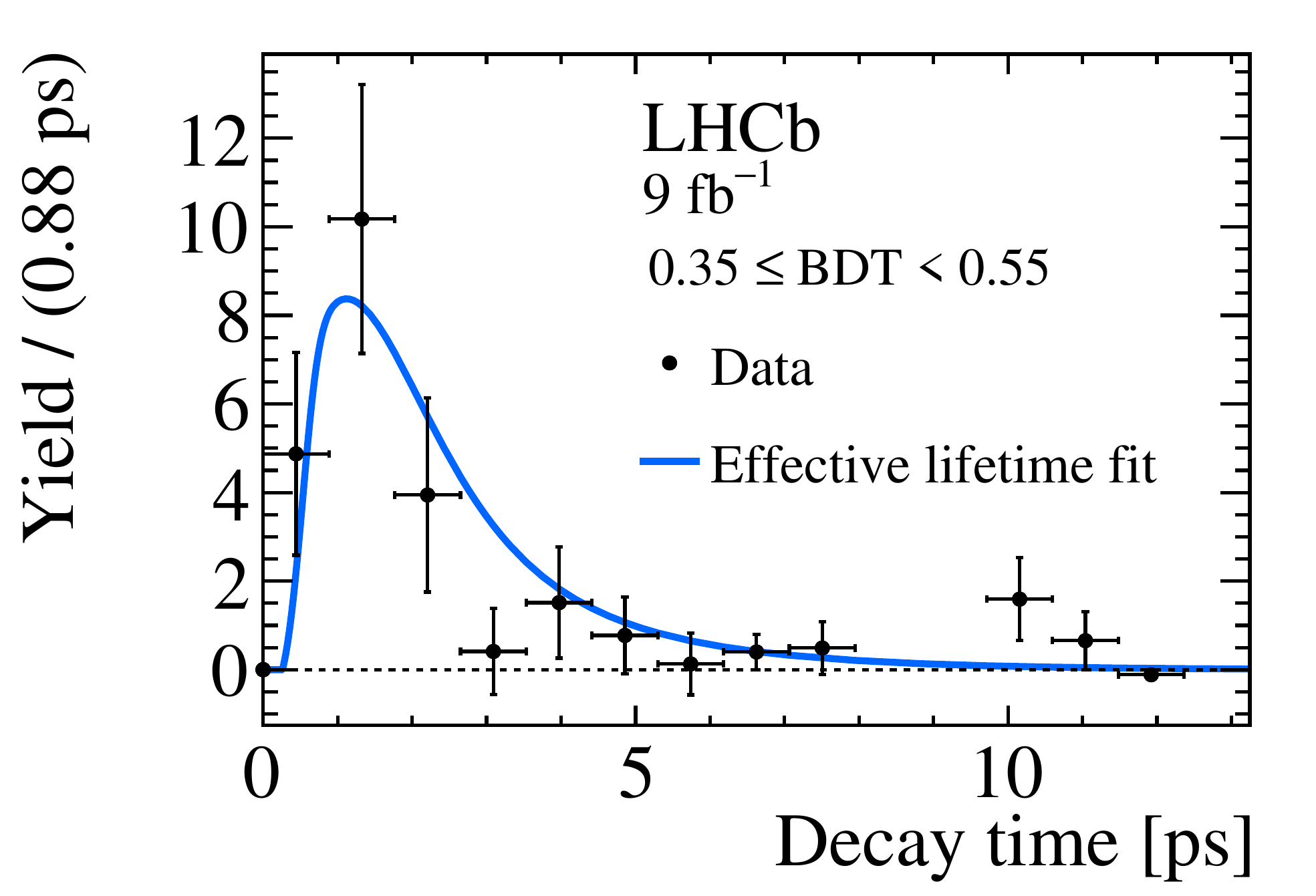}
     \includegraphics[width=0.48\linewidth]{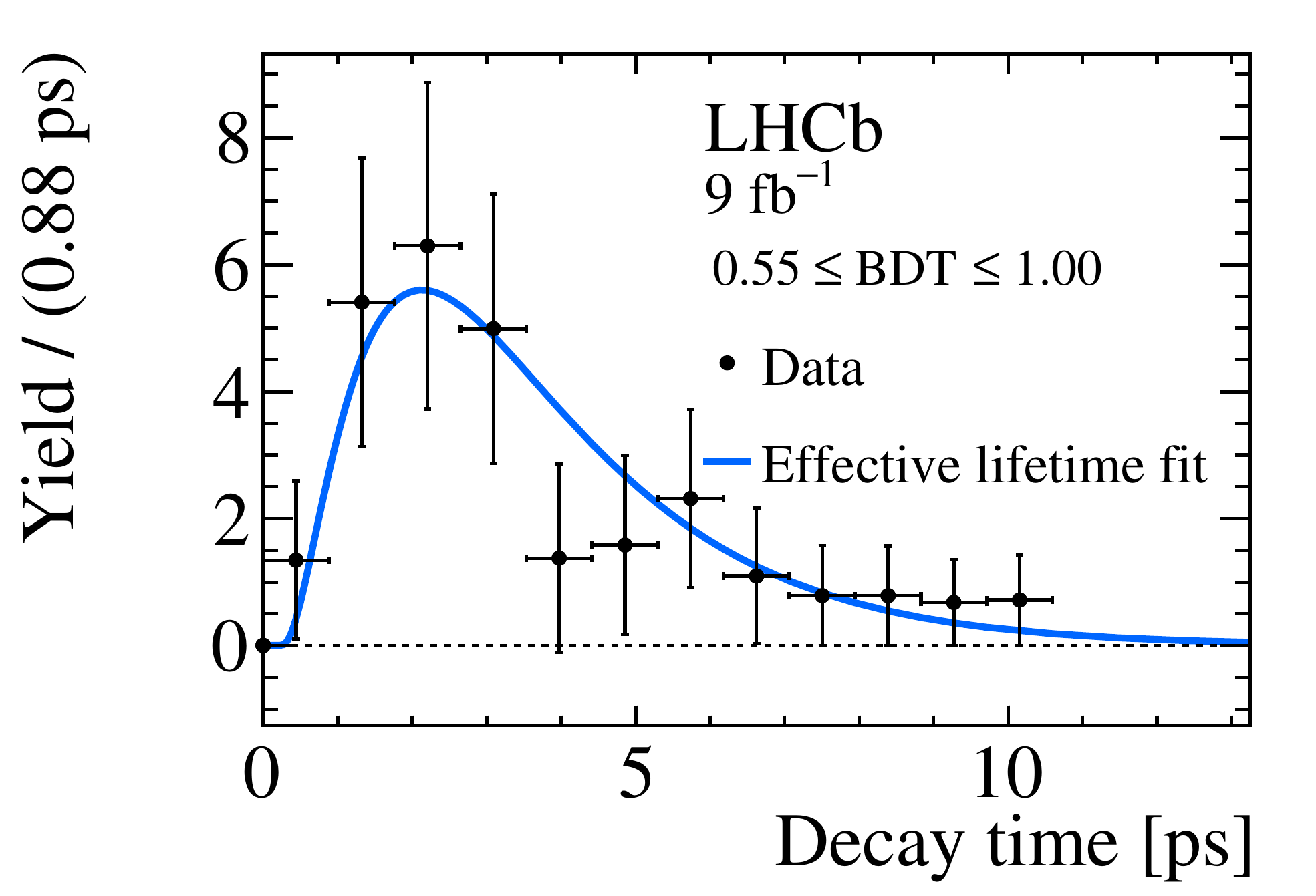}
\caption{Background-subtracted decay-time distributions with the fit model used to extract the \Bsmumu effective lifetime superimposed in the (left) low and (right) high BDT regions.}\label{fig:LT_fits}
\end{figure}

A summary of the systematic uncertainties is reported in Table~\ref{tab:eff_lifetime_syst}. 
\begin{table}[tbp]
    \begin{center}
        \caption{Summary of the systematic uncertainties for the measurement of the \Bsmumu effective lifetime.}
        \label{tab:eff_lifetime_syst}
            \begin{tabular}{cc}
            \toprule                             
            Source & systematic uncertainty on $\tau_{\mu^{+}\mu^{-}}$ (ps) \\ 
            \midrule
            Fit accuracy                         & 0.009 \\
            Background contamination             & 0.012 \\
            Background decay-time model          & 0.003  \\
            Production asymmetry                 & 0.002  \\
            Decay-time acceptance accuracy       & 0.026 \\
            \hline
            Total    &  0.030   \\  %\hline
            \bottomrule 
            \end{tabular}
    \end{center}
\end{table}
Finally, the effective \Bsmumu lifetime is measured with the fits to the decay-time distributions shown in Fig.~\ref{fig:LT_fits}, as
\begin{equation*}\label{eq:Bsmumulifetime}
\tau_{\mu^{+}\mu^{-}} = 2.07 \pm 0.29 \pm 0.03 \ps, \nonumber
\end{equation*}
where the first uncertainty is statistical and the second systematic. While this value lies above the physical range defined by the lifetimes of the light ($A_{\Delta\Gamma} = -1$) and heavy ($A_{\Delta\Gamma} = 1$, predicted by the SM) mass eigenstates, which are $\tau_{L} = 1.423 \pm 0.005\ps$ and $\tau_{H} = 1.620 \pm 0.007\ps$~\cite{PDG2020}, it is consistent with these lifetimes at $2.2\sigma$ and $1.5\sigma$, respectively.

\section{Conclusions}
\label{sec:Conclusions}

In summary, the full \runone and \runtwo data sample of the LHCb experiment was analysed to measure the \bsmumu branching fraction and effective lifetime and to search for the \bdmumu and \bsmumugamma decays. 

The branching fractions of the \Bsmm, \Bdmm and \Bsmumugamma decays are measured to be 
\begin{eqnarray*}\
\BRof \Bsmumu&=&\Bsbr, \\
\BRof \Bdmumu&=&\Bdbr, \\
\BRof \Bsmumugamma&=&\Bsmmgbr\mbox{ with } m_{\mu\mu} > 4.9\gevcc,
\end{eqnarray*}
where the first uncertainty is statistical and the second systematic. The systematic uncertainty on the \bsmumu decay is significantly reduced compared to previous measurements thanks to a new precise value of the hadronisation fraction ratio and a more precise  calibration of the BDT response and of the particle misidentification rate. The \bsmumu branching fraction is the most precise single-experiment measurement to date. 

The \bdmumu and \bsmumugamma signals are not statistically significant, and consistent with the background-only hypothesis at \Bdsigma and $\Bsmumugammasigma\,\sigma$ level, respectively. Therefore, upper limits on the branching fractions are set to 
\begin{eqnarray*}
\BRof \Bdmumu &<& \Bdobslimitnf\\
\BRof \Bsmumugamma &<& \Bsmmgobslimitnf
\end{eqnarray*}
at 95\% CL, the latter with $m_{\mu\mu} > 4.9\gevcc$. The limit on the \bdmumu decay is the most stringent to date from a single experiment. An upper limit on the \bsmumugamma branching fraction is determined for the first time. This limit only constrains the high-$q^2$ region of this decay and no attempt is made here to extrapolate the result to the full branching fraction.  

Using the same data sample, with a slightly different selection, the effective lifetime of the \bsmumu decay is found to be
\begin{equation*}
\tau_{\mu^{+}\mu^{-}} = 2.07 \pm 0.29 \pm 0.03  \ps ,
\end{equation*}
where the first uncertainty is statistical and the second systematic. 

All the results are compatible with the predictions of the SM and with previous measurements of these quantities. In particular, in the two-dimensional space of \mbox{\bsmumu} and \bdmumu branching fractions the compatibility is at one standard deviation level. These results significantly constrain possible contributions to these decays from new interactions that would cause effective scalar, pseudoscalar or axial-vector currents, and thus limit the parameter space of new physics models.

%\clearpage

% Do not include this in any draft (just for information in the template)
%\input{acknowledgements_intro}
% Comment this in for paper drafts; do not include this in analysis note, conference and figure reports
\section*{Acknowledgements}
%
% These Acknowledgements valid from 3-May-2019
%
\noindent We express our gratitude to our colleagues in the CERN
accelerator departments for the excellent performance of the LHC. We
thank the technical and administrative staff at the LHCb
institutes.
We acknowledge support from CERN and from the national agencies:
CAPES, CNPq, FAPERJ and FINEP (Brazil); 
MOST and NSFC (China); 
CNRS/IN2P3 (France); 
BMBF, DFG and MPG (Germany); 
INFN (Italy); 
NWO (Netherlands); 
MNiSW and NCN (Poland); 
MEN/IFA (Romania); 
MSHE (Russia); 
MICINN (Spain); 
SNSF and SER (Switzerland); 
NASU (Ukraine); 
STFC (United Kingdom); 
DOE NP and NSF (USA).
We acknowledge the computing resources that are provided by CERN, IN2P3
(France), KIT and DESY (Germany), INFN (Italy), SURF (Netherlands),
PIC (Spain), GridPP (United Kingdom), RRCKI and Yandex
LLC (Russia), CSCS (Switzerland), IFIN-HH (Romania), CBPF (Brazil),
PL-GRID (Poland) and NERSC (USA).
We are indebted to the communities behind the multiple open-source
software packages on which we depend.
Individual groups or members have received support from
ARC and ARDC (Australia);
AvH Foundation (Germany);
EPLANET, Marie Sk\l{}odowska-Curie Actions and ERC (European Union);
A*MIDEX, ANR, IPhU and Labex P2IO, and R\'{e}gion Auvergne-Rh\^{o}ne-Alpes (France);
Key Research Program of Frontier Sciences of CAS, CAS PIFI, CAS CCEPP, 
Fundamental Research Funds for the Central Universities, 
and Sci. \& Tech. Program of Guangzhou (China);
%Key Research Program of Frontier Sciences of CAS, CAS PIFI,
%Thousand Talents Program, and Sci. \& Tech. Program of Guangzhou (China);
RFBR, RSF and Yandex LLC (Russia);
GVA, XuntaGal and GENCAT (Spain);
the Leverhulme Trust, the Royal Society
 and UKRI (United Kingdom).

\newpage
%\input{appendix}
% This should be taken out in the final paper
%\input{supplementary-app}

\addcontentsline{toc}{section}{References}
%\setboolean{inbibliography}{true}
\bibliographystyle{LHCb}
\bibliography{main,standard,LHCb-PAPER,LHCb-CONF,LHCb-DP,LHCb-TDR}

\newpage
% LHCb collaboration author list
% Data extracted on August 4th, 2021 at 8:28pm for paper reference LHCb-PAPER-2021-008
\centerline
{\large\bf LHCb collaboration}
\begin
{flushleft}
\small
R.~Aaij$^{32}$,
C.~Abell{\'a}n~Beteta$^{50}$,
T.~Ackernley$^{60}$,
B.~Adeva$^{46}$,
M.~Adinolfi$^{54}$,
H.~Afsharnia$^{9}$,
C.A.~Aidala$^{86}$,
S.~Aiola$^{25}$,
Z.~Ajaltouni$^{9}$,
S.~Akar$^{65}$
% ...

{\footnotesize \it

% ...
$^{9}$Universit{\'e} Clermont Auvergne, CNRS/IN2P3, LPC, Clermont-Ferrand, France\\
% ...
$^{25}$INFN Sezione di Milano, Milano, Italy\\
% ...
$^{32}$Nikhef National Institute for Subatomic Physics, Amsterdam, Netherlands\\
% ...
$^{46}$Instituto Galego de F{\'\i}sica de Altas Enerx{\'\i}as (IGFAE), Universidade de Santiago de Compostela, Santiago de Compostela, Spain\\
% ...
$^{50}$Physik-Institut, Universit{\"a}t Z{\"u}rich, Z{\"u}rich, Switzerland\\
% ...
$^{54}$H.H. Wills Physics Laboratory, University of Bristol, Bristol, United Kingdom\\
% ...
$^{60}$Oliver Lodge Laboratory, University of Liverpool, Liverpool, United Kingdom\\
% ...
$^{65}$University of Cincinnati, Cincinnati, OH, United States\\
% ...
$^{68}$Syracuse University, Syracuse, NY, United States\\
% ...
$^{86}$University of Michigan, Ann Arbor, United States, associated to $^{68}$\\
\bigskip
% ...
\medskip
}
\end{flushleft}

%\clearpage
%\input{supplementary}
%\clearpage

\end{document}